\documentclass[longauth]{aa}  

\usepackage{graphicx}
\usepackage{txfonts}
\usepackage{siunitx}
\usepackage{natbib}
\newfont{\sfsl}{cmssqi8 scaled 1200}
\newfont{\sfslp}{cmssqi8 scaled 1250}
\newfont{\sfsls}{cmssqi8 scaled 900}
\newfont{\sfsln}{cmssqi8 scaled 1350}
\newfont{\sfslz}{cmssqi8 scaled 1500}
\newfont{\sfslzz}{cmssqi8 scaled 1150}
\newfont{\sfslms}{cmssqi8 scaled 1000}  
\newfont{\sfsla}{cmssqi8 scaled 3000}
\newfont{\sfslb}{cmssqi8 scaled 2200}

\newcommand{\ro}{{\it ROSAT}}

\newcommand{\xmm}{{\it XMM-Newton}}

\newcommand{\cha}{{\it Chandra}}
\newcommand{\suz}{{\it Suzaku}}

\newcommand{\athena}{{\it Athena}}

\newcommand{\rosi}{{\it eROSITA}}

\newcommand{\om}{\Omega_{\rm m}}

\newcommand{\ol}{\Omega_{\Lambda}}

\usepackage{appendix}

\usepackage[hidelinks]{hyperref}
\makeatletter
\renewcommand*\aa@pageof{, page \thepage{} of \pageref*{LastPage}}
\makeatother
\usepackage{color}

\begin{document} 

   \title{The Abell 3391/95 galaxy cluster system}

   \titlerunning{A3391/95}

   \subtitle{A 15 Mpc intergalactic medium emission filament, a warm gas bridge, infalling matter clumps, and (re-) accelerated plasma discovered by combining SRG/\emph{eROSITA} data with ASKAP/EMU and DECam data}

   \author{
   T.H. Reiprich\inst{1}
   \and
   A. Veronica\inst{1}
   \and
   F. Pacaud\inst{1}
   \and
   M.E. Ramos-Ceja\inst{2}
   \and
   N. Ota\inst{1,18}
   \and
   J. Sanders\inst{2}
   \and
   M. Kara\inst{1}
   \and
   T. Erben\inst{1}
   \and
   M. Klein\inst{3}
   \and
   J. Erler\inst{1,19}
   \and
   J. Kerp\inst{1}
   \and
   D.N. Hoang\inst{4}
   \and
   M. Br\"uggen\inst{4}
   \and
   J. Marvil\inst{5}
   \and
   L. Rudnick\inst{15}
   \and
   V. Biffi\inst{3}
   \and
   K. Dolag\inst{3}
   \and
   J. Aschersleben\inst{1}
   \and
   K. Basu\inst{1}
   \and
   H. Brunner\inst{2}
   \and
   E. Bulbul\inst{2}
   \and
   K. Dennerl\inst{2}
   \and
   D. Eckert\inst{6}
   \and
   M. Freyberg\inst{2}
   \and
   E. Gatuzz\inst{2}
   \and
   V. Ghirardini\inst{2}
   \and
   F. K\"afer\inst{2}
   \and
   A. Merloni\inst{2}
   \and
   K. Migkas\inst{1}
   \and
   K. Nandra\inst{2}
   \and
   P. Predehl\inst{2}
   \and
   J. Robrade\inst{4}
   \and
   M. Salvato\inst{2}
   \and
   B. Whelan\inst{1}
   \and
   A. Diaz-Ocampo\inst{7}
   \and
   D. Hernandez-Lang\inst{3}
   \and
   A. Zenteno\inst{8}
   \and
   M.J.I. Brown\inst{9}
   \and
   J.D. Collier\inst{10,13,16}
   \and
   J.M. Diego\inst{11}
   \and
   A.M. Hopkins\inst{12}
   \and
   A. Kapinska\inst{5}
   \and
   B. Koribalski\inst{10,13}
   \and
   T. Mroczkowski\inst{14}
   \and
   R.P. Norris\inst{10,13}
   \and
   A. O'Brien\inst{13}
   \and
   E. Vardoulaki\inst{17}
          }
   \institute{Argelander-Institut f\"ur Astronomie (AIfA), Universit\"at Bonn,
              Auf dem H\"ugel 71, 53121 Bonn, Germany\\
              \email{reiprich@astro.uni-bonn.de}
         \and
             Max-Planck-Institut f\"ur extraterrestrische Physik, Giessenbachstra{\ss}e 1, 85748 Garching, Germany
         \and
             Ludwig-Maximilians-Universit\"at M\"unchen, Scheinerstra{\ss}e 1, M\"unchen, Germany
         \and
             Universit\"at Hamburg, Hamburger Sternwarte, Gojenbergsweg 112, 21029 Hamburg, Germany
         \and
             National Radio Astronomy Observatory, P.O. Box O, Socorro, NM 87801, USA
         \and
             Astronomy Department, University of Geneva
             Ch.\ d'Ecogia 16, CH-1290 Versoix, Switzerland
         \and
             University of La Serena, Chile
         \and
            Cerro Tololo Inter-American Observatory, NSF's National Optical-Infrared Astronomy Research Laboratory, Casilla 603, La Serena, Chile
        \and
            School of Physics \& Astronomy, Monash University, Clayton, VIC 3800, Australia
        \and
            CSIRO Astronomy \& Space Science, P.O. Box 76, Epping, NSW 1710, Australia
        \and
            Instituto de Física de Cantabria (CSIC-UC), Avda. Los Castros s/n, 39005 Santander, Spain
        \and
            Australian Astronomical Optics, Macquarie University, 105 Delhi Rd, North Ryde, NSW 2113, Australia
        \and
            School of Science, Western Sydney University, Locked Bag 1797, Penrith, NSW 2751, Australia
        \and
            ESO - European Southern Observatory, Karl-Schwarzschild-Str.\ 2, D-85748 Garching b.\ M\"unchen, Germany
        \and
            Minnesota Institute for Astrophysics, University of Minnesota, 116 Church St. SE, Minneapolis, MN 55455, USA
        \and
            The Inter-University Institute for Data Intensive Astronomy (IDIA), Department of Astronomy, University of Cape Town, Private Bag X3, Rondebosch, 7701, South Africa
        \and
            Th{\"u}ringer Landessternwarte, Sternwarte 5, 07778 Tautenburg, Germany
        \and 
            Department of Physics, Nara Women's University, Kitauoyanishi-machi, Nara, 630-8506, Japan
        \and
            Deutsches Zentrum f\"ur Luft- und Raumfahrt e.V.\ (DLR) Projekttr\"ager, Joseph-Beuys-Allee 4, 53113 Bonn, Germany
             }

   \date{Received \dots; accepted \dots}

 
  \abstract
   {Inferences about dark matter, dark energy,
   and the missing baryons all depend on the accuracy of our model of large-scale structure evolution. In particular, with cosmological simulations in our model of the Universe, we trace the growth of structure, and visualize the build-up of bigger structures from smaller ones and of gaseous filaments connecting galaxy clusters.}
   {Here we aim to reveal the complexity of the large-scale structure assembly process  in great detail and on scales from tens of kiloparsecs up to more than 10 Mpc with new sensitive large-scale observations from the latest generation of instruments. We also aim to compare our findings with expectations from our cosmological model.}
   {We used dedicated SRG/\rosi\ performance verification (PV) X-ray, ASKAP/EMU Early Science radio, and DECam optical observations of a $\sim$15 deg$^2$ region around the nearby interacting galaxy cluster system A3391/95 to study the warm-hot gas in cluster outskirts and filaments, the surrounding large-scale structure and its formation process, the morphological complexity in the inner parts of the clusters, and the (re-)acceleration of plasma.
   We also used complementary Sunyaev-Zeldovich (SZ) effect data from the \textit{Planck} survey and custom-made Galactic total (neutral plus molecular) hydrogen column density maps based on the HI4PI and IRAS surveys. We relate the observations to expectations from cosmological hydrodynamic simulations from the Magneticum suite.}
   {We trace the irregular morphology of warm and hot gas of the main clusters from their centers out to well beyond their characteristic radii, $r_{200}$. Between the two main cluster systems, we observe  an emission bridge on large scale and with good spatial resolution. This bridge includes a known galaxy group but this can only partially explain the emission. Most gas in the bridge appears hot, but thanks to \rosi's unique soft response and large field of view, we discover some tantalizing hints for warm, truly primordial filamentary gas connecting the clusters. Several matter clumps physically surrounding the system are detected. For the ``Northern Clump,'' we provide evidence that it is falling towards A3391 from the X-ray hot gas morphology and radio lobe structure of its central AGN. Moreover, the shapes of these X-ray and radio structures appear to be formed by gas well beyond the virial radius, $r_{100}$, of A3391, thereby providing an indirect way of probing the gas in this elusive environment. Many of the extended sources in the field detected by \rosi\ are also known clusters or new clusters in the background, including a known SZ cluster at redshift $z=1$. We find roughly an order of magnitude more cluster candidates than the SPT and ACT surveys together in the same area. We discover an emission filament north of the virial radius of A3391 connecting to the Northern Clump. Furthermore, the absorption-corrected \rosi\ surface brightness map shows that this emission filament extends south of A3395 and beyond an extended X-ray-emitting object (the ``Little Southern Clump'') towards another galaxy cluster, all at the same redshift. The total projected length of this continuous warm-hot emission filament is 15 Mpc, running almost 4 degrees across the entire \rosi\ PV observation field. The Northern and Southern Filament are each detected at $>$4$\sigma$. The \textit{Planck} SZ map additionally appears to support the presence of both new filaments. Furthermore, the DECam galaxy density map shows galaxy overdensities in the same regions. Overall, the new datasets provide impressive confirmation of the theoretically expected structure formation processes on the individual system level, including the surrounding warm-hot intergalactic medium distribution; the similarities of features found in a similar system in the Magneticum simulation are striking. Our spatially resolved findings show that baryons indeed reside in large-scale warm-hot gas filaments with a clumpy structure.
   }
  {}
   \keywords{
Galaxies: clusters: individual: Abell 3391, Abell 3395;
Galaxies: clusters: intracluster medium;
intergalactic medium;
X-rays: galaxies: clusters;
large-scale structure of Universe}
   \maketitle
%
%
\section{Introduction}
\label{intro}
The statistical properties of large-scale structure (LSS) have been studied through the spatial distribution of galaxies, ranging from scales of individual superstructures \citep[e.g.,][]{2018MNRAS.481.1055H} up to the full sky \citep[e.g.,][]{2004PASA...21..396J}. Furthermore, the evolution of the galaxy component and that of the gaseous component have been followed directly on small and large scales through cosmological hydrodynamical simulations
\citep[e.g.,][]{Dolag2005,2017MNRAS.465.2936M,cui2018,2019MNRAS.486.3766M,2019MNRAS.486.4001R,2020MNRAS.tmp.2378P}. This understanding of LSS evolution is a crucial part of our view of the Universe as a whole, and is the basis for all cosmological tests based on the growth of structure, such as the galaxy cluster mass function \citep[e.g.,][]{sr17b,pab19}, cosmic shear \citep[e.g.,][]{2018PhRvD..98d3528T,2020arXiv200715633A}, and the spatial and redshift distribution of galaxies \citep[e.g.,][]{2020arXiv200708991E}. These tests are powerful and complementary to geometric dark energy and dark matter tests, and they are required for us to be able, in principle, to distinguish between modified gravity and dark energy \citep[e.g.,][]{abc06,aab13}. However, clear observational evidence showing the details of this LSS evolution in action in the gaseous component is hard to obtain, as the hot and tenuous intergalactic gas tracing the LSS is difficult to measure over large sky areas with high spatial resolution and sensitivity.

Moreover, our current LSS view allows us, in principle, to understand where much of the normal matter (``baryons'') in the Universe should reside. When we compare the amount of baryons contained in stars and gas in galaxies and galaxy clusters in the nearby Universe, we find they contain only a small fraction ($\sim$10\%) of them \citep[e.g.,][]{fp04}. Most of the ``missing baryons'' are expected to reside in LSS filaments that connect the highest density peaks in the Universe; that is, galaxy clusters \citep[e.g.,][]{co99,2019MNRAS.486.3766M}. The low temperature ($\sim$$10^5$ K) portion of this ``warm-hot intergalactic medium'' (WHIM) has indeed been found through UV absorption spectroscopy of background quasars \citep[e.g.,][]{rpk08}. Also, a new method employing fast radio bursts to constrain WHIM densities  shows promise \citep[e.g.,][]{2020Natur.581..391M}.
Nevertheless, the high-temperature  ($\sim$$10^6$ K) fraction of these missing baryons remains elusive, especially when trying to trace it in  spatially resolved emission because of the lack of sensitive instruments in the soft-X-ray band that can cover large areas efficiently and are simultaneously able to resolve out unrelated background X-ray sources like active galactic nuclei (AGNs).
\begin{figure}[ht]
\centerline{
\includegraphics[width=0.99\columnwidth]{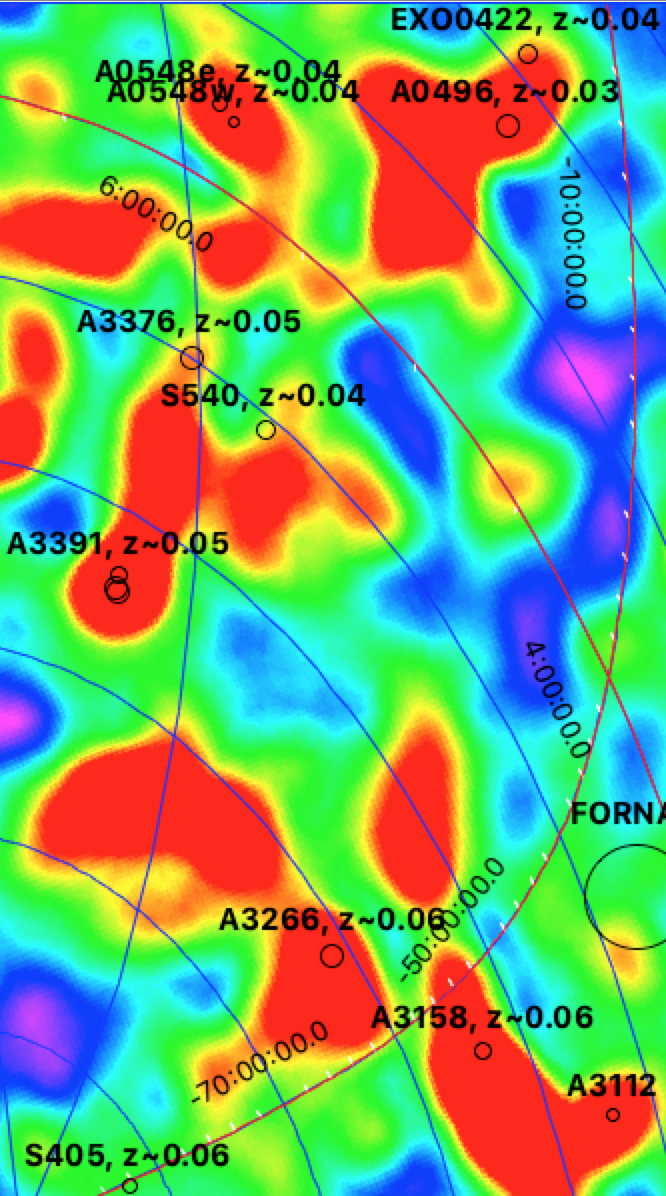}
}
\caption{Smoothed large-scale 2MASS extended source map \citep{jcc00}. At RA=06H this map covers more than 60 deg in Dec. High galaxy number density is shown in red, with low density in blue. Overlaid are positions and apparent sizes of X-ray-selected clusters from the HIFLUGCS sample \citep{rb01,rss03}. The A3391/95n/s system is in the middle-left part of the image (three circles, only A3391 is labeled). Similar overdensities to those visible in this distribution of a large number of near-infrared selected galaxies are also very prominently traced by just a handful of X-ray-selected clusters. There is an indication of a large north--south galaxy filament going right through A3391/95.
\label{fig:2MASS}}
\end{figure} 

There have been a number of claimed hot-phase WHIM detections in X-ray absorption that have been challenged by other teams \citep[e.g.,][]{nme05,wmn06,rkp06,kwd06,klt03,bl06,bzf09,fbh10}. X-ray absorption measurements mostly rely on oxygen absorption; they depend on the presence of bright background X-ray-emitting quasars or gamma-ray bursts with known intrinsic spectra and therefore do not allow us to study the WHIM in a spatially resolved manner. One of the primary science drivers for the upcoming ESA Large Mission \athena\ \citep[e.g.,][]{nbb13} is to characterize the WHIM in detail through a large number of such absorption measurements \citep[e.g.,][]{kfn13}. 
Recently, also UV absorption measurements of iron have been proposed to trace the hot WHIM phase \citep{2020arXiv200911346F}.
According to our expectations from detailed simulations, the best places to search for LSS activity such as infalling substructures and the hot dense fraction of the WHIM is close to nearby galaxy clusters, in their outskirts, and in the presumed warm filaments connecting them. Therefore, searching in the direction of superclusters, between galaxy clusters, or within large-scale galaxy structures or galaxy pairs using X-ray and and Sunyaev-Zeldovich (SZ) observations seems to be the currently most promising way forward (e.g., \citealt{fth07,wfk08,fbh10,dwc12,pla13,ntl15,ejs15,2016ApJ...818..131B,2017MNRAS.470.3742P,afa17,arb18,thm19,dch19,2020MNRAS.491.2605P,tab20}; Ghirardini et al., subm.).

In this paper, we provide a detailed spatially resolved view of LSS evolution, employing new \rosi\ X-ray data together with new DECam optical data, new ASKAP/EMU radio data,
custom-made total hydrogen column density maps based on 21cm (HI4PI) and IRAS data, and also \textit{Planck} SZ data.
All of these datasets provide a unique combination of high sensitivity and high spatial resolution over large sky areas.
We bring the power of these data sets to bear on a well-known nearby ($z\approx0.05$) triple galaxy cluster merger system, Abell 3391/Abell 3395n/Abell 3395s (e.g., \citealt{rb01}; A3391/95 from now on). The large-scale galaxy environment is shown in Fig.~\ref{fig:2MASS}, indicating that A3391/95 is part of a very large structure that runs from north to south in projection.  Below, observations are compared to our expectations from detailed hydrodynamical cosmological simulations from the Magneticum project.

The A3391/95 system has been studied in detail at a range of wavelengths and with a number of instruments.
Contrary to early ASCA studies \citep{th01}, the more recent \xmm, \cha, and \suz\ studies of this system concluded that the apparent emission excess between A3391 and A3395 is likely due to cluster gas that has been tidally stripped from the clusters because of the interaction process \citep[][]{sti17,arb18}. This conclusion is predominantly based on the lack of detection of warm (k$T<1$ keV) gas in the emission ``bridge.'' While only poorly resolved, the presence of an emission bridge in the \textit{Planck} data of this system \citep[][]{pla13} also indicates hotter gas temperatures rather than lower temperature true primordial WHIM. \rosi\ has a larger effective area and higher energy resolution for extended sources at soft X-rays than any other past or present mission \citep[][]{mpb12,pab14,paa20}. Moreover, \rosi's large field of view (FoV; $\sim$1 deg diameter) and scanning mode allow us not only to efficiently cover the entire system, but crucially also to characterize the foreground emission from our galaxy in this area, which can be expected to have a similar temperature as the possible WHIM. Therefore, \rosi\ should allow us to either detect a true WHIM filament or to unambiguously rule out the presence of significant amounts of warm gas in the bridge region. Both findings would have a significant impact on our understanding of the formation and evolution of the dense WHIM phase. Furthermore, the large field covered deeply in \rosi's scanning mode may reveal serendipitous LSS discoveries in the environment of the A3391/95 system, such as for example low surface filaments in other directions.

Unless stated otherwise, $\om=0.3$, $\ol=0.7$, and $H_0=70$ km/s/Mpc are assumed throughout, and $h$ is defined by $H_0=100h$ km/s/Mpc. At the A3391 redshift, $z=0.0555$, and 1 arcsec therefore corresponds to 1.078 kpc.

\section{Data reduction and analysis}
\label{data}

\subsection{\rosi}
\label{rosi}

\rosi\ is a new X-ray telescope launched on July 13, 2019, onboard the Spectrum-Roentgen-Gamma (SRG, Sunyaev et al., in prep.)\ satellite. It will perform eight all-sky surveys with a total sensitivity improved by at least a factor of 20 compared to the only existing imaging all-sky survey, performed by the \ro\ satellite 30 years ago. A comprehensive description of \rosi\ is provided by \citet{paa20}. 
The science enabled by \rosi\ has been summarized by \citet{mpb12,2020NatAs...4..634M}.
The  main science driver of \rosi\  is the study of the nature of dark energy by discovering and characterizing about 100 000 galaxy clusters. Forecasts for expected cosmological constraints, for the number of clusters to be detected, for the precision of cluster temperature measurements and X-ray redshifts, for the selection function, and so on,\ are available \citep{ppr12,brm14,hsc17,prp18,crr18,zfp18,gmd18,2020arXiv200808404C}.

\subsubsection{Data preparation}
\label{sect-general-properties}
The A3391/95 field was targeted during the \rosi\ Performance Verification (PV) phase in October 2019.
There are four \rosi\ PV phase observations of A3391/95:  one pointed observation and three raster scan observations. During each raster scan, \rosi's aim point scanned a square region. Given the FoV of 1 deg across and the slightly different centers and orientations of the scans, a total area of about 15 deg$^2$ was covered with at least 30 s exposure ($\sim$10 deg$^2$ with at least 1000 s).
Scan III was performed with all cameras of all seven Telescope Modules (TMs) switched on while for the other observations TMs 5--7 were on. Information about these observations is listed in Table \ref{tab:obsID}.

\begin{table}[ht]
\renewcommand\thetable{1}
\centering
\caption{\rosi\  observations of the A3391/95 field}
\label{tab:obsID}
\begin{tabular}{@{}cccc@{}}
\hline
\hline
\bf Name & \bf ObsID & \bf TM & \bf Comment\\
\hline
A3391/3395\_Point & 300014  & 5--7 & 40 ks pointing \\
A3391/3395\_I & 300005  & 5--7 & 60 ks scan\\
A3391/3395\_II & 300006  & 5--7 & 60 ks scan\\
A3391/3395\_III & 300016  & 1--7 & 60 ks scan\\
\hline
\end{tabular}
\end{table}

Data reduction of each TM in each observation (16 data sets in total) was done using the extended Science Analysis Software System (eSASS, Brunner et al., in prep.). The first step
was to generate clean event files and images using the \texttt{evtool} task. This was done by specifying \texttt{flag=0xc00fff30}, which removes bad pixels and the strongly vignetted corners of the square CCDs, and \texttt{pattern=15} to select single, double, triple, and quadruple patterns.

The next step was to filter out any possible soft proton flares in the observations. While SRG/\rosi\ is in orbit around the Sun--Earth Lagrange point 2 (L2), in principle, soft proton flares similar to those observed by \xmm\ \citep[e.g.,][]{rsk04} could be present. We extracted light curves using the eSASS task \texttt{flaregti} in the energy range 6--10 keV. This energy range was chosen to limit the fraction of focussed X-rays in the events; the effective area above 6 keV is rather small ($\lesssim$15 cm$^2$ per TM). This is important because  the sources in the FoV vary from time bin to time bin during
the scans.

No flares were detected with this procedure. However, count rate drops were found in the beginning half of the observations for some TMs; that is, TM6 in observation 300005 and 300006, and TM1, 2, and 6 in observation 300016 (the individual light curves are shown in Appendix \ref{App_A}). These drops signify unaccounted-for exposure loss and were found to occur during the scans spatially close to the star Canopus, the optically
second brightest star in the sky,  with $-$0.5 mag. This exposure drop seemed to be caused by event rates exceeding the telemetry threshold, presumably due to optical and/or UV photons penetrating through the optical blocking filters (Canopus is an A star, and is therefore not an intrinsically strong X-ray emitter). Once this threshold is exceeded for a given camera, no image frame data are transmitted to ground for a short time period. This exposure loss was apparently not propagated through the subsequent eSASS tasks in the version available at the time of data reduction, including the light curve and the exposure map tasks. To ensure that the actual exposures are properly taken into account, we simply excluded the time intervals when these drops occurred from all further steps. These exposure losses were not observed in TM5 and TM7 because their low-energy ADU thresholds had been increased before to alleviate the impact of the light leak present in these two cameras without on-chip filter (more details about light leaks in Section~\ref{image} below and in \citealt{paa20}).

New good time interval files (GTIs) were generated and combined with the eSASS-generated GTIs. Afterwards, the \texttt{evtool} task with argument \texttt{gti="GTI"} was run to apply the combined GTIs to the event lists.

\subsubsection{Image extraction}
\label{image}

To investigate possible emission from warm gas, we focussed on the lowest energies available from \rosi,\ which for most TMs means the 0.3--2 keV band.
After launch, it was found that optical photons from the Sun can reach the cameras, creating apparent soft X-ray photon events through pile-up, with the number of events depending on orbital parameters, pointing direction, and detector position. While those optical photons are filtered out by the on-chip filters of the TM1--4 and TM6 cameras, this is not the case for TM5 and TM7, as these do not possess optical blocking filters directly on their chips (only in their filter wheels). As a result, fake X-ray events with energies $\lesssim$0.5 keV are generated in many observations with TM5 and TM7, in rare cases also up to $\sim$0.8 keV. This source of contamination is usually referred to as `light leak.' As we search for low-surface brightness features in the final images, all image processing for TM5 and TM7 was done in the energy range 1.0--2.0 keV, instead of 0.3--2.0 keV, to prevent the light leak from affecting our conclusions. Hereafter, we refer to this TM-dependent energy selection as the ``soft band.''

Soft-band images and exposure maps were generated for each TM individually using the combined GTIs described in Section~\ref{sect-general-properties}. The CCD corners were excluded from the exposure maps by specifying \texttt{withdetmap=yes} in the \texttt{expmap} task and we created both vignetted and
nonvignetted maps. The exposure maps are weighted; that is, all of the exposure maps from  individual TMs were divided by 7. In the following, we use a specific notation for the combination of different telescope modules, namely TM8=sum(TM1,2,3,4,6), TM9=sum(TM5,7), and TM0=TM8+TM9.

\subsubsection{Particle background subtraction}
\label{data:PIB}
In order to create a PIB-subtracted image, the PIB was first modeled for each TM in each observation. The modeling is based on some key elements concluded from our extensive studies of \rosi\ Filter-Wheel-Closed (FWC) observation data as follows.

As the spatial variation of the PIB appears to be small ($\lesssim$$10\%$),
the modeling was done using the nonvignetted/flat exposure map. Moreover, the temporal variability of the PIB spectral shape seems very small, and therefore the count rate in a hard band strongly dominated by PIB events, for example 6--9 keV, can be converted to the PIB count rate in any other band.\footnote{Currently, it is advisable to limit the upper energy for the purposes here to $\lesssim$9 keV, as the basic upper energy thresholds to limit telemetry volume are applied in ADU and due to gain variations this can result in different thresholds in units of keV for different columns of the same camera, for  different cameras, and possibly at different times.}
Hence, the estimation of the PIB for a given TM in a given observation was done in the following steps: Firstly, the number of PIB counts contributed in a given soft band $S_{\rm obs}$, for example 0.3--2.0 keV, was calculated by multiplying the hard band counts $H_{\rm obs}$, for example 6.0--9.0 keV, in the observation by the ratio $R$ of the number of counts in the 0.3--2.0 keV band, $S_{\rm PIB}$, and in the 6.0--9.0 keV band, $H_{\rm PIB}$, of the FWC data; that is, $S_{\rm obs}=H_{\rm obs}S_{\rm PIB}/H_{\rm PIB}\equiv H_{\rm obs}R$. The resulting number was then spatially distributed by multiplying it to the nonvignetted exposure map, which was normalized to 1 by dividing each pixel by the sum of all pixel values, sum(\text{nonvig.\ exposure map}). The values of $H_{\rm obs}$ and $R$ are listed in Table \ref{tab:HR} in Appendix~\ref{App_B0}.
The calculation is then summarized by 
\begin{equation}
\text{Background Map} = \frac{\text{nonvig.\ exposure map}}{\text{sum}(\text{nonvig.\ exposure map})} \times H_{\rm obs} \times R\;.
\end{equation}
Based on the FWC data available at this time, the temporal variation of the PIB normalization is only about $6\%$. With about 20000 hard-band counts available per observation, this rescaling procedure allowed us to 
further improve and constrain the amplitude of the PIB to $<$$1\%$.

This step gave rise to PIB maps for each TM and observation, which were then co-added to obtain the complete PIB map. The PIB-subtracted image was acquired by subtracting the complete PIB map from the combined photon image.

\subsubsection{Exposure correction}
\label{data:expo}

The field features an inhomogeneous coverage by the different TMs, especially at the edges of the field and around the star Canopus. Furthermore, TM8 and TM9 have different soft responses because of the different filter setups: TM8 has 200 nm Pl in the filter wheel plus 200 nm Al on-chip and TM9 has 200 nm Pl plus 100 nm Al both in the filter wheel.
More importantly, we used different energy bands for TM8 and TM9 to simultaneously maximize the signal and avoid any impact from the light leak: 0.3--2.0 keV for the former and 1.0--2.0 keV for the latter. The inhomogeneous coverage together with the different responses and energy bands made it necessary to determine and apply an additional correction factor when combining all TMs and all observations together.

\begin{figure*}[ht]
\centering
\centerline{\includegraphics[width=2.0\columnwidth]{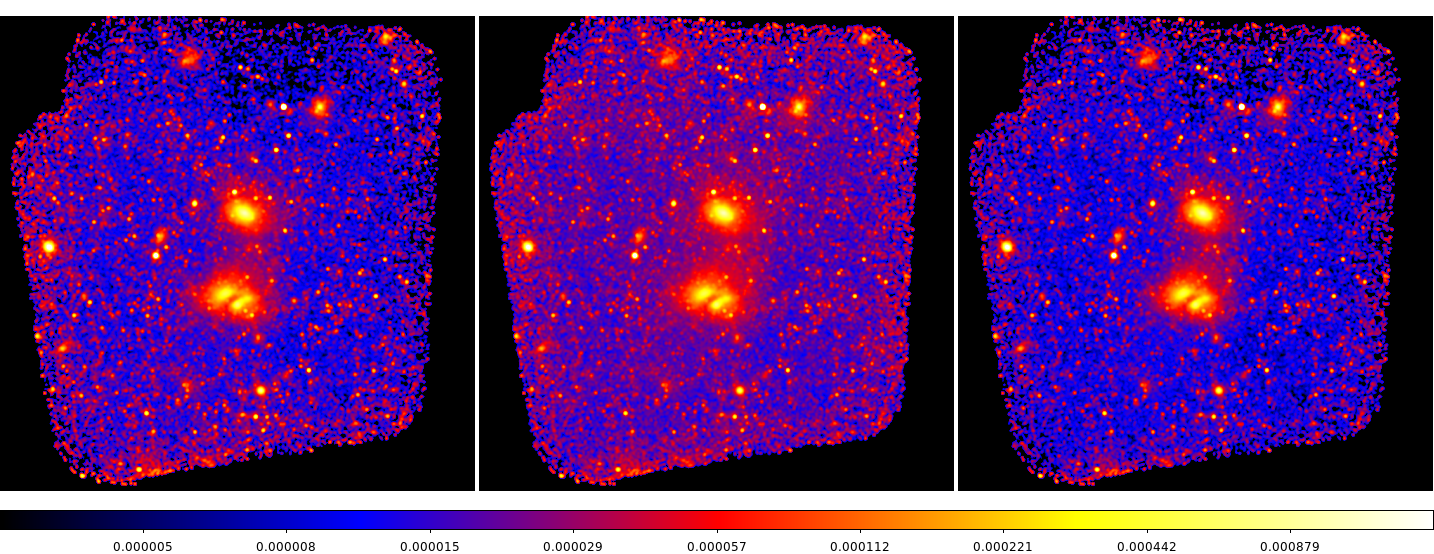}}
\caption{Vignetting-corrected count rate images of combined observations in energy band 0.3--2.0 keV for TM8 and 1.0--2.0 keV for TM9. \textit{Left}: PIB-subtracted but without correction factor applied to the TM9 exposure map. For visualization purposes, this image has been normalized such that the peak values of A3391 in this image and the final image (right) have the same color. \textit{Middle}: Without PIB-subtraction but with TM9 exposure correction. \textit{Right}: PIB-subtracted and with TM9 exposure correction. We note that, after correction, edge effects and exposure loss effects due to the optically bright star Canopus (the brightest spot in the north) are strongly reduced.
The image on the right is the final image after all corrections. The enhanced emission towards the lower left is real and due to a galaxy cluster just outside the covered field, but clearly visible in eRASS:1.
\label{fig:CLCR}}
\end{figure*}

The expected count rate for TM9, for which we used a narrower energy band, is lower and introduces an underestimation of the count rate in the Canopus region which is only covered by TM9 of 300005 and 300006, and an overestimation in the left and bottom sides of the combined observations image which are only covered by 300016 (for which the contribution from TM9 is smallest; see left panel of Figure~\ref{fig:CLCR}).

As the source and X-ray background spectra vary across the FoV, this count rate bias is spatially variable. More importantly, the exposure maps for TM8 and TM9 differ, creating a strong spatially variable bias.

Accounting for these effects in full precision requires a very elaborate method. Here, we implemented a first-order correction by treating the combined exposure maps of TM8 and TM9 separately and by using the TM9-to-TM8 ratio of full FoV (excluding the main clusters) PIB-subtracted count rates for estimating the needed correction factor. This can be formulated as %
\begin{equation}
\text{CORR} = \frac{\text{BGSUB count rate TM9}}{\text{BGSUB count rate TM8}}\;,
\end{equation}
where CORR is the correction factor and BGSUB stands for PIB-subtracted. The effective correction and exposure map calculation imply that the count rates of the final combined image correspond to an effective area given by seven TMs with on-chip filter in the energy band 0.3--2.0 keV.

The combined exposure map of all observations with all TMs (TM0) was then calculated as
\begin{equation}
\text{exp. TM0} = \text{exp. TM8 }  +  \text{ CORR}\times\text{exp. TM9}\;.
\end{equation}

The correction factor CORR for the combined energy bands 0.3--2.0 keV for TM8 and 1.0--2.0 keV for TM9 is
0.36.
The final exposure map for all observations and TMs after all corrections is shown in the right panel of Fig.~\ref{fig:expmap_TM0} in Appendix \ref{App_B}; unless noted otherwise, this is the one used for all \rosi\ images shown in this paper.
Also, further details and exposure maps showing different features for different TMs and observations are provided in Appendix \ref{App_B}.

The PIB-subtracted and exposure-corrected image was generated by dividing the combined PIB-subtracted photon image by the corrected exposure map and is shown in the right panel of Figure \ref{fig:CLCR}. Figure \ref{fig:CLCR} also shows the exposure-corrected count rate image before PIB subtraction (middle panel).

\subsection{DECam}
\label{DEC}
The A3391/95 system and its surroundings were optically observed with the Dark Energy
Camera \citep[DECam; e.g.,][]{Flaugher_2015} as part of the Dark
Energy Survey \citep[DES; e.g.,][]{10.1093/mnras/stw641}.  Within
the DeROSITAS project (PI A. Zenteno), we  obtained additional DECam observations
in the five optical filters $u$,$g$,$r$,$i$, and $z$
on December 12, 2018, 
January 23, 2019, 
January 24, 2019, 
December 11, 2019, 
December 12, 2019, 
and
January 31, 2020. 
These data cover
primarily the region $\SI{94.0}{\degree} \leq R.A. \leq
\SI{99.0}{\degree}$ and $\SI{-55.5}{\degree} \leq Dec \leq
\SI{-53.0}{\degree}$. The main purpose of the additional data is to
have deeper coverage around the core-cluster region of A3391/95 and
to include the $u$-band for the estimation of photometric redshifts.

Our reduction of DECam data starts from raw data provided by the NOAO
archive.\footnote{\url{https://astroarchive.noao.edu}} Most
of the processing algorithms used are similar to those initially
developed for the wide-field imager on the ESO 2.2-m telescope at La
Silla, as described in \citet{2005AN....326..432E}.  A more in-depth
description with tests on the current \textsc{Theli} DECam data
products
will be published in Kara et al.~(in prep.).

\begin{figure}[ht]
  \centering
  \includegraphics{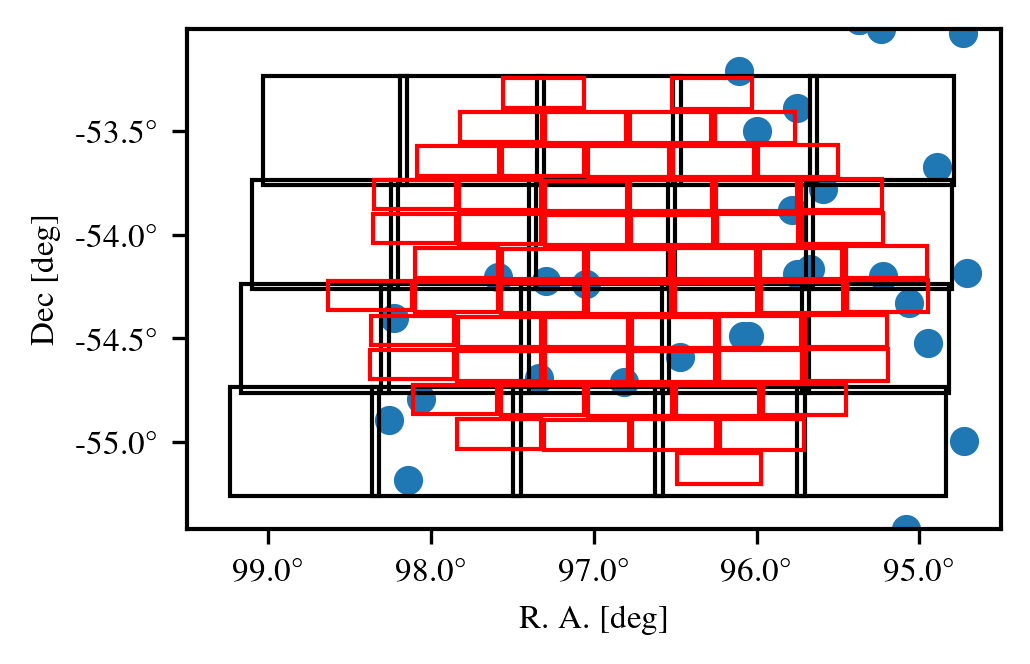}
  \caption{Layout of our DECam tiles with an area of $0.25$
    square degrees (black squares), the location of individual
    $r-$band exposures (blue dots), and the chips of one 62-chip
    raw DECam image (some chips are missing due to very bright
    sources or technical problems).
    \label{fig:DECam_layout}}
\end{figure}

In the following, we provide only a brief overview of the processing steps of
our DECam data reduction procedure.
\begin{enumerate}
 \item In addition to our own observations, the basis for our
    processing is formed by {all} publicly available DECam data
    around A3391/95 at the time of processing. We use additional
    archival data to allow for better flatfield correction and
    photometric calibration. All data are retrieved from the NOAO data
    archive.
  \item Science data are corrected for nonlinearity and crosstalk
    effects. Coefficients for these corrections are provided by
    Frank Valdes on the NOAO Web pages.\footnote{\url{http://www.ctio.noao.edu/noao/content/DECam-Calibration-Files}} We
    also correct our data for the so-called brighter-fatter effect
    \citep{2015JInst..10C5032G}. A correction scheme, code, and necessary
    data products were kindly provided by Daniel Gruen.
  \item The characterization and removal of the instrumental signature
    (bias, flat field, and illumination correction) is performed
    simultaneously on all data from a two-week period around each
    new-moon and full-moon phase. Each two-week period of dark or
    bright time defines a DECam processing run (see also Section 4 of
    \citealt{2005AN....326..432E}), over which we assume that the
    instrument configuration is stable.  The processing run definition
    by moon phase also naturally corresponds to the observations with
    different filters ($u$,$g$,$r$ in dark time and $i$,$z$
    in bright time).
  \item Photometric zero-points, atmospheric extinction coefficients,
    and color terms are estimated per complete processing run.  These
    are obtained by calibration of {all} science observations
    in a run that overlap with SDSS Data Release 12 
    \citep{2015ApJS..219...12A}. Individual exposures with a high
    atmospheric extinction of more than
    0.2 magnitudes are excluded
    from all further processing.
  \item If necessary we correct individual DECam CCDs for artefacts caused
    by very bright stars (saturated regions on the edges of CCDs).
  \item As the last step of the run processing, we subtract the sky
    from all individual chips.
  \item DECam is a 62-chip camera with a hexagonal layout. We subdivide the sky
    in regions of
    0.25 square degrees. We henceforth refer to regions whose neighbors
    have a slight overlap of about \SI{10}{\percent} as DECam tiles.
    Figure~\ref{fig:DECam_layout} visualizes this setup.
  \item All science CCDs belonging to a given DECam tile are
    astrometrically calibrated against the GAIA-DR2 catalog
    \citep{refId0}.
  \item After rejecting problematic images (high PSF ellipticity, high seeing),
    the astrometrically calibrated data are co-added with a
    weighted mean algorithm.  The identification of pixels that should
    not contribute, for example those affected by cosmic rays, and
    weighting of usable pixels is determined as described in
    \citet{2013MNRAS.433.2545E}.
  \item Finally, \textsc{SExtractor} \citep{1996A&AS..117..393B} is
    run on the co-added image to generate a source catalog for
    matched-aperture photometry and later photometric redshifts.
\end{enumerate}

Due to the very complex relation between tiles and individual DECam
exposures, it is not possible to find extended areas of homogeneous
data quality. Table~\ref{tab:characteristics} lists average parameters
for image seeing (Gaussian fit to point-like sources) and the limiting
magnitude of our tiles. The limiting magnitude is defined as the
5$\sigma$ detection limit in a \SI{2.0}{\arcsec} aperture via $m_{\rm
  lim}=ZP-2.5\log(5\sqrt{N_{\rm pix}}\sigma_{\rm sky})$, where $ZP$ is
the magnitude zeropoint, $N_{\rm pix}$ is the number of pixels in a
circle with radius \SI{2.0}{\arcsec}, and $\sigma_{\rm sky}$ is the sky
background noise variation.
\begin{table}
  \caption{Average quality parameters of our final DECam science tiles (see text
    for an explanation of the columns).}           
\label{tab:characteristics}      
\centering          
\begin{tabular}{llll}    
\hline\hline       
\multicolumn{1}{c}{Filter}  &
\multicolumn{1}{c}{$m_{\rm lim}$ [AB mag]} & \multicolumn{1}{c}{seeing [$''$]}\\ 
\hline
$u$ & 23.50 & 1.10\\ 
$g$ & 25.40 & 1.10\\ 
$r$ & 25.30 & 0.95\\ 
$i$ & 25.27 & 0.95\\ 
$z$ & 25.00 & 0.85\\ 
\hline                  
\end{tabular}
\end{table}

The resulting optical color image is shown and described in Section~\ref{LSS}.

For the construction of the large-scale galaxy density map shown at the end of Section~\ref{filaments}, in order to ensure a more homogeneous depth over the full area, we make use of the publicly available DECam based catalog from the DECaLS survey, which is part of the Legacy Survey data release 8 \citep[DR8;][]{2019AJ....157..168D} and the galaxy density map tool as part of the multi component matched filter cluster confirmation code \citep[MCMF;][]{2018MNRAS.474.3324K, 2019MNRAS.488..739K}. The DECaLS survey  makes use of existing public  DECam data as well as its own in the $g$, $r$, $z$ bands. Particularly for the region here, this means data from the Dark Energy Survey \citep{10.1093/mnras/stw641}.
MCMF is designed to identify optical counterparts to X-ray- or SZ-detected cluster candidates. It estimates the probability of being a chance superposition  for those clusters and derives precise redshifts ($\Delta z/[1+z]=0.006$). It makes use of the characteristic redshift-dependent color--magnitude relation of cluster member galaxies known as the red sequence. In particular, it calculates the distance of each galaxy from the red sequence in multiple colors and assigns weights to them according to agreement with a red sequence model for a given redshift. We use these weights to create galaxy density maps for galaxies brighter than $19.7$ mag in the $z$ band at the redshift of the clusters. The map is further smoothed by a Gaussian kernel of 150 kpc in size. Although only using $g$, $r$, and $z$ bands, the applied weights effectively suppress the signal of structures at higher redshifts than that of A3391/95. The signal in the galaxy density map of a cluster at $z=0.07$ in the A3391/95 field is already reduced by a factor two compared to the signal in a galaxy density map at the cluster redshift. Other clusters with $z>0.12$ are approximately  consistent with the noise in the galaxy density map.

\subsection{ASKAP/EMU}
\label{EMU}
The A3391/95 field was observed with the Australian Square Kilometre Array Pathfinder (ASKAP; e.g., \citealt{2008ExA....22..151J}; Hotan et al., subm.)\
during the Early Science period in collaboration with the Evolutionary Map of the Universe (EMU; \citealt{nha11}) survey. The observations took place on 22 March 2019 with an integration time of 10 hours using 35 antennas and achieving a resolution of $\sim$$10$ arcsec over an instantaneous field of view of approximately 30 deg$^2$. The data were taken in a pseudo-continuum mode consisting of 1 MHz channels, with a total correlated bandwidth of 288 MHz centered at 990 MHz. A preceding observation of the source PKS B1934-638 was used for flux density, phase, delay and bandpass calibration. 

Data reduction was carried out using the ASKAPsoft\footnote{\url{https://www.atnf.csiro.au/computing/software/askapsoft/sdp/docs/current/}} software package running on the \textit{Galaxy} cluster at the Pawsey Supercomputing Centre. Calibration, flagging (to excise radio-frequency interference and instrumental issues), and imaging operations were performed separately on each of the 36 beams of the ASKAP phased array feed. Imaging employed standard wide-field (i.e., w-projection) and wide-band (i.e., multi-term deconvolution) techniques and included one loop of phase-only self-calibration. The resulting 36 images were combined using the ASKAPsoft linear mosaic utility.
The measured rms in areas free of artefacts is $\sim$30 microJy/beam. Further details about the radio observations, data processing, and data characteristics are presented in the accompanying paper focussed on these data
\citep{brb20}.

\subsection{Planck}
\label{SZ}

The hot electrons of the intracluster medium (ICM) can interact with the photons of the cosmic microwave background (CMB) through inverse Compton scattering, giving rise to a CMB spectral distortion known as the SZ effect \citep[e.g.,][]{zs69,Sunyaev70, Sunyaev72, Birkinshaw99, Carlstrom02, Mroczkowski19}. The SZ effect is commonly broken down into several components, the most important of which is the thermal SZ (tSZ) effect. The tSZ effect is caused by the isotropic scattering of CMB photons by the thermal population of ICM electrons.
The tSZ effect is observed at millimeter and submillimeter wavelengths as a decrement in observed CMB temperature below $\sim$$217 \, \mathrm{GHz}$ and an increment at frequencies above. The tSZ effect has been observed directly for a few thousand galaxy clusters \citep[e.g.,][]{Hasselfield13, Bleem15, Planck_PSZE, Planck_PSZE2,2020ApJS..247...25B,2020arXiv200911043H}.
The CMB temperature anisotropy caused by the tSZ effect is commonly written as
\begin{equation}
 \frac{\Delta T_\mathrm{tSZ}}{T_\mathrm{CMB}} = f(x, T_\mathrm{e}) \, y,
\end{equation}
where $T_\mathrm{CMB}$ is the CMB temperature, $x$ is the scaled frequency $x=(h_\mathrm{p}\nu)/(k_\mathrm{B}T_\mathrm{CMB})$, with $h_\mathrm{p}$ the Planck constant and $k_\mathrm{B}$ the Boltzmann constant, $f(x, T_\mathrm{e})$ is the relativistic tSZ spectrum \citep[e.g.,][]{Wright79, Itoh98, Chluba12}, and $y$ is the Comptonization parameter, which is proportional to the line-of-sight integral of the ICM pressure $P_\mathrm{e}$
\begin{equation}
y(r) = \frac{\sigma_\mathrm{T}}{m_\mathrm{e}c^2}\int_\mathrm{l.o.s.} \mathrm{d}l \, \underbrace{n_\mathrm{e}(r)\, k_\mathrm{B}\, T_\mathrm{e}(r)}_{P_\mathrm{e}(r)}\,,
\end{equation}
where
$\sigma_\mathrm{T}$ is the Thomson cross-section, $m_\mathrm{e}$ is the electron rest mass, and $n_\mathrm{e}$ and $T_\mathrm{e}$ are the number density and temperature of the electrons in the ICM.  

One of the most powerful observatories for the study of nearby extended galaxy clusters through the SZ effect is the \textit{Planck} satellite \citep{Planck_overview}. \textit{Planck} observed the microwave sky for four years from the second Sun--Earth Lagrange point (L$_2$) in nine frequency bands from $30 \, \mathrm{GHz}$ up to $857 \, \mathrm{GHz}$, covering the full spectral range of the SZ effect. We extract a $4^\circ \times 4^\circ$ map of the Comptonization parameter $y$ of Abell 3391/95 from the Modified Internal Linear Combination Algorithm \citep[MILCA;][]{Hurier13} all-sky $y$-map which was published by the \citet{Planck_YMAPS}. The MILCA algorithm computes an optimal linear combination of the \textit{Planck} multifrequency data with individual weights for each frequency channel, which are tuned to preserve the characteristic spectrum of the tSZ in its nonrelativistic approximation, while minimizing the variance of the desired Compton $y$-map \citep[ILC;][]{Bennett03, Remazeilles11_CILC, Remazeilles11_NILC}. The optimal linear combination is performed in 11 optimized spatial windows in spherical harmonic space, allowing for both spectral and spatial localization of the tSZ effect of galaxy clusters. The resulting map has a spatial resolution of $10 \, \mathrm{arcmin}$ and is shown 
at the end of Section~\ref{filaments}.
   \begin{figure*}
   \centering
    \includegraphics[width=1.9\columnwidth]{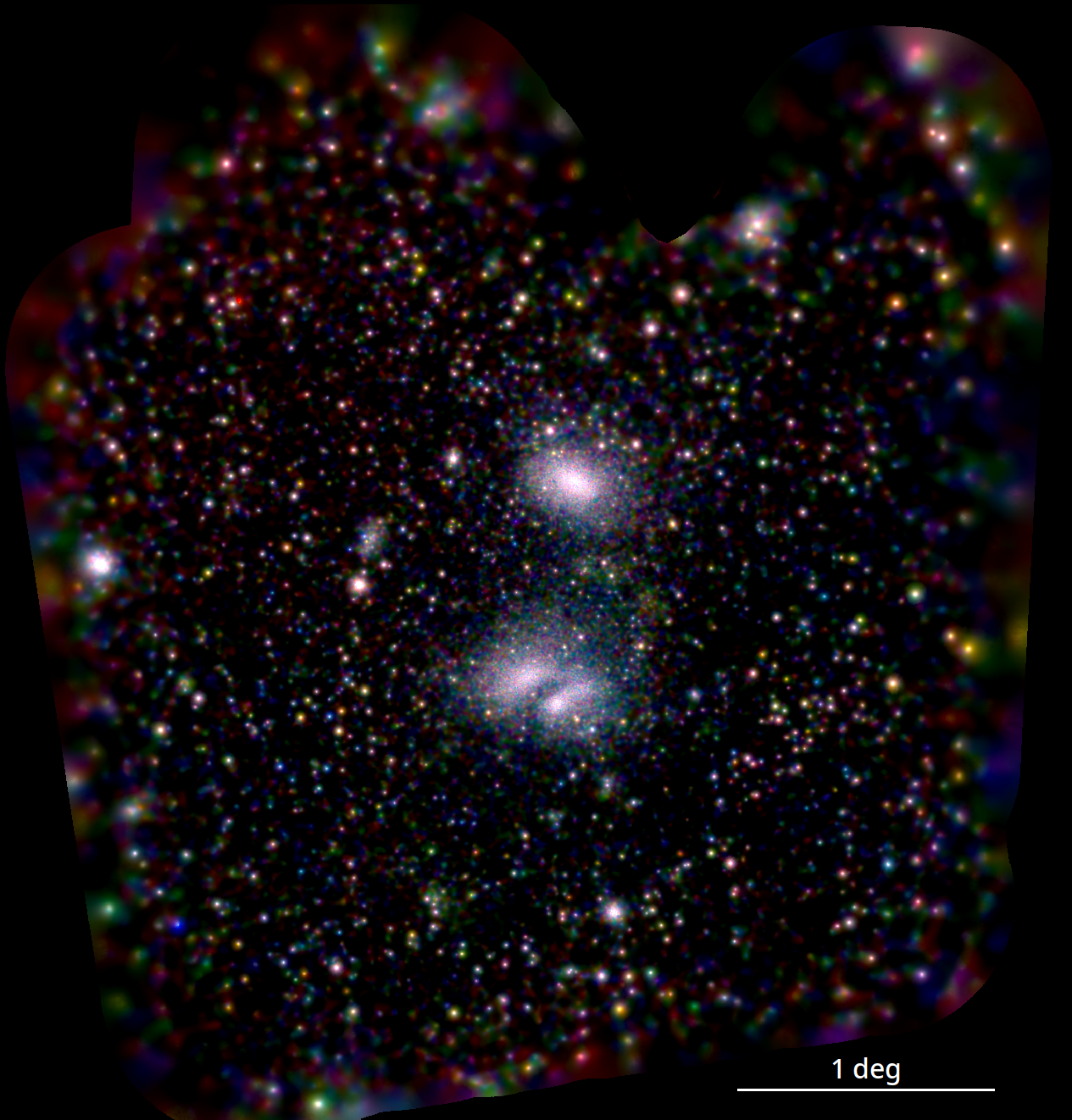}
   \caption{\rosi\ RGB image of the A3391/95 system. The red channel corresponds to 0.3--0.75 keV, green to 0.75--1.2 keV, and blue to 1.2--2.3 keV. Since energies below 1 keV are required, only TM8 has been used here. Scans I and II, therefore, have only TM6 included, which has little exposure north of Canopus. Regions with less than 10 s exposure have been excluded here, hence the black area in the north. All shown \rosi\ images from now on are exposure- and vignetting-corrected and the particle-induced background has been subtracted. A Gaussian smoothing has been applied which is scaled by the size of a top hat kernel with 25 photon counts in it. In the center of the field, A3391 is the northern large cluster and A3395n and A3395s are the two southern large clusters. One deg at $z=0.0555$ (A3391 redshift) corresponds to $\sim$3.9 Mpc.}
   \label{fig:eROSITA_RGB}%
    \end{figure*}

   \begin{figure*}
   \centering
    \includegraphics[width=1.9\columnwidth]{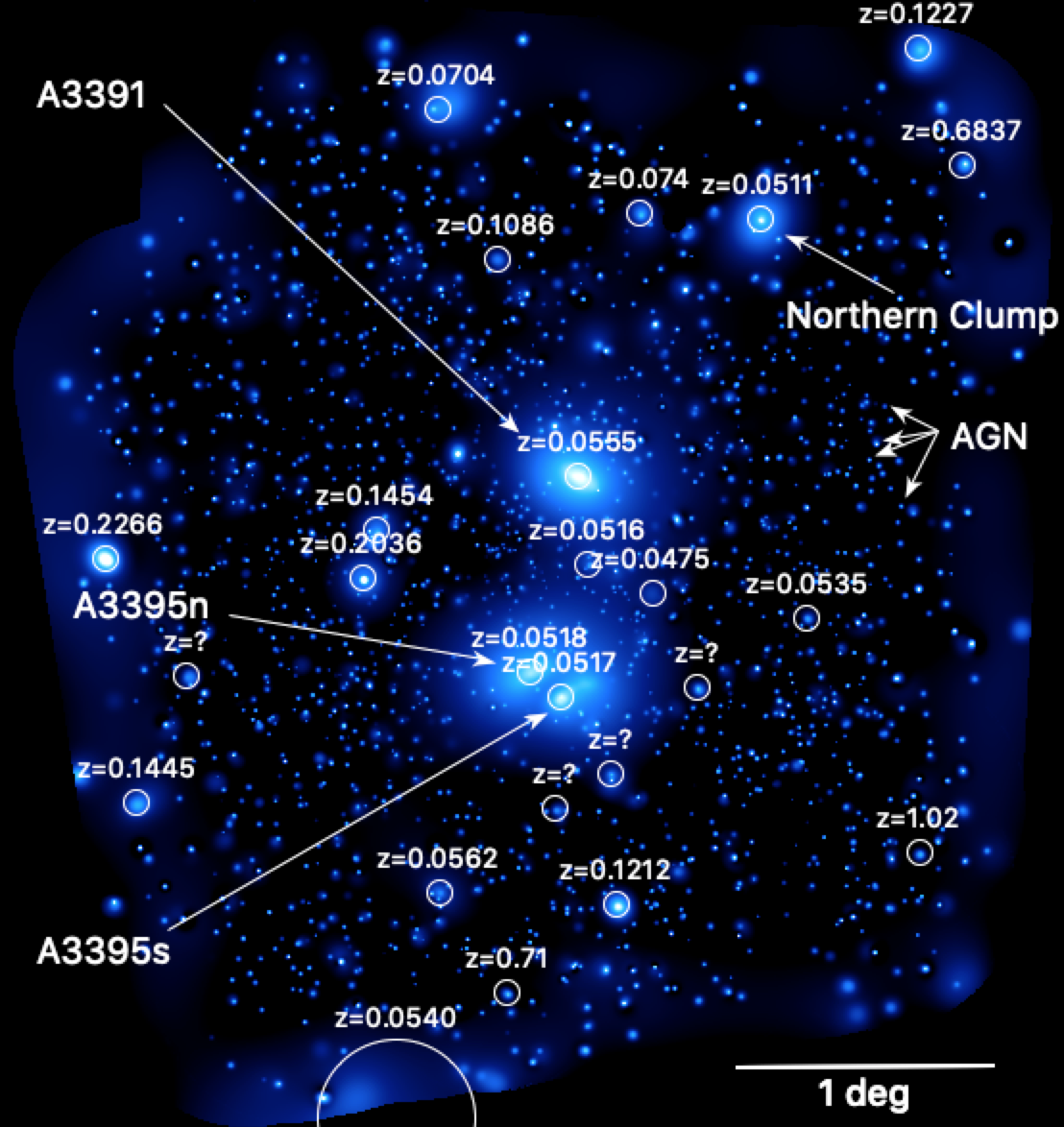}
   \caption{Wavelet-filtered \rosi\ image (TM0) of the A3391/95 system. The main clusters are marked. Most of the point-like sources are background AGNs.
   A small selection of brighter ``clumps'' and background clusters in the field are marked with circles, some of which have spectroscopic redshifts compiled from NED galaxies. The redshifts of the two ACT clusters with $z>0.7$ are based on photometric cluster redshifts. We  also note the galaxy group ESO 161-IG 006 at $z=0.0516$ in the bridge region and the ``Little Southern Clump'' that coincides with a galaxy at $z=0.0562$. The large white circle at the bottom indicates $r_{200}$ of the cluster MCXC J0631.3-5610.
   One deg corresponds to $\sim$3.9 Mpc at the redshift of A3391.}
     \label{fig:eROSITA}%
    \end{figure*}

\subsection{Hydrogen column density map}
\label{NH}
X-rays interact with the gaseous phases of the interstellar and intergalactic medium by photoelectric absorption along their way to the observer. Essentially, their photon energies are sufficiently high to free electrons out of the inner shells of atoms and molecules. Consequently, the X-ray spectrum is altered depending on the amount of gas along the line of sight. 
The brightness decrease due to photoelectric absorption depends exponentially on the column density of absorbing material and the photoelectric absorption cross-section, $\sigma \propto E^{-3}$, which is a strong function of energy \citep[e.g.,][]{2000ApJ...542..914W}; 
it therefore leads to a hardening of the observed X-ray spectrum.
This dominance of photoelectric absorption at soft X-ray energies is due to the cosmological abundances of hydrogen and helium in space. At very soft X-ray energies ($\lesssim$$0.28$ keV), the photoelectric absorption cross-section is independent of its location along the line of sight, either extragalactic or interstellar; it is determined by the hydrogen and (especially) helium abundances as a result of primordial nucleosynthesis. At higher X-ray energies ($\gtrsim$$0.28$ keV, as relevant for the analysis here) ``metals'' formed by stellar fusion and explosions need to be accounted for because they are accumulated in the interstellar medium of the  Milky Way. As is common practice, we use the hydrogen column density along the line of sight as a proxy for the amount of absorbing material together with the assumption of solar metallicity. This requires determination of the {total} hydrogen column density, independently of its state (neutral, molecular, ionized).

A3391/95 is located towards the high Galactic latitude sky. Due to its proximity to the Magellanic Cloud System, the HI distribution is unusually complex in its spatial structure. We observe  $N_\mathrm{HI} = 4.4\cdot 10^{20}\,\mathrm{cm^{-2}}$ on average towards the area of interest. 
According to \citet[][]{1996A&A...312..256B} a certain threshold $N_\mathrm{HI}$/$E_\mathrm{B-V}$ of $5\cdot 10^{20}\,\mathrm{cm^{-2}}$/0.08\,mag exists above which significant amounts of molecular gas ($N_\mathrm{H_2}$) might be present. While towards the central region of the A3391/95 system the $N_\mathrm{HI}$ is compatible with a pure atomic medium, the optical extinction with $<E_\mathrm{B-V}> = 0.21$\,mag is clearly in excess of the 0.08\,mag threshold  \citep{1998ApJ...500..525S, 2011ApJ...737..103S}, demanding an inspection for molecular gas.

For this task we follow the ansatz of \citet[][]{ 1996A&A...312..256B}, that is,
\begin{equation}
    I_\mathrm{FIR} = \epsilon \cdot N_\mathrm{H} + R\,,
    \label{EqNH}
\end{equation}
with $I_\mathrm{FIR}$ the far-infrared (FIR) emissivity, $\epsilon$ the FIR emission per hydrogen nucleus, and $R$ the Milky Way unrelated diffuse FIR background \citep[][]{ 2019ApJ...883...75L}. $I_\mathrm{FIR}$ is a quantitative measure for the number of hydrogen nuclei $N_\mathrm{H}$. In case of a phase transition, from atomic to molecular hydrogen, $N_\mathrm{H} > N_\mathrm{HI}$. To quantify $N_\mathrm{H}$ we study the linear correlation between the IRAS 100$\mu$m \citep{2005ApJS..157..302M} and the HI4PI data \citep{2016A&A...594A.116H}. For this aim we re-sample the IRIS and HI4PI on the same grid and smooth it to the angular resolution of HI4PI data.
We find a correlation coefficient between both of 0.85 and $\epsilon = (0.71 \pm 0.11)\cdot 10^{-20}\,\mathrm{cm^{-2}}$ which is marginally in excess in comparison to a similar sized area towards the northern Galactic hemisphere \citep{2014A&A...564A..71R}. $R = 0.22 \pm 0.65\,\mathrm{MJy\,sr^{-1}}$ is consistent with zero. The high degree of linear correlation between both gas tracers implies that across the whole field of interest the dominant gaseous phase consists of neutral atomic hydrogen.

Because the clusters' extent is narrowly confined relative to the analyzed field of interest, field averaged values might smooth out the small-scale molecular structures. We need to image the spatial structure of the FIR excess/high extinction gas.
For this aim we calculate a difference map according to eq.~(\ref{EqNH}) disclosing the spatial distribution of molecular gas, because $N_\mathrm{H_2} = (N_\mathrm{H} - N_\mathrm{HI})/2$. As displayed in Fig.~\ref{fig:NH2map} in Appendix~\ref{App_C} the distribution of the $N_\mathrm{H_2}$ is rather patchy. The peak $N_\mathrm{H_2}$ column densities are about a quarter of the $N_\mathrm{HI}$ column density. Implying that towards these sight lines half of the absorption would not be accounted for when using only neutral hydrogen as tracer.

Unaccounted so far is the extragalactic gas contribution to the soft X-ray photoelectric absorption. According to observations \citep[][]{2007MNRAS.382.1657K} and numerical simulations \citep[][]{2018MNRAS.476.3716R} the number of Ly-$\alpha$ absorbers is a strong function of redshift, with $N_\mathrm{HI}^{Ly-\alpha} \propto (1+z)^{3}$. With respect to the rather low redshift of A3391/95 the superposed column density of Ly-$\alpha$ absorbers,  $N_\mathrm{HI}^{Ly-\alpha}$, is estimated to sum up to a few $10^{19}\,\mathrm{cm^{-2}}$, which is insignificant compared to the Milky Way foreground. A quantitative estimate within that redshift range can be expected from the MeerKAT absorption line survey during the next two years \citep{2016mks..confE..14G}.

For the construction of the absorption-corrected surface brightness map in Section~\ref{filaments} we use a spatially resolved total $N_{\rm H}=N_{\rm HI}+2N_{\rm H_2}$ map, determined as described above.

   \begin{figure*}
   \centering
    \includegraphics[width=6cm]{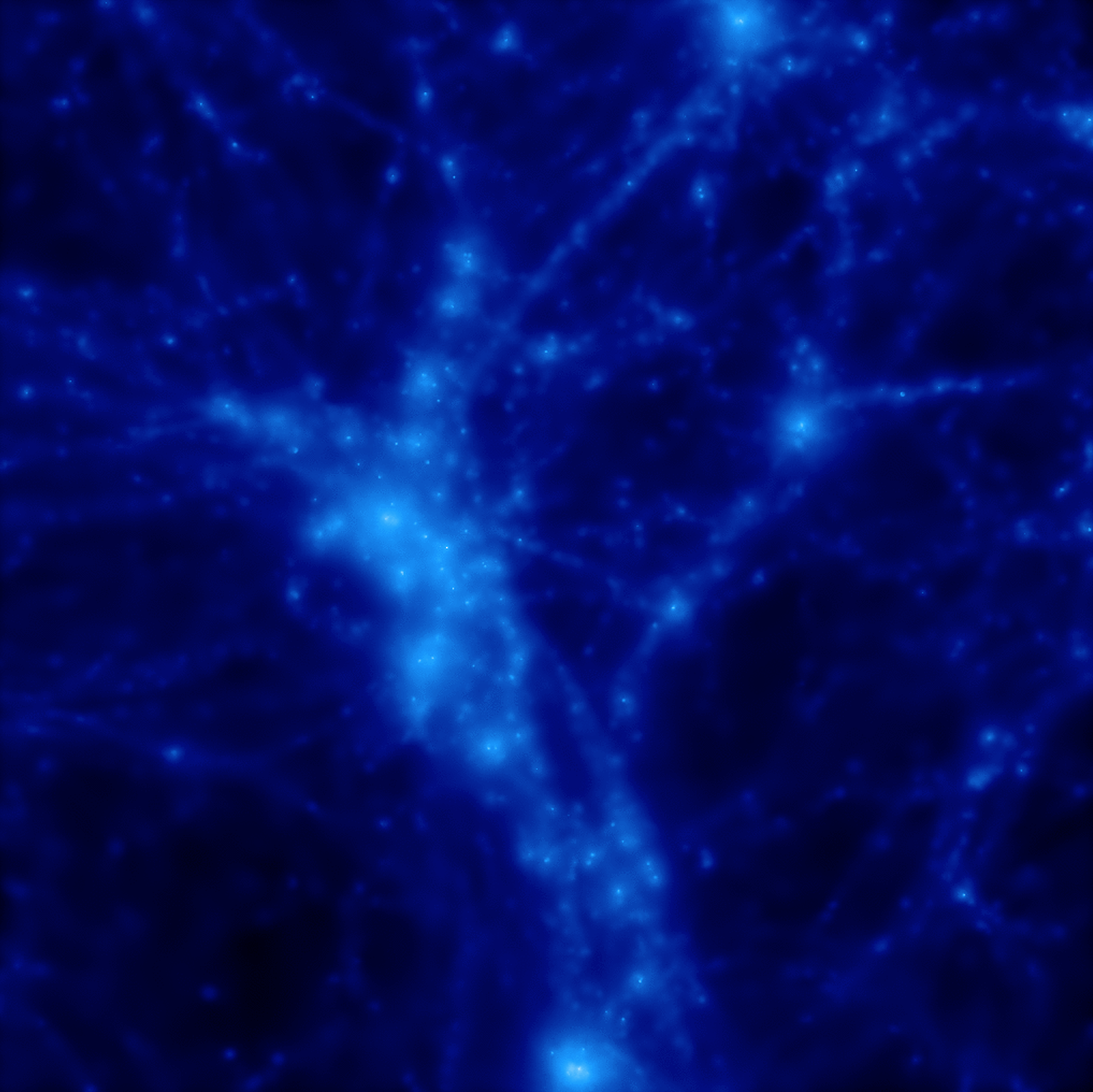}
    \includegraphics[width=6cm]{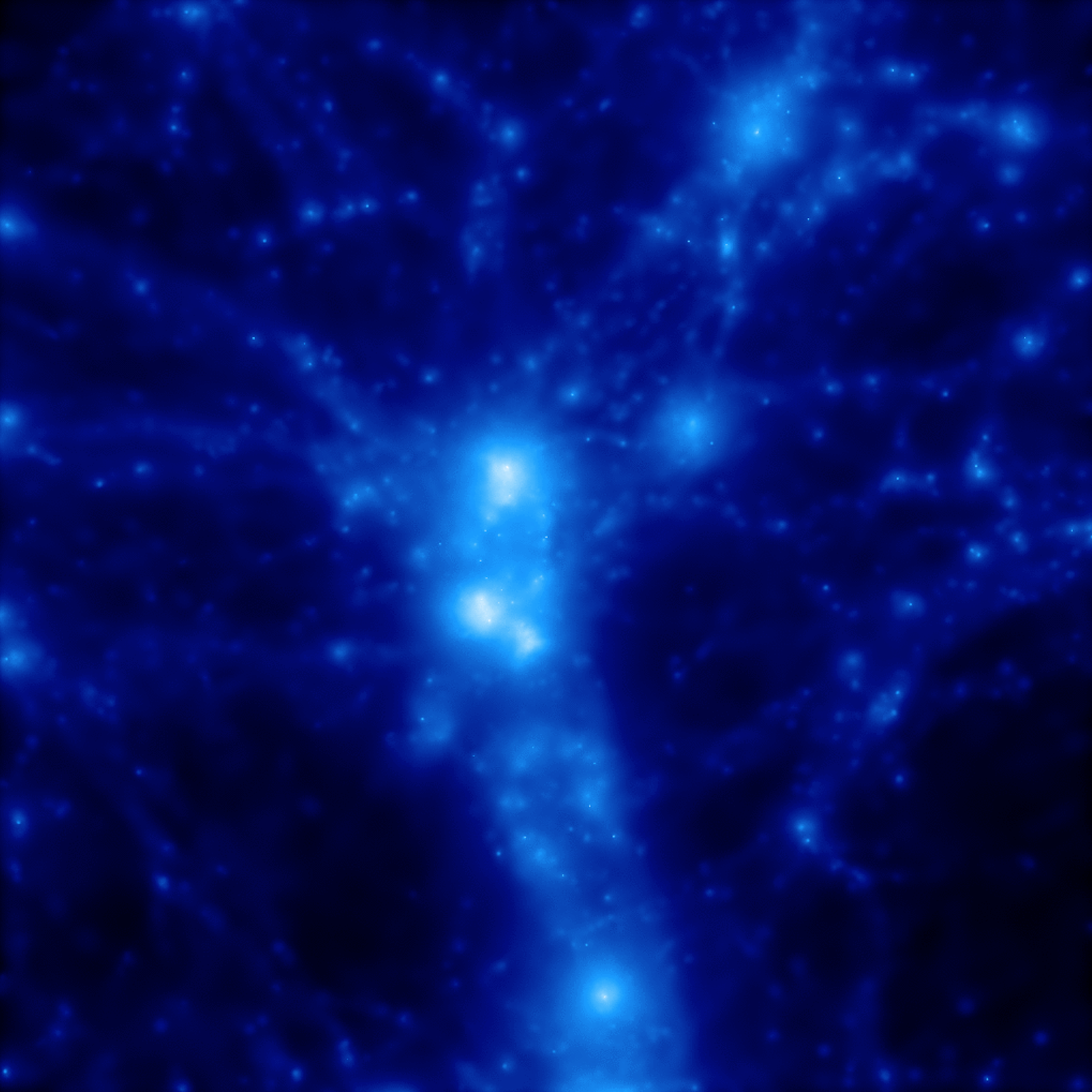}
    \includegraphics[width=6cm]{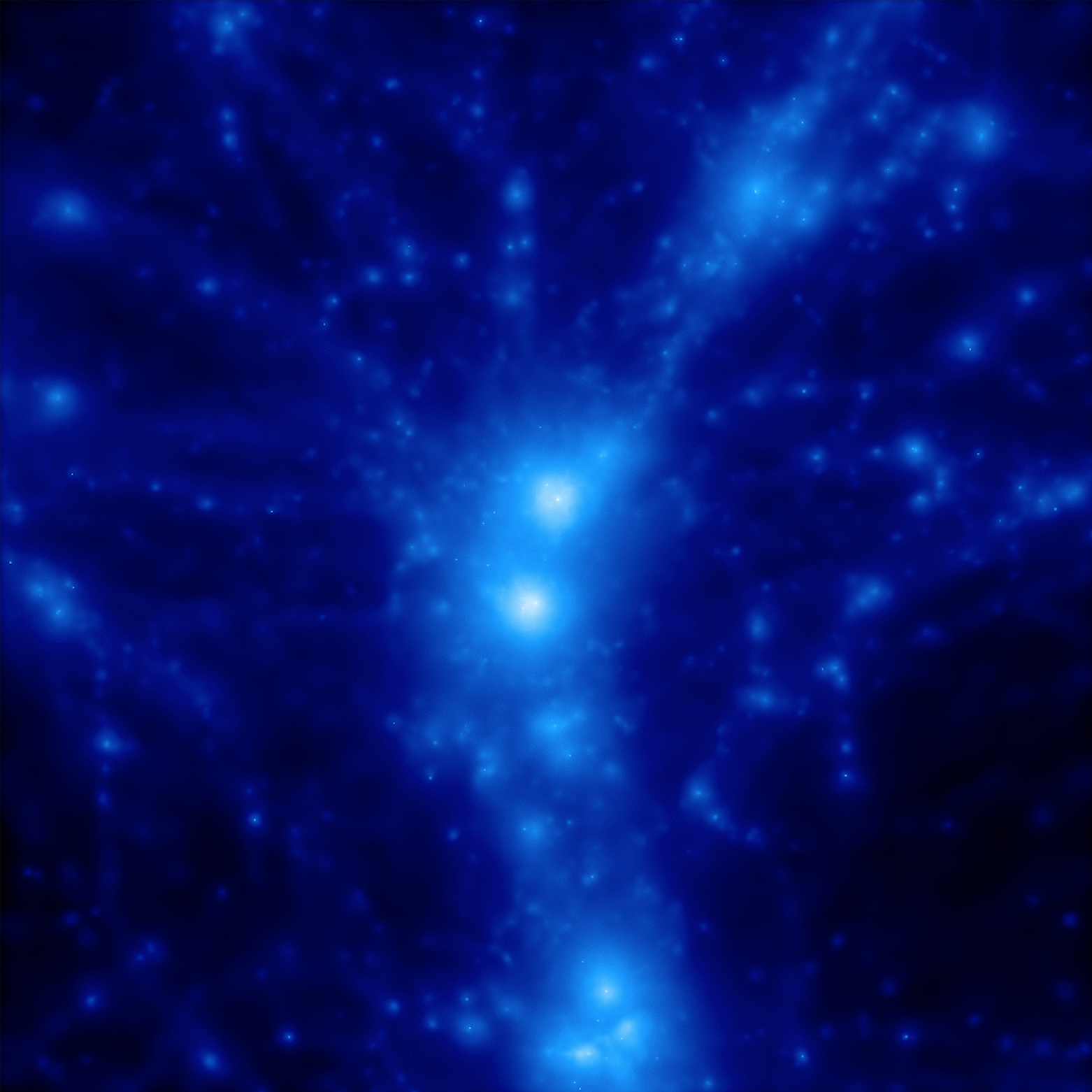}
   \caption{Large-scale evolution of the A3391/95 analog found in the Magneticum simulation. The gas density distribution is shown in a
   cubic cutout region of $20$\,Mpc$/h$ per side around the redshift of the  main system in comoving coordinates from redshift $z\approx1$ (left) to $z\approx0.34$ (middle) and $z\approx0.07$ (right). We note the clumps falling in along the large-scale filaments, merging with the main systems to form ever larger, denser, and hotter structures. The corresponding movie with more redshift snapshots is available online at \url{https://astro.uni-bonn.de/~reiprich/A3391_95/}\,.
   }
    \label{fig:magneticum_evol}%
    \end{figure*}

\subsection{Magneticum simulation}
\label{Mag}

A theoretical counterpart of the A3391/95 system is presented in
Section~\ref{LSS},
extracted from the Magneticum Pathfinder
simulations\footnote{\url{http://www.magneticum.org}}, a set of state-of-the-art
cosmological hydrodynamical simulations comprising boxes of different
volumes and resolution.
The simulations were performed with an extended version of the
TreePM/SPH code Gadget-2 \citep[][]{springel2005}, including the
treatment of a large variety of baryonic processes.  These are comprised of,
among others, metal-dependent gas cooling \citep[][]{wiersma2009} and
star formation \citep[][]{springel2003}, metal enrichment from SNIa,
SNII, and AGB stars following stellar evolution models
\cite[][]{tornatore2004,tornatore2007}, energy feedback from stellar
sources \cite[][]{springel2003} and from AGNs due to gas accretion
onto super massive black holes
\cite[][]{springeldimatteo2005,fabjan2010}.
In particular, we consider the ``Box2'' cosmological volume of
$(500\,{\rm Mpc})^3$ at high resolution
\cite[i.e.,\ $m_{\rm DM}=6.9\times 10^8\,M_\odot/h$ and $m_{\rm gas}=1.4\times 10^8M_\odot/h$,
  for dark matter and gas particles respectively; see][for more details]{biffi2018},
which contains $\sim$450 cluster-size haloes with
$M_{500} > 10^{14}\,M_{\odot}$ at redshift $z\approx 0.07$.
At this redshift, which is fairly similar to the  one observed,
we identify the A3391/95 analog as a close pair of two
clusters with a mass ratio of $\sim$1.2. The two pair members have
masses $M_{500}=1.96\times 10^{14}\,M_{\odot}$ and
$M_{500}=1.67\times 10^{14}\,M_{\odot}$, and are separated by a
projected distance of $2.6$\,Mpc in the plane of the sky. Physically,
this system has not yet merged; the two clusters are  separated by a
3D distance of $4.5$\,Mpc.

   \begin{figure*}
   \centering
     \includegraphics[width=2.1\columnwidth]{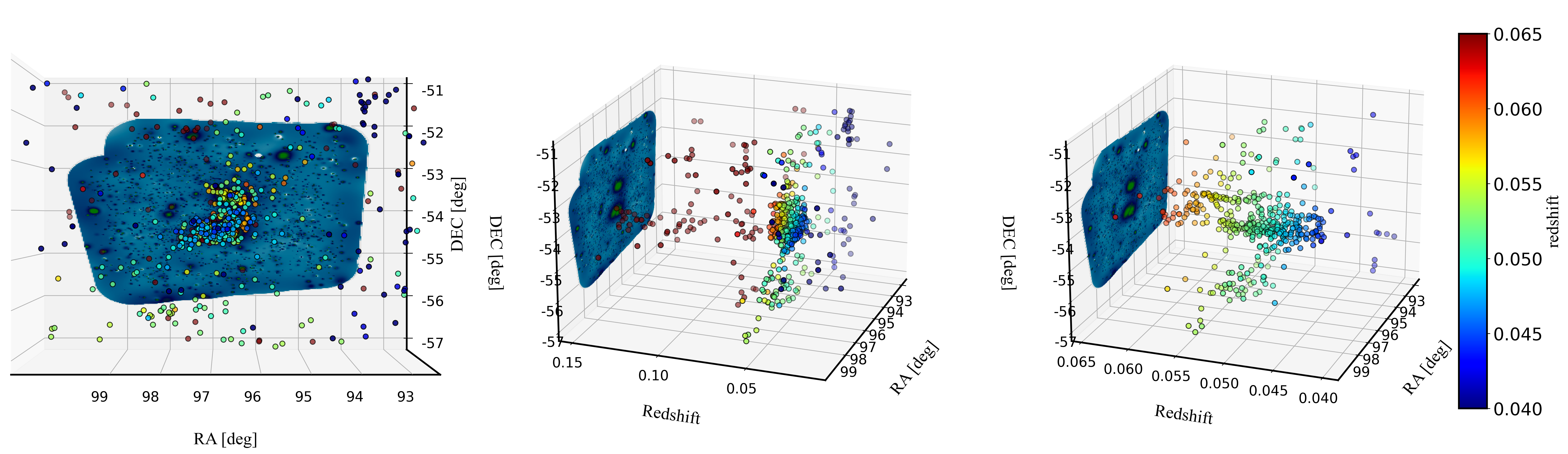}
   \caption{Spectroscopic redshift distribution of galaxies in the ranges $0.01<z<0.15$ (left and middle) and $0.04<z<0.065$ (right) in two different projections. The background shows the \rosi\ wavelet-filtered image. The redshift color bar applies to all three plots. Animated versions of the 3D plots are available at \url{https://astro.uni-bonn.de/~reiprich/A3391_95/}\,.}
    \label{fig:NED}%
    \end{figure*}

The evolution of the system in its large-scale environment from $z\approx1$ to $z\approx 0.07$ is also shown
in Section~\ref{LSS}.
We show surface gas density maps centered on the center of mass
of the simulated cluster pair, covering an area of
$20$~comoving~Mpc$/h$ per side and projecting for $20$~comoving~Mpc$/h$
along the line of sight.
At the final redshift $z\approx 0.07$, where the pair system is selected,
the map encompasses a physical volume of $(\sim$$26.5$\,Mpc$)^3$.
A dedicated study of the A3391/95 simulation analog investigating
the properties and evolution of the main system structures and of the
diffuse gas in the filaments will be pursued in a separate
paper (Biffi et al., in prep.).
We note that the results from the Magneticum simulation shown here do not differ drastically from those of other cosmological hydrodynamical simulations, and therefore conclusions drawn here from comparisons between observations and simulations can be considered to be general.

\section{Results and Discussion}
\label{results}

\subsection{Large-scale structure}
\label{LSS}

An RGB image of the TM8-only \rosi\ data is displayed in Fig.~\ref{fig:eROSITA_RGB}.
Most of the 100s of point-like sources are background supermassive black holes actively accreting matter from their host galaxies (AGNs). Furthermore, there is likely a significant population of stars in the foreground (a small area around  the optically second-brightest star in the sky, Canopus, has been masked here and in subsequent images). The large range of colors seen illustrates the variety of spectral shapes encountered in the point source population. Stars are generally expected to have a softer spectrum than AGNs, but also differing amounts of X-ray absorption along the line of sight for  stars and intrinsic to the AGNs likely contribute to the spectral variation. A detailed study of the point source population will be provided by Liu et al.\ (in prep.).
Moreover, the image shows the extended emission from nearby galaxy groups and clusters in green and white.

In order to assess the significance of extended emission in this image, we ran a wavelet filtering algorithm designed to rigorously model and suppress Poisson noise. Initially developed for the XXL survey \citep[]{fps18}, the software is a new implementation of the \textit{mr\_filter} task of the \texttt{MR/1} multiresolution package \citep{smb98} with an improved treatment of exposure correction. It first computes a stationary wavelet transform of the signal, sometimes referred to as the `\`a trous' algorithm, using a cubic B-spline wavelet. At each position and wavelet scale, it estimates the significance of a signal excess compared to a flat distribution  using the method of wavelet histogram auto-convolutions \citep{slb93} and applies a hard thresholding to remove insignificant wavelet coefficients. A reconstructed image can be obtained applying the inverse wavelet transform to the filtered coefficients. However, since this filtering process is nonlinear, it iteratively improves on the result by detecting significant signal in the wavelet transform of the residuals and adding them to the reconstructed signal. In order to rigorously subtract the PIB and correct for exposure variations across the field, while preserving Poisson statistics, it estimates the significance of wavelet coefficients at each scale and position based on the raw photon images, but reconstructs the signal based on the wavelet transform of the corrected count rate images. For \rosi, we filtered the signal of wavelet scales 1 to 8 (i.e., up to $\sim$$27^{\prime}$), incorporating the smoothed unfiltered background on larger scales into the results. As the number of effective pixels decreases with increasing scale, we applied a scale-dependent threshold (from $5\sigma$ at scale 1, to $2.8\sigma$ for the largest scales) in an attempt to control the occurrence of false positives. Furthermore, all areas with less than 30 s exposure were masked, which only affects a very small area around the outer edges (see Fig.~\ref{fig:expmap_TM0}, right).

In Fig.~\ref{fig:eROSITA} we present the wavelet-filtered \rosi\ image taking into account all the information that went into producing the fully reduced \rosi\ image shown in Fig.~\ref{fig:CLCR} (right). Recall (Section~\ref{rosi}) that data from the five TMs with on-chip filter (TM8) cover the energy range 0.3--2.0 keV while that from the two other TMs (TM9) cover 1.0--2.0 keV.
The three main clusters of the A3391/95 system dominate the central part.
Most of the other fuzzy blobs are groups and clusters of galaxies, some at the same redshift, which we then call ``clumps.'' This includes the ``Northern Clump,'' which is listed in the NASA/IPAC Extragalactic Database\footnote{NED, \url{https://ned.ipac.caltech.edu}} as MCXC J0621.7-5242.
Other extended sources are groups and clusters in the background. 
The image therefore provides an impressive high-contrast projection of the nearby large-scale structure.

We searched for a system with a similar appearance in the Magneticum simulation described in Section~\ref{Mag}. It is important to note that this is {not} a constrained realization specifically designed to reproduce the properties of the A3391/95 system. We found one such system, and the evolution of the gas density distribution around  it  is shown in Fig.~\ref{fig:magneticum_evol}.

In Fig.~\ref{fig:magneticum_evol} we notice a large number of additional structures
(namely clusters, groups, and galaxies in the same redshift slice; i.e., clumps) embedded in the large-scale network
and connected through lower-density filaments.
Going from $z\approx 1$, through $z\approx 0.34$, to the final configuration
at $z\approx 0.07$, some of these structures have merged with the
main systems.

Comparison of Fig.~\ref{fig:eROSITA} to Fig.~\ref{fig:magneticum_evol} reveals a number of similarities. First, both regions are dominated by a complicated system of a few nearby galaxy clusters. Second, surrounding clumps are apparently being attracted to the main system; at least in the simulation we know they are at a similar redshift; for the real data, this is discussed in more detail in Section~\ref{clumps}.
Third, an emission ``bridge'' connects the main clusters. Fourth, WHIM filaments away from the main system run from north to south, at least in the simulation, including an alignment of larger infalling  clumps. The bridge and filaments are discussed in more detail in Section~\ref{filaments}. 

Similar structures are apparent in the galaxy distribution. Figure~\ref{fig:NED} shows the spectroscopic redshift distribution of galaxies found in NED. We note the filamentary structure in the galaxy distribution, especially the bridge between A3391 and A3395n/s but also the extensions of ``greenish'' points running from north to south, and that some of the surrounding clumps appear to be at the redshift of the main system; i.e., they are likely infalling. Finger-of-god effects are clearly visible for both A3391 and A3395n/s. A3391 also appears to be slightly further away from us in redshift space than A3395n and A3395s, either because of a configuration space offset, or a peculiar velocity offset, or a combination of both. This is also quantified in Table~\ref{tab:z}. We  also note there is indication of a background structure just beyond $z=0.1$ projected almost exactly onto the position of A3395n.

   \begin{figure*}
   \centering
    \includegraphics[width=1.8\columnwidth]{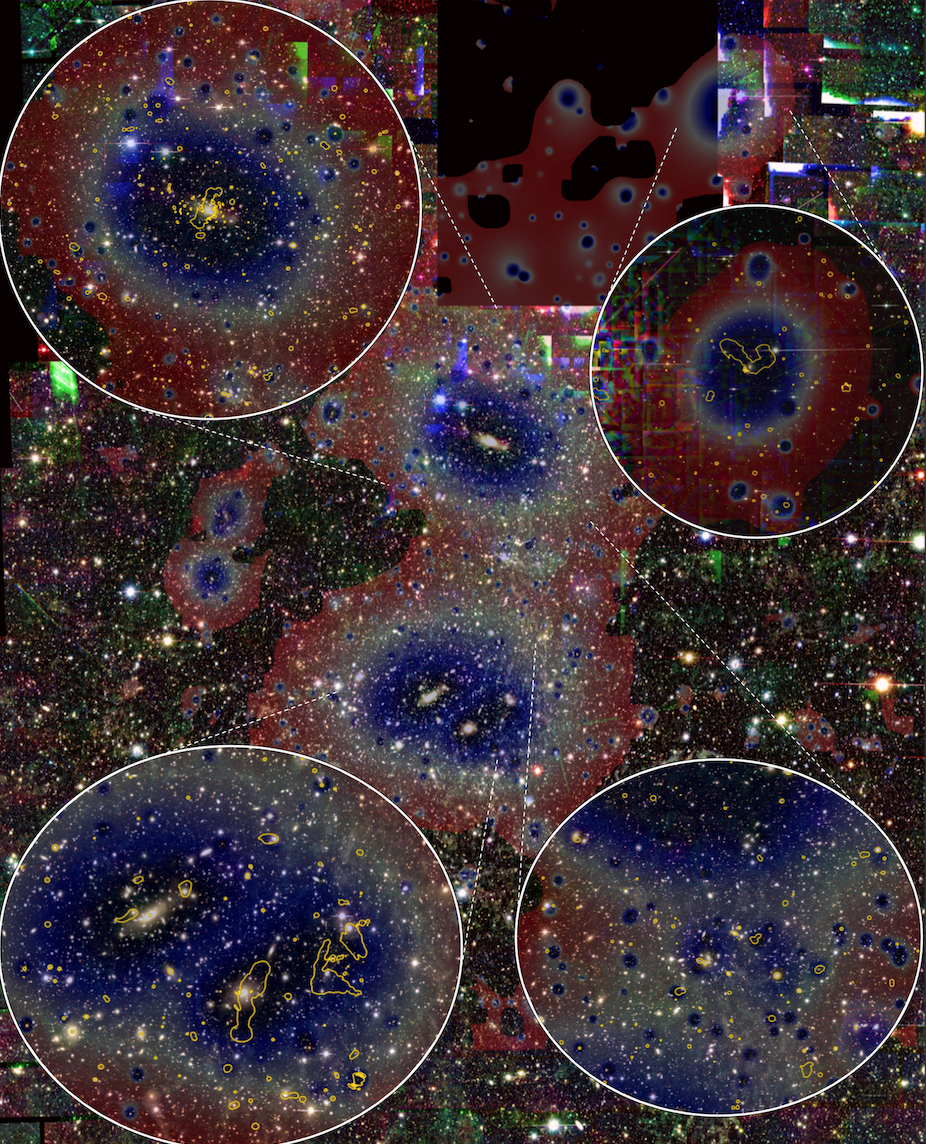}
   \caption{RGB image based on $i,r,g$ DECam observations of the system with \rosi\ wavelet-filtered surface brightness overlaid (darker/more transparent corresponds to higher X-ray  surface brightness; assuming gas traces mass, the darker regions are also those with higher dark matter density). The insets also show ASKAP/EMU radio contours (from Fig.~\ref{fig:EMU}). The inset in the upper right depicts the Northern Clump. There are some artefacts remaining in the optical image due to the bright star Canopus in the north.}
 \label{fig:DECam}%
\end{figure*}

The selection of galaxies with spectroscopic redshifts from NED is not homogeneous, and so the apparent large-scale structure may be partially caused by selection effects. In order to homogeneously sample the full galaxy distribution we acquired additional DECam data and also employed existing archival data (Section~\ref{DEC}). The resulting RGB image is presented in Fig.~\ref{fig:DECam} with X-ray surface brightness from the \rosi\ wavelet-filtered image overlaid. While the star Canopus creates significant artefacts in the northern part of the image (the black square part is masked), a good overall correspondence between the galaxy and gas distributions on large scales is indicated. A detailed galaxy density map based on DECam data is shown and discussed later in Section~\ref{filaments}.

   \begin{figure*}
   \centering
    \includegraphics[width=1.9\columnwidth]{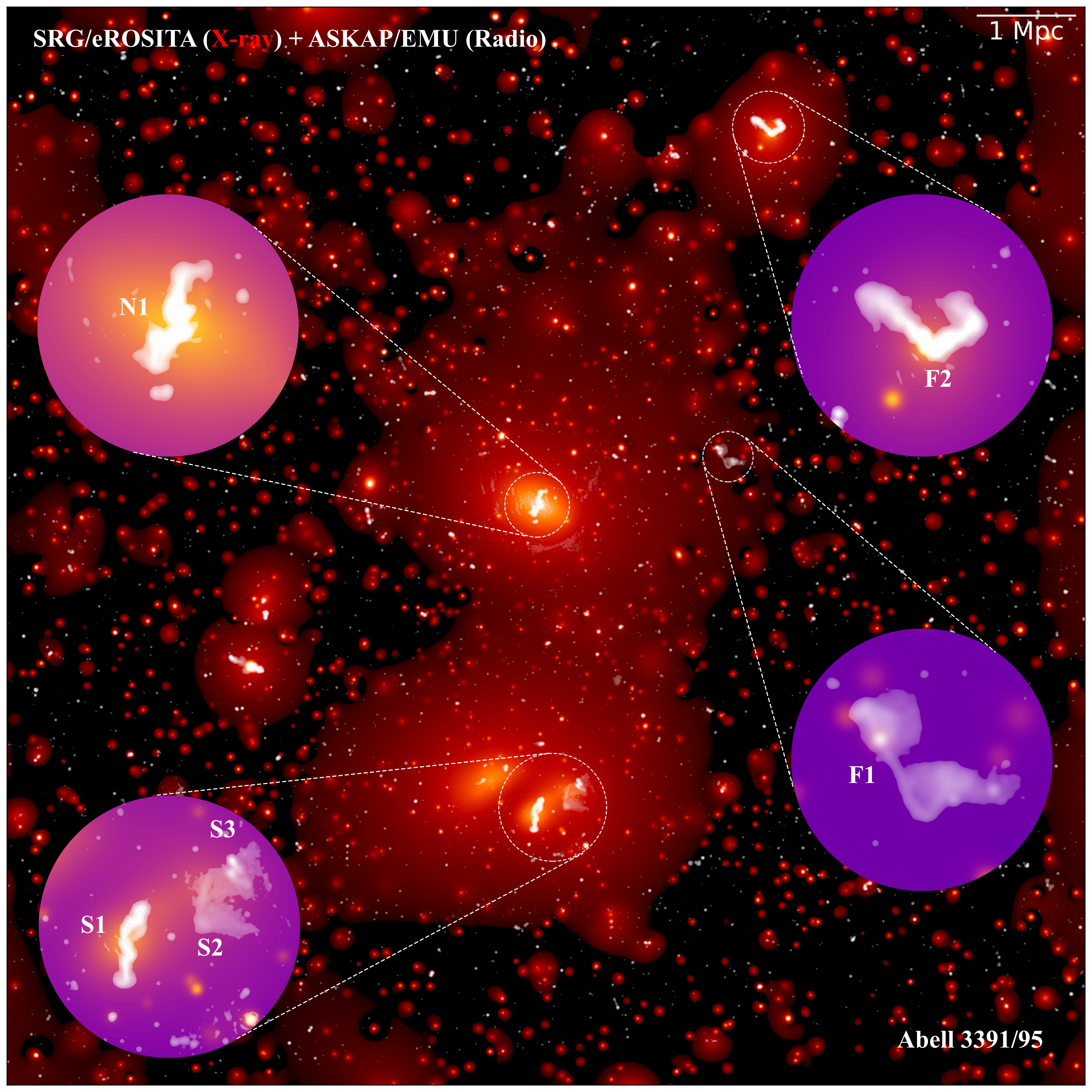}
      \caption{ASKAP/EMU Early Science observation of the A3391/95 system overlaid on the \rosi\ wavelet-filtered image. We note the bright radio galaxies and the diffuse S2/S3 structure.
      All of them are at the  redshift of the main system. 
      }
              \label{fig:EMU}%
    \end{figure*}
Figure~\ref{fig:EMU} shows the ASKAP/EMU radio image of the system on top of the \rosi\ wavelet-filtered image, constructed as described in Section~\ref{EMU}. A few individual giant radio galaxies are highlighted, including the BCGs of A3391 (N1) and A3395s (S1). 
The morphology of the wide angle tail galaxy (F2) in the Northern Clump indicates that it is moving towards A3391; that is, it is infalling;  this is similarly observed in the Magneticum simulations. We note that this Northern Clump is way outside the virial radius, $r_{100}$, of A3391, and so the bending is likely due to WHIM beyond the cluster borders. The projected separation of the X-ray emission peaks of the Northern Clump and A3391 is 1.217 deg = 4.722 Mpc = 1.929 $r_{100}$; assuming both systems lie at the redshift of A3391. The radio galaxy in the western part (F1) shows clear signs of interaction with the ICM of the outskirts of  A3391, and
its redshift, $z=0.0567$, is consistent with this hypothesis. Its projected separation from A3391 is 0.5203 deg = 2.019 Mpc = 1.122 $r_{200}$. The projected distances are lower limits to the 3D distances; that is, the interaction processes shed light on the gas properties well beyond $r_{200}$. This is consistent with statistical analyses of radio galaxy morphologies \citep[e.g.,][]{2019AJ....157..126G}.
We also note the interesting faint extended radio source to the west of A3395s (S2/S3). This could, for example, be a radio relic or be due to re-accelerated relativistic plasma related to a radio AGN.
A detailed investigation of the radio properties of the system, including upper limits on diffuse radio emission in the bridge region and details about the radio galaxies is provided in \citet{brb20}.

In summary, what we witness here is the formation of LSS in action: massive clusters merging at the nodes of filaments, smaller clumps (themselves mature groups or even clusters of galaxies) falling towards those centers of mass along filaments. The \rosi\ data show this directly for the gas distribution in a single region of the sky and not just statistically by stacking a large number of galaxy overdensity regions. This is impressive confirmation of the expectation from simulations.

All wavelengths used in this study show the same dynamics, but in contrast: gas and galaxies trace the dark matter, requiring large-scale observations; detailed gas morphology and wide-angle tail radio galaxies (WATs) or even potential radio relics indicate interaction and flow directions, both requiring observations with good spatial resolution. For the X-ray band, \rosi\ is currently the only available instrument that provides these capabilities.

\subsection{Clusters, groups, and clumps}
\label{clumps}

   \begin{figure*}
   \centering
    \includegraphics[width=1.0\columnwidth]{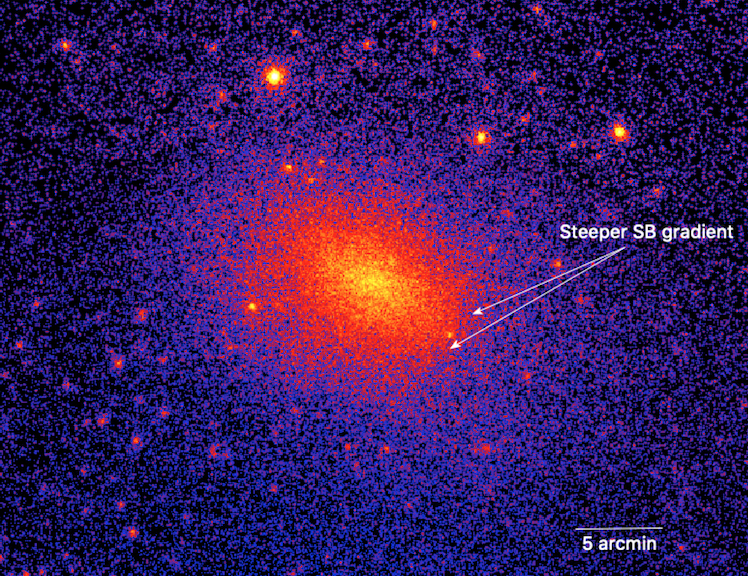}
    \includegraphics[width=1.0\columnwidth]{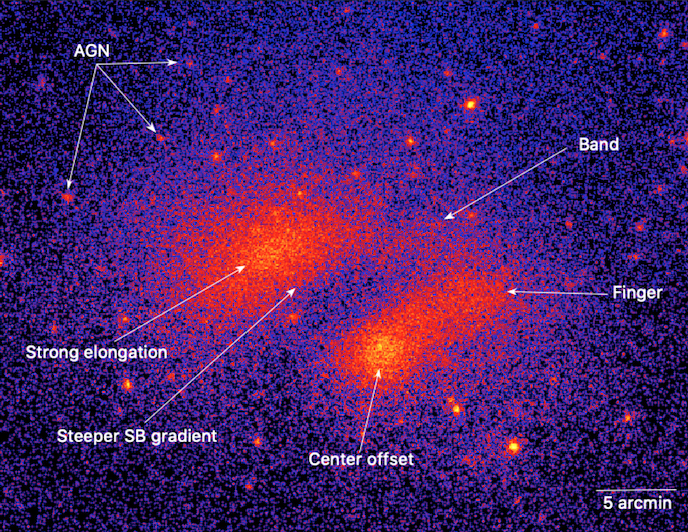}
   \caption{\rosi\ close-up views of the main clusters A3391 (left) and A3395n/A3395s (right). The morphologies and their interpretations are described in Section~\ref{clumps}.
   Most of the point-like sources are expected to be background AGNs.}
   \label{fig:A3391}%
    \end{figure*}
Following the overall large-scale view provided in the previous section, we focus here on the individual systems in the field.
Zoom-ins to A3391 and A3395n/s are shown in Fig.~\ref{fig:A3391}, left and right, respectively, illustrating their complicated morphologies; these images also demonstrate the strength of \rosi\ to efficiently cover large fields but to simultaneously provide fine detail with good spatial resolution. The disturbed morphologies, especially of the A3395n/s system, clearly indicate dynamical activity. Towards the southwest of A3391, a slightly stronger gradient in the surface brightness compared to the north west can be noted. This may indicate a motion relative to the gas in the emission bridge between A3391 and A3395n/s, or a sloshing motion. Even stronger features are apparent for A3395n/s and some have been described before \citep[e.g.,][]{2001ApJ...562..254D,2011ApJ...743...78L}. Both A3395n and A3395s show a very elongated shape and an apparent offset of the brighter central regions from their overall center of mass. A3395s exhibits a very prominent emission peak and a ``Finger'' of emission extending towards the northwest. The Finger ends right were the brightest part of the interesting diffuse radio structure can be seen (Fig.~\ref{fig:EMU}, source S3; also Fig.~\ref{fig:oxygen} later) and this region is discussed in detail in the accompanying paper \citep{brb20}. Moreover, there appears to be an emission ``Band'' west of A3395n and north of A3395s connecting both systems. This could be plasma displaced due to an ongoing merger process between these two clusters. On the other hand, the strong central radio AGN in A3395s (S1) with its north--south jets may also contribute to pushing some of the hot plasma towards the north. Furthermore, A3395n shows a steeper surface brightness gradient towards the southwest compared to the northeast, indicating gas compression due to interaction with A3395s. 

The insets in Fig.~\ref{fig:DECam} zoom into A3391, A3395n/s, the galaxy group ESO 161-IG 006, and the Northern Clump, and additionally show contours obtained from the ASKAP/EMU radio data. On this overlay, there are a number of obvious smaller-scale features that we have not yet discussed. First, the X-ray surface-brightness elongation position angle roughly matches the brightest cluster galaxy (BCG) position angle for both A3391 and A3395n. This consistency of elongations could be interpreted as an indication of the absence of strong dynamical disturbances. Second, the jet/lobe extension of A3391's central AGN (N1) is perpendicular to this position angle. Third, there appear to be a number of small head--tail galaxies in the central A3395n/s area, which could be interpreted as indicating dynamical activity. At least in the case of A3395n, we now have one indication for the absence of dynamical activity (similar X-ray and BCG elongations) and one indication for the presence of it (head--tail galaxies). This is apparently contradictory. However, this may be explained either by coincidence; that is, the disturbance happens such that it enhances the elongation in the right angle, or if such position angle correlations are hard to destroy during a merger process.

Based on spectroscopic galaxy redshifts available in NED, we (re-) determined the redshifts of A3391, A3395n, A3395s, and two clumps. Using the \rosi\ X-ray emission peak as center and a circle with $\sim$0.25 deg radius, we determined the cluster and clump redshifts from the median of all $N$ galaxies with $z<0.2$ in this area (Table~\ref{tab:z}). We note that for A3395n and A3395s the search cones overlap. In order to count each galaxy redshift only once, we therefore assigned each galaxy in the overlap region to the cluster whose center is closer. Also listed in the table are the comoving distances, $D_{\rm c}$, assuming zero peculiar velocities. Furthermore, to illustrate an alternative extreme scenario, we define approximate peculiar velocities as $v_{{\rm pec},i}\equiv c(z_{\rm A3391}-z_{i})$, assuming all clumps and clusters, $i$, to be at the distance of A3391, which in turn is assumed to follow the Hubble flow exactly.
Table~\ref{tab:radii} then lists relevant radii, which were determined by taking the
$r_{500}$ values from \citet{rb01} and calculating the other radii as in \citet[][Section~2]{rbe13}.
A more detailed multi-wavelength study of the central and outer parts of the main clusters is carried out in separate investigations (Sanders et al., in prep.; Veronica et al., in prep., respectively).

One of the more exciting discoveries from the \rosi\ observations is the large number of clumps (and background clusters); that is, extended X-ray-emitting sources in the field. There are more than 50 in the 15 deg$^2$ that appear to be associated with galaxy overdensities. For many of them, we can determine redshifts either from the \rosi\ X-ray data directly or from optical data (spectroscopic galaxy redshifts). Some are at a similar redshift to the main A3391/95 system, and so can be interpreted as clumps that are presumably infalling. Other extended sources are at different, mostly higher redshifts, and are therefore likely background galaxy groups and clusters. The analysis of these extended sources is the subject of a detailed study (Ramos-Ceja et al., in prep.). Here, we show a small selection of some of the brighter clumps and background clusters in Fig.~\ref{fig:eROSITA}. The redshifts were determined by using the \rosi\ emission peaks as centers and calculating the median of spectroscopic galaxy redshifts found in NED. The presence of several clumps at the same redshift as the main A3391/95 system qualitatively confirms the expectations from cosmological simulations; compare  to Figs.~\ref{fig:magneticum_evol} and \ref{fig:eROSITA2} (left) for example. This has not been observed as clearly in an individual system before. Furthermore, many of the extended sources are background clusters. This shows the great improvement of \rosi\ over the ACT \citep[][]{2020arXiv200911043H} and SPT \citep{bds18,2020AJ....159..110H,2020ApJS..247...25B} SZ surveys, which have both covered the field. These latter detect only four  clusters in this area (SPT: 2; ACT: 4, including both SPT clusters). Interestingly, the huge clusters A3391 and A3395 are not part of their catalogs, presumably because of highpass filtering and the point source masking here; combination with the lower resolution Planck data may help
\citep[e.g.,][]{2019A&A...632A..47A}.
These four SZ clusters are to be compared to tens of clusters from \rosi\, including all four ACT/SPT clusters, even out to $z=1.0$.
Thus, \rosi\ finds all the high-redshift massive systems in the field, as ACT and SPT do, but additionally finds about an order of magnitude more groups and clusters, presumably mostly with lower mass and at lower redshift.
%
   \begin{figure*}
   \centering
    \includegraphics[width=1.0\columnwidth]{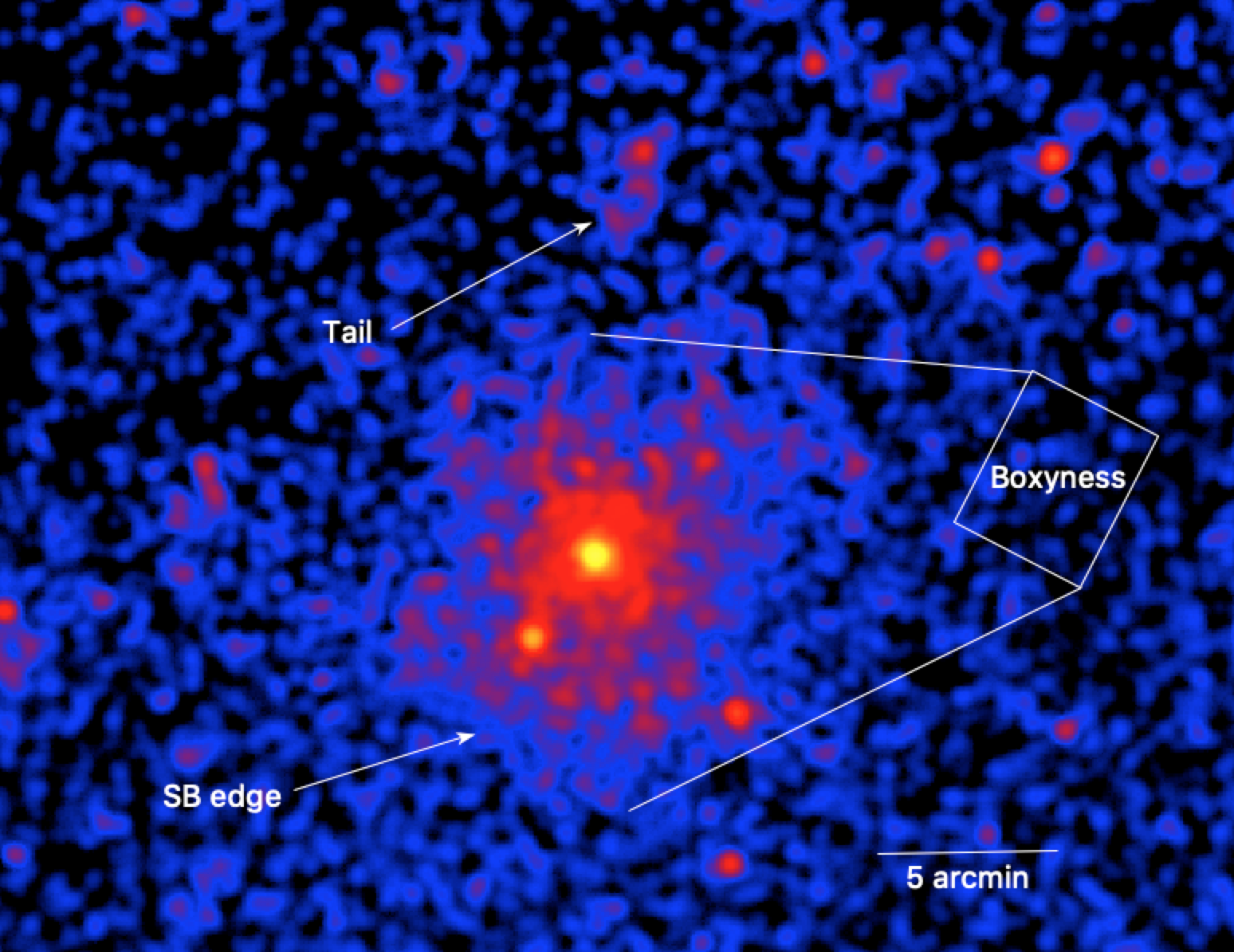}
    \includegraphics[width=1.0\columnwidth]{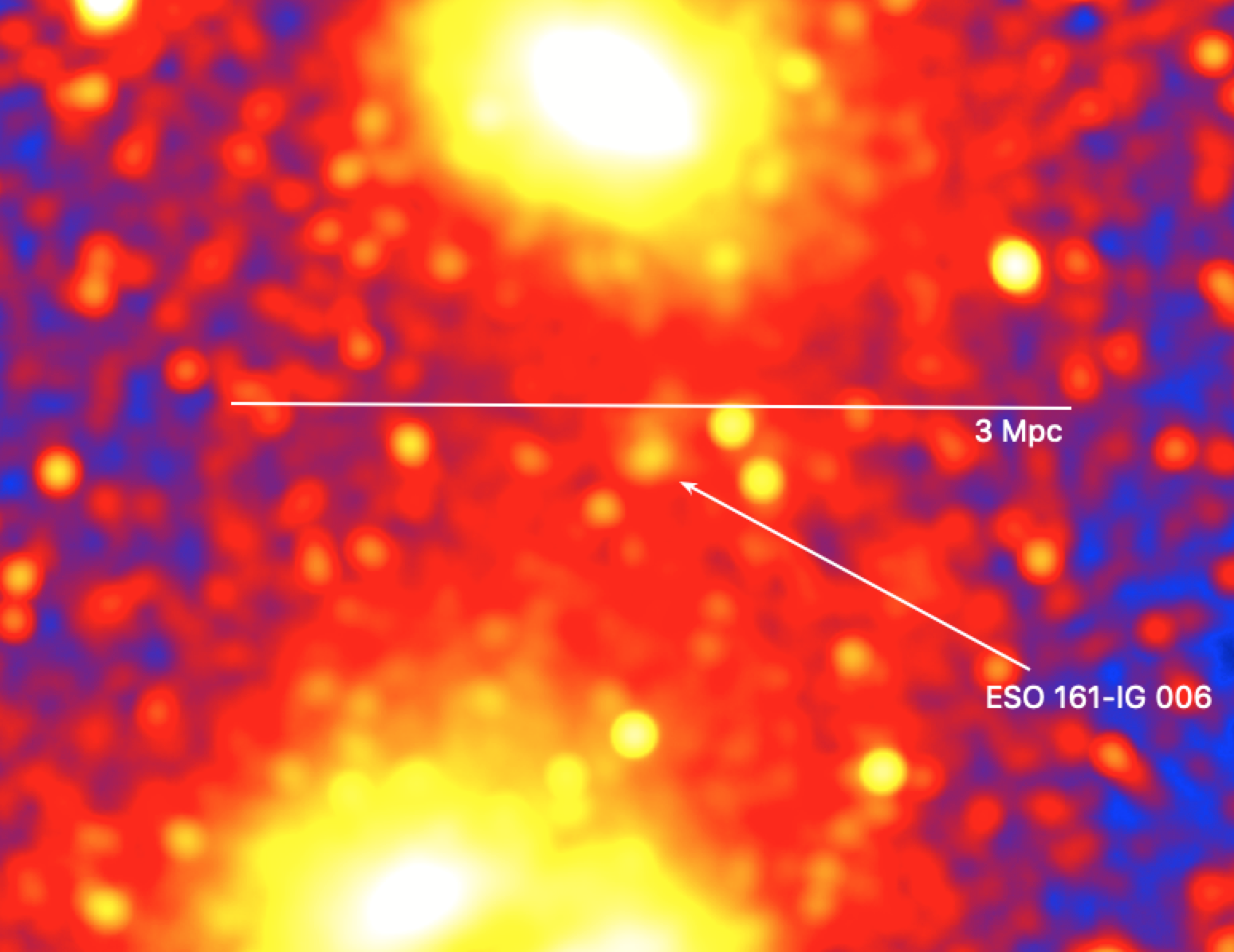}
   \caption{Gaussian smoothed versions of the \rosi\ count rate image. Left: Zoom in to the apparently infalling Northern Clump, hosting a bright WAT (F2, see Fig.~\ref{fig:EMU}).
   Right: Zoom in to the emission bridge between A3391 and A3395. Three Mpc correspond to 46$'$.}
     \label{fig:nclump}%
    \end{figure*}

The left panel of Figure~\ref{fig:nclump} shows an \rosi\ zoom into the Northern Clump. We note features in the surface brightness morphology: a boxyness, an apparent surface brightness edge to the south, and a possible tail to the north. While other interpretations are possible, these features are consistent with the Northern Clump being an infalling galaxy cluster experiencing ram pressure due to its motion relative to the very outskirts of A3391; that is, the infalling Northern Clump appears to feel the relative motion of gas located at about twice the virial radius, $r_{100}$, of A3391, presumably from primordial filament gas. However, considering this further, a large relative motion might actually be surprising as one should expect this potential filament gas to be infalling into A3391 as well; that is, the relative motion between the Northern Clump and this gas might be relatively small as they would both be moving in the same direction. On the other hand, gas in the potential filament may also be attracted by the Northern Clump itself, especially if they are  close to each other, resulting again in a larger relative motion. A more detailed investigation is required; the Northern Clump features will be discussed in an upcoming paper (Veronica et al., in prep.).

\begin{table}
\caption{Main cluster, Northern Clump, and Little Southern Clump redshifts, comoving distances, and approximate peculiar velocities (see Section~\ref{clumps}). $N$ denotes the number of constituent galaxies with spectroscopic redshift.}             \label{tab:z}      
\centering                          
\begin{tabular}{c c c c c}        
\hline\hline                 
Cluster & $N$ & $z$ & $D_{\rm c}$ & $v_{\rm pec}$\\    
\hline                        
A3391 & 57 & 0.0555 & 234.7 Mpc & 0 km/s\\
A3395n & 133 & 0.0518 & 219.2 Mpc & 1109 km/s\\
A3395s & 64 & 0.0517 & 218.8 Mpc & 1139 km/s\\
Northern Clump & 3 & 0.0511 & 216.3 Mpc & 1319 km/s\\
L. S. Clump & 1 & 0.0562 & 237.6 Mpc & $-$209.9 km/s\\
\hline                                   
\end{tabular}
\end{table}

\begin{table}
\caption{Main cluster centers (emission peaks) and relevant cluster radii.}             
\label{tab:radii}      
\centering                          
\begin{tabular}{l r r r}        
\hline\hline                 
Center/Radius / cluster & A3391 & A3395n & A3395s\\    
\hline
RA (J2000) & 96.587 & 96.906 & 96.702 \\
Dec (J2000) & $-$53.692 & $-$54.447 & $-$54.546 \\
\hline
$r_{2500}$ [$'$] & 7.79 & 9.82 & 9.88 \\
$r_{2500}$ [Mpc] & 0.504 & 0.595 & 0.598 \\
\hline
$r_{500}$ [$'$] & 18.09 & 22.81 & 22.94 \\
$r_{500}$ [Mpc] & 1.170 & 1.383 & 1.388 \\
\hline
$r_{200}$ [$'$] & 27.83 & 35.09 & 35.29 \\
$r_{200}$ [Mpc] & 1.800 & 2.128 & 2.136 \\
\hline
$r_{100}$ [$'$] & 37.85 & 47.72 & 47.99 \\
$r_{100}$ [Mpc] & 2.448 & 2.893 & 2.904 \\
\hline
$3r_{200}$ [$'$] & 83.49 & 105.27 & 105.87 \\
$3r_{200}$ [Mpc] & 5.400 & 6.383 & 6.407 \\
\hline                                   
\end{tabular}
\end{table}

\subsection{Bridge and filaments}
\label{filaments}

Before we show direct evidence for bridges and filaments in the form of diffuse X-ray emission,
we remind the reader that some of the infalling clumps discussed in the previous sections, for example, Figs.~\ref{fig:EMU} and \ref{fig:eROSITA}, trace gas well beyond $r_{200}$ or even beyond $r_{100}$ of A3391. Therefore, these clumps may indeed be considered to provide indirect evidence of filaments and, moreover, to trace the properties of filamentary regions along which matter falls towards the main clusters.

On the right of Fig.~\ref{fig:nclump}, a close-up view of the emission bridge between the main clusters is shown. This is a simple Gaussian smoothed image, and so the photon noise is reduced but not as much as in the wavelet-filtered images, and therefore the true underlying emission is not expected to be as clumpy as it appears here. We also note the X-ray emission from the known galaxy group to the south of A3391, ESO 161-IG 006 ($z=0.0516$). The emission from this group is limited to a small area; the whole bridge region has a much larger horizontal extent of at least 3 Mpc. Therefore,  this emission bridge is clearly not dominated by this group but is due to  gas from the cluster outskirts well beyond $r_{500}$ or gas from a filament between A3391 and A3395n/s. The numbers shown in Table~\ref{tab:z} make it clear that we cannot decipher the true physical separation between A3391 and A3395n/s simply from the redshifts, as the measured redshifts are caused by both the cosmological redshift and possible peculiar motions. We only know that the projected separation of their centers ($\sim$50$'$) is larger than the sum of the $r_{500}$ radii of A3391 and A3395n ($\sim$40$'$, Table~\ref{tab:radii}), but the 3D separation may be much larger, which could imply that we are looking down a significant column density of true primordial filament gas \citep[e.g.,][]{th01}.
\citet[][]{sti17} and \citet[][]{arb18} argued that the physical separation is not much larger than the projected one, a conclusion mostly based on a lack of warm emission, as they find a temperature in the bridge region of $\sim$3.5 keV and $\sim$4.5 keV, respectively.
From true primordial filament gas one may expect much cooler emission (unless it was heated by compression). \citet[][]{arb18} interpret the excess emission as being due to cluster outskirts gas that has been moved further away from their host clusters due to tidal forces during the interaction.
%
   \begin{figure}
   \centering
    \includegraphics[width=\columnwidth]{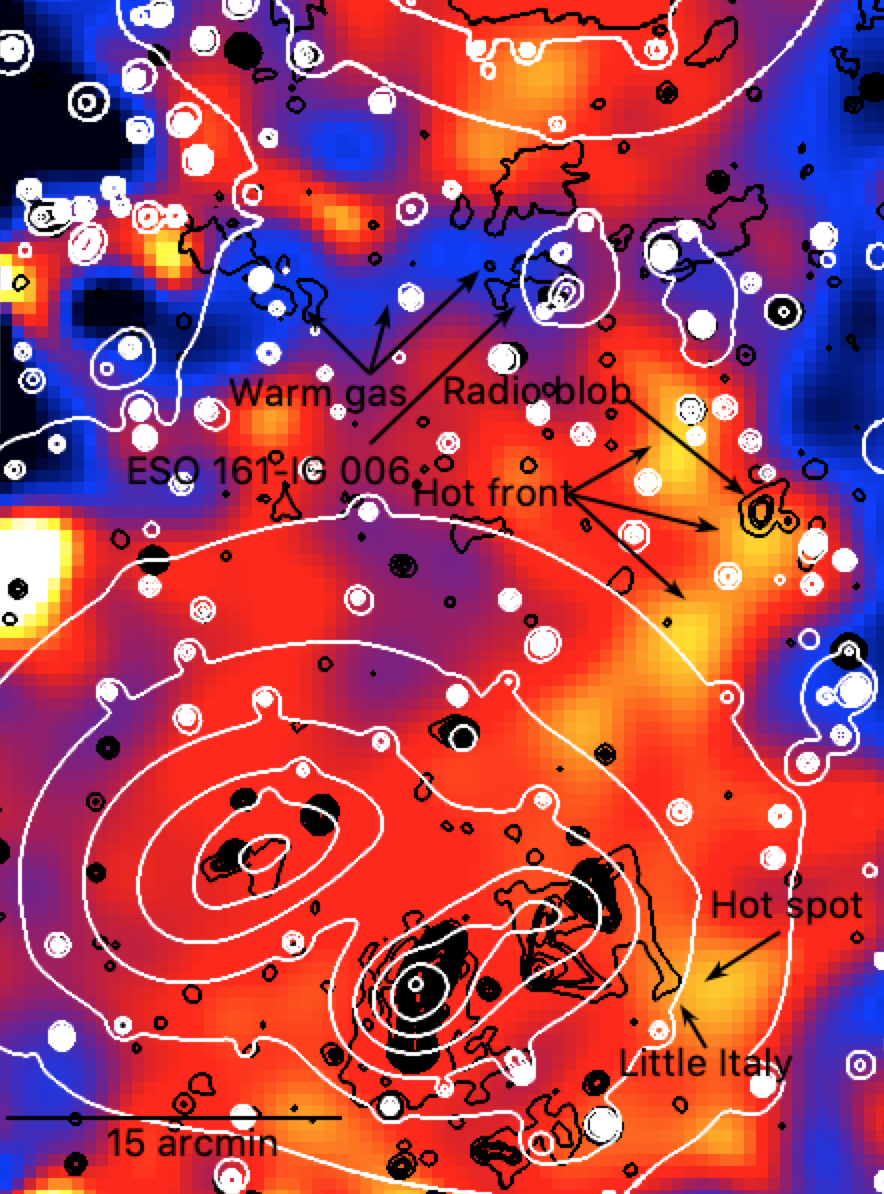}
   \caption{Oxygen band image of the A3391/95 emission bridge region. White contours are from the \rosi\ wavelet-filtered image and black contours from the ASKAP/EMU radio image. We note the increased OVII+OVIII (warm) emission in blue in the middle of the main emission bridge. We also note the regions of apparently hotter gas (yellow), some of which are close to potentially interesting radio features (the southern ``Little Italy'' extension of S3 and the ``Radio blob'' near the ``Hot front'' which has no optical counterpart).}
    \label{fig:oxygen}%
    \end{figure}

Plasma at temperatures $\gtrsim$3.5 keV is not expected to emit strong oxygen emission lines as most oxygen should be completely ionized. One way to search for the presence of warm ($\lesssim$1 keV) plasma is therefore to search for oxygen emission lines, in particular OVII (574 eV) and OVIII (654 eV).
A spatially resolved search can be conducted by creating an oxygen-to-soft-band-ratio (``oxygen band'') image where fore- and background emission has been subtracted.
\rosi\ has higher effective area and, most importantly, a better energy resolution around the relevant photon energy, $\sim$0.6 keV ($<$70 eV, \citealt{paa20}), than any other past or present X-ray imaging telescope. Furthermore, the large area covered by \rosi\ allows us to properly take into account the soft foreground emission coming from within the Milky Way.
Therefore, we constructed an \rosi\ oxygen band image in the following way.
Given the approximate redshift of the system, $z\approx 0.05$, the emission lines of OVII and OVIII are shifted to 547 eV and 623 eV, respectively.
Assuming an energy resolution not worse than 70 eV at those energies, we chose the band 512--658 eV as the ``oxygen band'' while for the continuum soft band we used 0.5--2.0 keV. We then calculated the oxygen band image as $[(I_{\rm ox}-B_{\rm ox})/E_{\rm ox}]/[(I_{\rm soft}-B_{\rm soft})/E_{\rm soft}-(I_{\rm ox}-B_{\rm ox})/E_{\rm ox}]$
while applying a smoothing to the main numerator and denominator.
Here $I$ is a photon image, $B$ a (particle+X-ray) background image, and $E$ an exposure map.
Given that images with only photons $<$1 keV are required in this procedure, it is clear that TM9 images were not used here. A demonstration of the effectiveness of this method to identify warm gas with \rosi\ is provided through SIXTE-simulated maps of the Magneticum clusters in Appendix \ref{App_E}.

The resulting oxygen band image is displayed in Fig.~\ref{fig:oxygen}. To the east of the ESO 161-IG 006 group we indeed find tantalizing suggestions of the presence of warm emission (blue) in the bridge region. This may indicate that we have detected true primordial filamentary gas, which in turn may be expected if the 3D separation of A3391 and A3395n/s were significantly larger than the projected one, contrary to recent findings \citep[e.g.,][]{arb18}. We will quantify the significance of the warm gas detection in a full spectral analysis of the \rosi\ data, taking into account variable foreground absorption and emission (Ota et al., in prep.).
Furthermore, there appear to be some particularly hot (yellow) regions, not strongly correlated with the surface brightness structure. We note that the \xmm\ temperature map of A3395 appears to show hotter gas in similar regions to those found here \citep[][]{2011ApJ...743...78L}. These hotter regions may indicate gas being heated by compression or even shocks from ongoing merger activity. While the merger between A3391 and A3395 proceeds in the north--south direction, equatorial shocks moving east and west may be possible, especially in the early phases of the merger. Alternatively, shocks and compression due to east--west accretion onto the bridge/filament could give rise to hotter regions. Interestingly, parts of the hot regions coincide with diffuse radio emission features that do not have an optical counterpart: for example, the ``Little Italy'' extension south of S3 (the exact shape of which may be determined by noise though) and the small diffuse ``Radio blob''. This raises the question of whether this relativistic plasma could be due to (re-) acceleration processes related to merging activity and/or accretion.

   \begin{figure*}
   \centering
    \includegraphics[width=2.0\columnwidth]{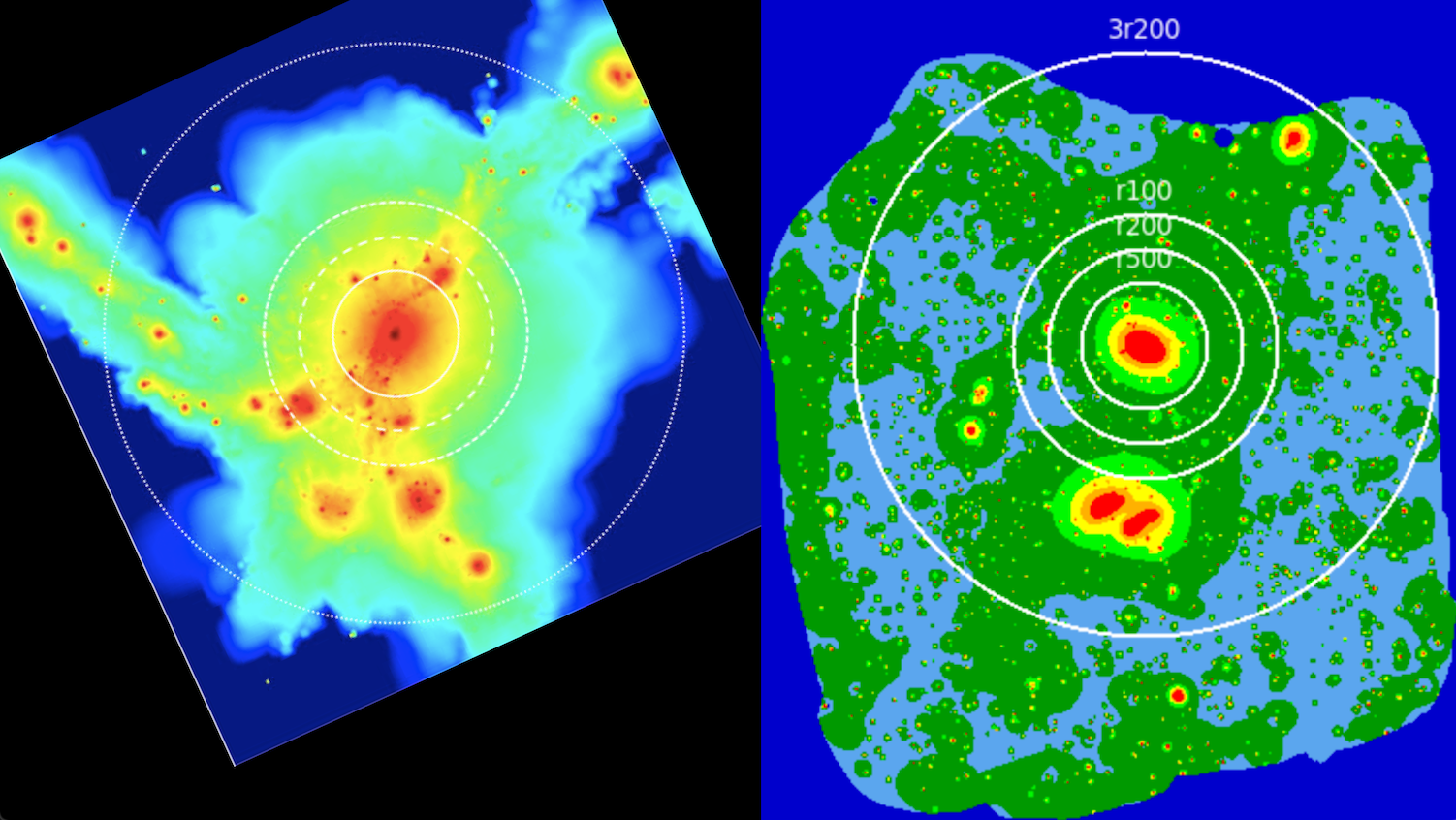}
   \caption{Left: Illustration of various radii of interest overlaid on the gas density distribution of a merging cluster from a hydrodynamic simulation \citep[rotated to appear similar to the A3391/95 structures, from][]{rbe13,red06}. Right: \rosi\ wavelet-filtered image as in Fig.~\ref{fig:eROSITA} but removing all areas with less than 1 ks effective exposure and using a color scheme to highlight low-surface-brightness features.
Overlaid are the same radii for A3391 as shown on the left for the simulated cluster.}
              \label{fig:eROSITA2}%
    \end{figure*}
%
Let us now focus our attention away from the emission bridge between the clusters but instead on the general cluster outskirts and also on potential large-scale structure emission filaments. 
The right panel of Figure~\ref{fig:eROSITA2}  shows a representation of the wavelet-filtered \rosi\ image with some radii of interest overplotted for A3391 (Table~\ref{tab:radii}). We note that for this image all areas with less than 1 ks effective exposure were masked during the wavelet run.
 The emission of A3391 is traced out well beyond $r_{200}$, and even beyond $r_{100}$ where the emission becomes strongly irregular. This is a major step forward from \xmm\ and \cha\ observations of clusters where emission is typically traced out to only $r_{500}$ or at most $r_{200}$. Moreover, there is an indication for an emission filament beyond $r_{100}$ connecting A3391 with the cluster in the north at the same redshift (the Northern Clump, see Section~\ref{clumps}), which we discuss further below.

\begin{figure*}
   \centering
    \includegraphics[width=1.0\columnwidth]{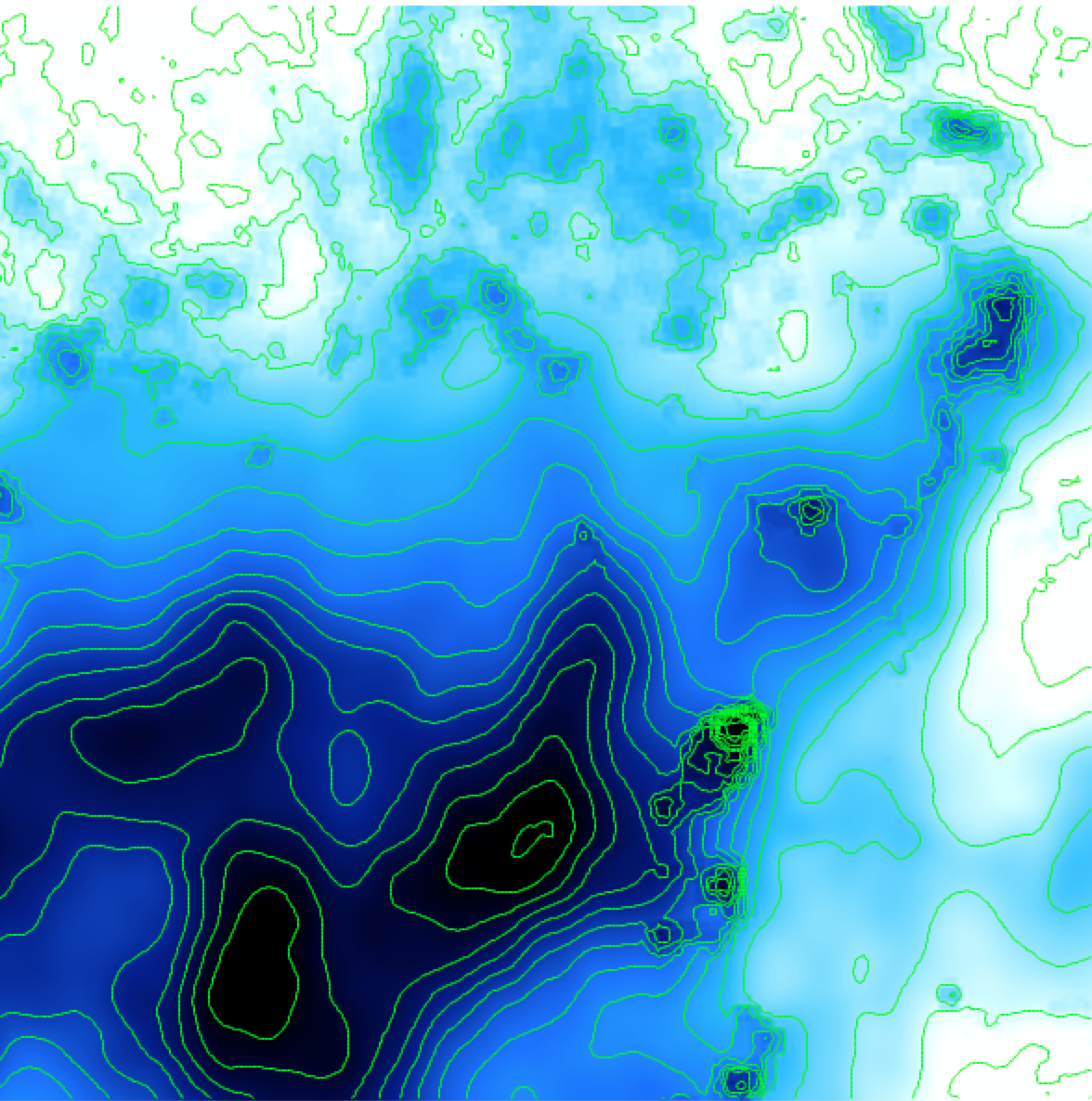}
    \includegraphics[width=0.99\columnwidth]{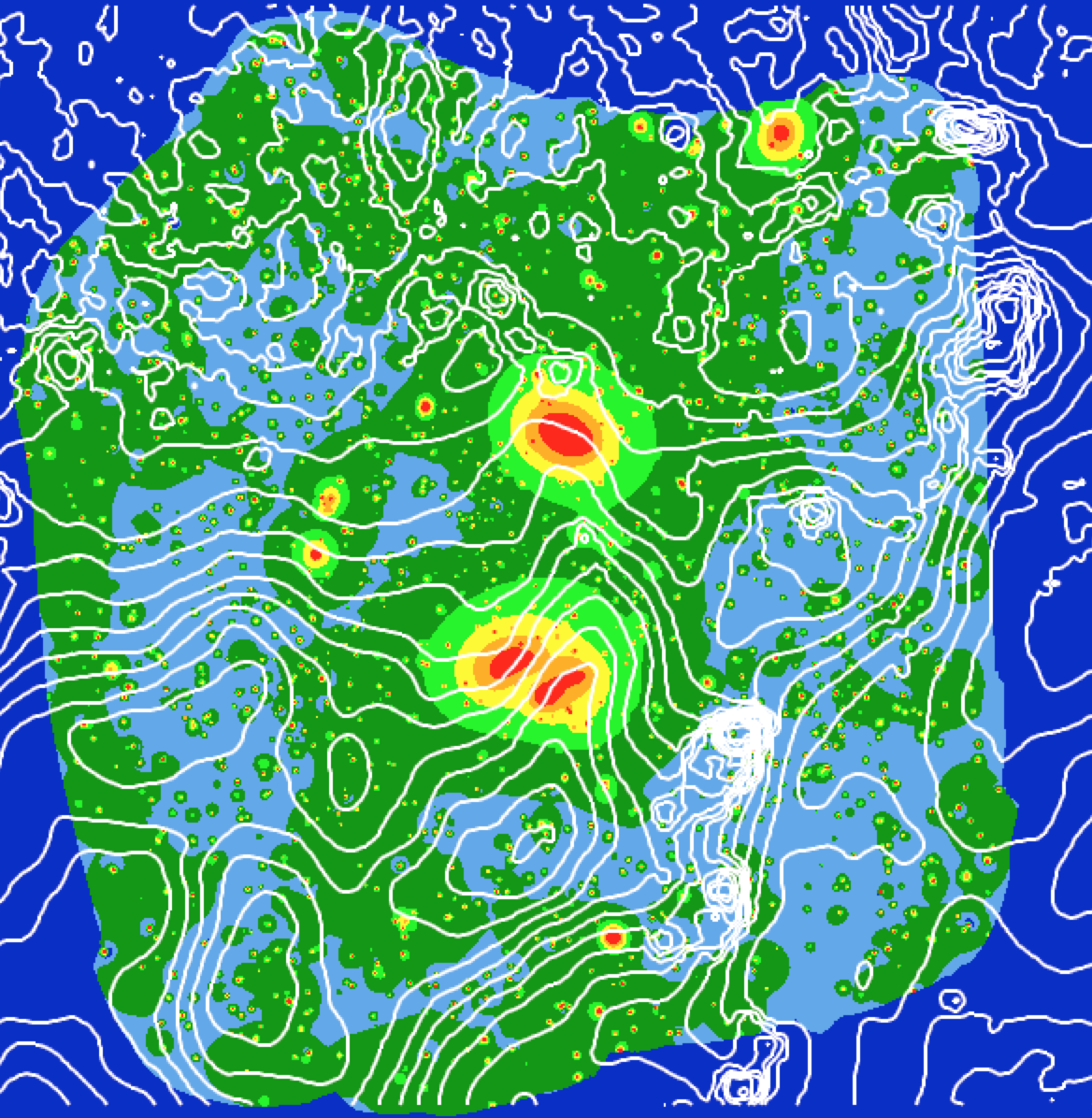}
   \caption{Left: Total $N_H$ map taking into account the contribution from both neutral and molecular hydrogen. The maximum variation in the field is from $3\times10^{20}$ cm$^{-2}$ to $1\times10^{21}$ cm$^{-2}$. Right: $N_H$ contours overlaid onto the wavelet-filtered \rosi\ image.}
              \label{fig:NH}%
    \end{figure*}

Such an emission filament would be a significant discovery. As we are tracing soft emission over large angular scales, we do need to worry about structure in foreground absorption. Some of the structures in the surface brightness map may be due to spatially varying foreground absorption. In order to test for absorption, we constructed  a total $N_{\rm H}$ map  as described in Section~\ref{NH}; see  Fig.~\ref{fig:NH} (left).
In some regions, an apparent anti-correlation between X-ray surface brightness and $N_{\rm H}$ column density is indicated, as expected if one were to assume a uniform X-ray background. However, the apparent emission filament towards the Northern Clump actually lies in a region with slightly {increased} hydrogen column density compared to neighboring regions east and west that show less emission. Therefore, this apparent filament may indeed be due to a real emission excess stemming from a filament connecting A3391 with the Northern Clump.

%
%
   \begin{figure}
   \centering
\includegraphics[width=1.0\columnwidth]{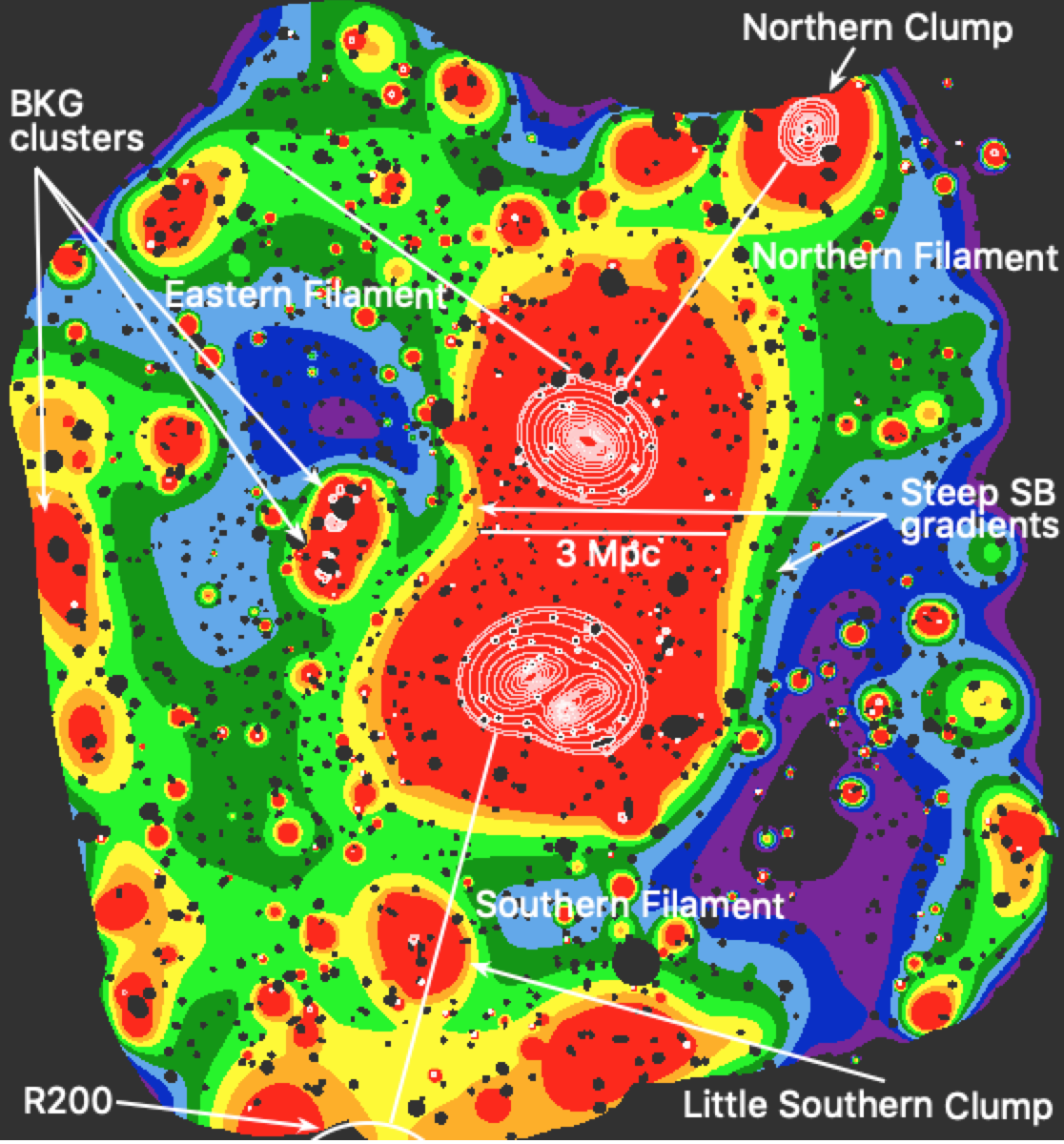}
   \caption{Wavelet-filtered absorption-corrected \rosi\ surface brightness image restricted to areas with at least 1~ks exposure. Point sources have been excised. We note the apparent very long ($\sim$15 Mpc) continuous emission filament running from the Northern Clump through A3391 and A3395 down beyond the Little Southern Clump where cluster MCXC J0631.3-5610 lies outside the \rosi\ FoV (its $r_{200}$ just visible at the bottom). There is an indication for an additional, weaker eastern filament. We also note the steep surface brightness gradients to the east and west of the 3 Mpc-wide emission bridge between A3391 and A3395. Emission from a few unrelated background clusters is also marked.}
    \label{fig:unabs}%
    \end{figure}
%

   \begin{figure}
   \centering
    \includegraphics[width=1.0\columnwidth]{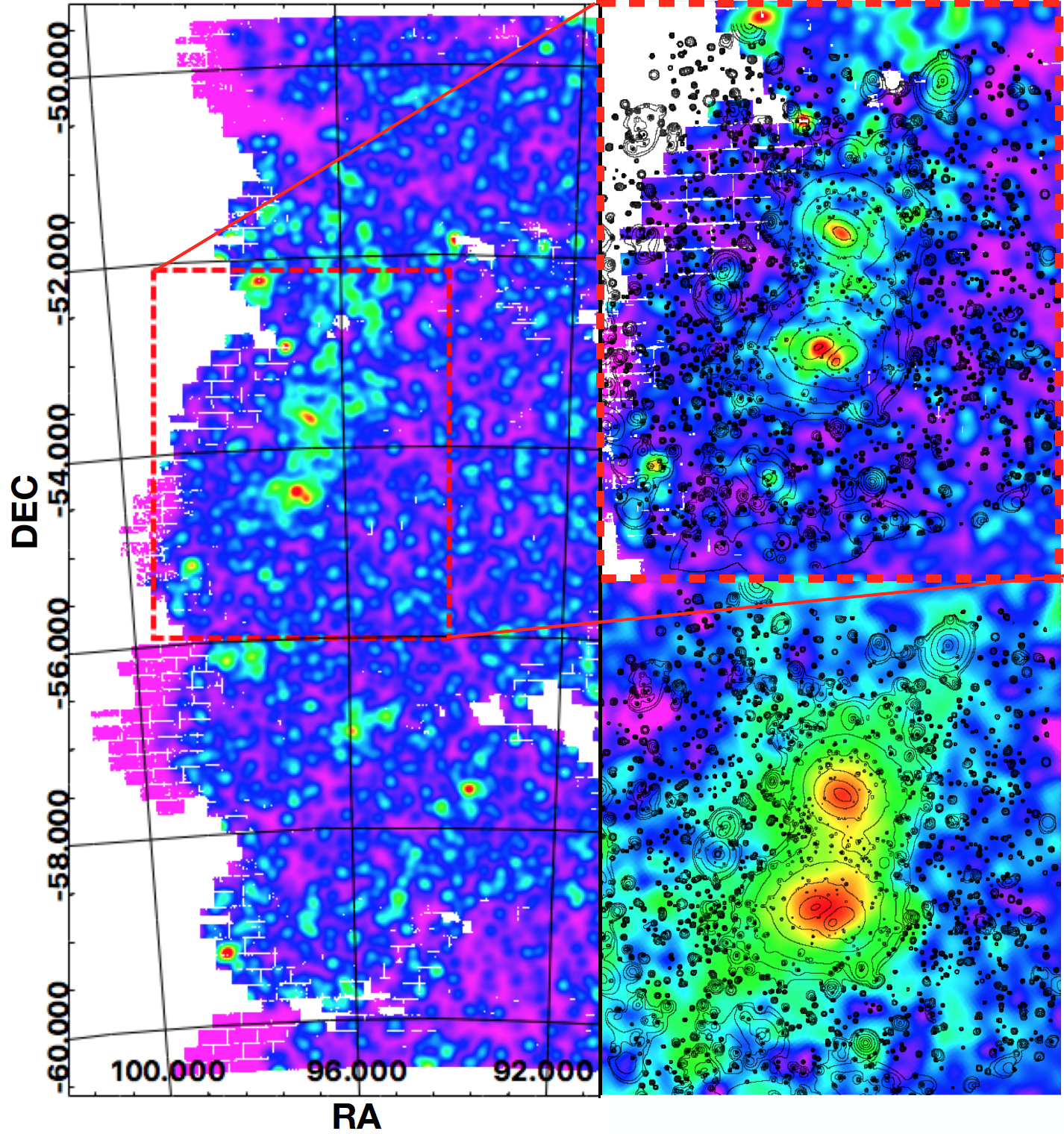}
   \caption{Left: Large-scale DECam galaxy density map of the A3391/95 system. Right/top: Zoom into the DECam galaxy density map. Overlaid are contours from the \rosi\ wavelet-filtered image.
   Right/bottom: \textit{Planck} $y$-map on the same spatial scale as in the top image, also with \rosi\ contours.}
    \label{fig:Planck}%
    \end{figure}

To investigate this in more detail, we constructed an absorption-corrected surface brightness map.
For this, we chose a simple model, that is, we assumed an X-ray fore- and background emission spectrum including one unabsorbed and one absorbed diffuse thermal component plus an absorbed power-law component due to unresolved background AGNs. We then determined the expected \rosi\ count rates once when assuming the column density at the respective position and once when assuming the median hydrogen column density, which resulted in a correction map of the ratio of the former to the latter. Subsequently, we divided the exposure map by this ratio map and produced a new exposure-, vignetting-, and absorption-corrected and PIB-subtracted image. The correction is typically $\lesssim$10\%.

In Fig.~\ref{fig:unabs} we show this new \rosi\ image, where we also ran the \textsc{SExtractor} source detection algorithm and excised detected point sources.
This absorption-corrected surface brightness map not only corroborates the Northern Filament towards the Northern Clump but an additional ``Southern Filament'' south of A3395n/s going through the ``Little Southern Clump'' towards MCXC J0631.3-5610 (at $z=0.0540$ and just outside the border of the \rosi\ image) becomes apparent.
The outskirts of MCXC J0631.3-5610 are visible as enhanced emission south of the Little Southern Clump, extending further along the filament. The projected separation between the Northern Clump and the center of MCXC J0631.3-5610 is 3.751 deg = 14.55 Mpc; that is, the total projected length of continuous emission along this filament is at least 15 Mpc.
This has never been seen before in observations of a single system.

To investigate the apparent excess emission and its significance quantitatively, we placed square boxes with 0.5 deg side length on the absorption-corrected and point source-excised \rosi\ count rate image (not the wavelet-filtered image)
and determined their surface brightnesses and standard deviations. From this, we derived excess surface brightnesses and significances for three potential filament regions: the Northern and Southern Filaments described above and also the apparent weak emission excess leading from A3391 towards the east (Table~\ref{tab:fils}). Details of the method are described in Appendix~\ref{App_F}.
The significance for the Eastern Filament is low, $<$3$\sigma$, but the Northern and Southern Filaments show $>$4$\sigma$. Therefore, we consider the Northern and Southern filaments to be clear detections and the Eastern Filament  a tentative detection. For completeness, we also provide excess surface brightness and significance for the bridge region in Table~\ref{tab:fils}, although this region clearly also contains cluster outskirt gas and not only filament gas (again, details are given in Appendix~\ref{App_F}).

\begin{table}[ht]
\centering
\caption{Filament excess surface brightnesses and significances}
\label{tab:fils}
{\small
\begin{tabular}{@{}lcc@{}}
\hline
\hline
Filament & Surface brightness excess & Significance \\
\hline
Northern Filament &
$\left(23.5_{-2.6}^{+2.7}\right)\%$ &
$\left(4.6_{-1.3}^{+1.5}\right)\sigma$ \\
Southern Filament &
$\left(20.7_{-2.3}^{+2.4}\right)\%$ &
$\left(4.1_{-1.1}^{+1.3}\right)\sigma$ \\
Eastern Filament &
$\left(14.4_{-2.4}^{+2.5}\right)\%$ &
$\left(2.8_{-0.8}^{+1.0}\right)\sigma$ \\
Bridge & 
$\left(137.0_{-6.7}^{+7.2}\right)\%$ & $\left(26.9_{-7.0}^{+8.2}\right)\sigma$ \\
\hline
\hline
\end{tabular}
}
\end{table}
We note that the apparent north--south emission ``stripe'' at the eastern border of the image is more difficult to interpret. There are unrelated background clusters at that position and also it is a region that might be affected by remaining low-level systematic errors due to exposure- and vignetting correction and background subtraction. Therefore this region is not further discussed here. The detailed structure, significance, and redshift of all apparent emission filaments will be quantified in detail by Veronica et al.\ (in prep.).

In addition to the 15 Mpc emission filament, we can identify further features that resemble what we see in the Magneticum simulations (Fig.~\ref{fig:magneticum_evol}). The emission bridge connecting A3391 and A3395 is very wide, about 3 Mpc. Moreover, there are strong surface-brightness drops east and west of the bridge. This is consistent with the expectation that gas surrounding the nodes (clusters) and filaments of the cosmic web gets accreted by these denser structures. There may even be accretion shocks forming by gas falling onto the bridge, resulting in  jumps in gas density and temperature. There is indeed some indication of a very hot region in the bridge in the oxygen-to-soft-band ratio image (Fig.~\ref{fig:oxygen}) but detailed spectral analysis is required to test for and trace possible temperature jumps in detail, which is beyond the scope of this paper.

If we see such a warm gas emission filament in X-rays it may  also be visible in the submm band through the Sunyaev-Zeldovich effect and in the galaxy distribution, as we now know where to look for it.
The left panel of Figure~\ref{fig:Planck} shows the galaxy density distribution in the very large-scale environment of the A3391/95 system, generated as described in Section~\ref{DEC}. A3391, A3395n, A3395s, and the bridge connecting them all clearly show up, corroborating the visual impression from Fig.~\ref{fig:DECam}. Furthermore, we also clearly see a north--south large-scale structure where galaxy densities are higher than in the east--west direction. Zooming further in and overlaying the \rosi\ contours (Fig.~\ref{fig:Planck}, right/top) shows a remarkable resemblance of X-ray emission excess and galaxy overdensity. This is exactly what is expected for a true filament.

In the bottom-right panel of Fig.~\ref{fig:Planck}, we show the \textit{Planck} $y$-map obtained as described in Section~\ref{SZ}.
The \rosi\ filament discovery towards the Northern Clump and Little Southern Clump seems to be supported by \textit{Planck} as the $y$-map reveals excess pressure in similar directions. The Little Southern Clump itself is not very obvious in the $y$-map, which seems consistent with it being only a small clump at the A3391/95 redshift and not a massive background cluster. The fact that the center of the Northern Clump also does not appear very prominent in the $y$-map could possibly be due to the strong central AGN and/or masking.
We also note that \textit{Planck} alone cannot resolve smaller structures like the individual components A3395n and A3395s; a combination with higher resolution SZ maps, for example from SPT or ACT, may improve this \citep[e.g.,][]{2019A&A...632A..47A}.

In summary, with \rosi\ we discovered a continuous 15 Mpc WHIM filament of significant X-ray emission including a number of clusters, groups, and clumps. The presence of this filament is supported by the Planck $y$-map, indicating excess pressure, and by the optical galaxy density map, indicating a galaxy overdensity. This filament reaches the edges of the \rosi\ FoV in both north and south directions, meaning that it might be even longer. When the continuing \rosi\ all-sky survey has acquired sufficient depth, we may be able to trace this new X-ray emission filament even further beyond the current raster scan observations from \rosi's PV phase.

\section{Conclusions}
\label{conclu}
We provide an overview of new sensitive large-scale \rosi\ and additional multi-wavelength  observations of the A3391/95 galaxy cluster system and its surroundings. Here, we report a range of new discoveries that will certainly trigger further detailed in-depth studies. Our most important conclusions are as follows.

   \begin{enumerate}
      \item We present a homogeneous soft X-ray image of $\sim$15 deg$^2$ taken with \rosi\ in scanning mode, providing the deepest large-scale X-ray image ever taken of this complex system.
      \item We characterize the complex morphology of the central parts of the main clusters combining radio, optical, and X-ray data, providing evidence for ongoing dynamical activity.
      \item We trace X-ray emission beyond the virial radii, $r_{100}$, of the main clusters, illustrating strong irregularity in the outer X-ray surface brightness and therefore gas density distribution.
      \item We discover a number of extended X-ray-emitting clumps surrounding the system; we use the X-ray and radio morphologies to suggest that the Northern Clump is falling towards A3391.
      \item With the \rosi\ data we detect about one order of magnitude more (unrelated) background galaxy group and cluster candidates in this field than the ACT and SPT SZ surveys together, including a known SZ cluster at $z=1$.
      \item The \rosi\ image reveals that the bridge region connecting A3391 and A3395
      has an extent of at least 3 Mpc perpendicular to their line of separation; the X-ray emission from the galaxy group ESO 161-IG 006 therefore only accounts for a small fraction of the bridge emission.
      \item Based on an oxygen band image, exploiting \rosi's superior soft response, large FoV, and good spatial resolution, we find hints of emission from warm gas in addition to the known hot gas emission in the bridge region that went unnoticed by past observations. If further confirmed by full spectral analysis, this could imply that the 3D separation of A3391 and A3395 is much larger than the projected one and that we are looking down a filament and not just at a bridge.
      \item The oxygen band image also indicates some particularly hot regions in the bridge area and southwest of A3395s, possibly related to detected diffuse radio emission.
      \item The \rosi\ absorption-corrected surface brightness map reveals an impressive $\sim$15 Mpc long continuous X-ray emission filament. The presence of this new WHIM filament is corroborated by the \textit{Planck} Sunyaev-Zeldovich map and the DECam galaxy density map.
      \item The new large-scale intergalactic gas observations of this system are strikingly similar to those found in one of the systems in the Magneticum cosmological hydrodynamic simulation. This corroborates the underlying assumptions going into simulations and adds further credibility to predictions, such as for example those concerning the WHIM distribution.
   \end{enumerate}

In the study of LSS, and in particular of the warm-hot gas in emission as traced by X-rays, individual systems will inevitably reveal a high degree of complexity, imprinted by the various merging geometries and histories of the LSS knots. One of the main assets of the ongoing \rosi\ all-sky survey will be the large number of systems discovered and characterized, which will hopefully allow robust statistical inferences as to the physical properties of the general population of such systems.

\begin{acknowledgements}
We would like to thank the referee for suggestions that helped improve the presentation of the paper. We thank H. Andernach for comments on the manuscript.
NO acknowledges support by a Fund for the Promotion of Joint International Research (JSPS KAKENHI Grant Number 16KK0101).
KD acknowledges support by the Deutsche Forschungsgemeinschaft (DFG, German Research Foundation) under Germany’s Excellence Strategy – EXC-2094 – 3907833. VB acknowledges support by the DFG project nr.\ 415510302 (BI 1699/2-1).
Partial support for LR comes from U.S.\ National Science Foundation grant AST17-14205 to the University of Minnesota.
This work is based on data from \rosi, the primary instrument aboard SRG, a joint Russian-German science mission supported by the Russian Space Agency (Roskosmos), in the interests of the Russian Academy of Sciences represented by its Space Research Institute (IKI), and the Deutsches Zentrum f\"ur Luft- und Raumfahrt (DLR). The SRG spacecraft was built by Lavochkin Association (NPOL) and its subcontractors, and is operated by NPOL with support from the Max Planck Institute for Extraterrestrial Physics (MPE).

The development and construction of the \rosi\ X-ray instrument was led by MPE, with contributions from the Dr. Karl Remeis Observatory Bamberg \& ECAP (FAU Erlangen-N\"urnberg), the University of Hamburg Observatory, the Leibniz Institute for Astrophysics Potsdam (AIP), and the Institute for Astronomy and Astrophysics of the University of T\"ubingen, with the support of DLR and the Max Planck Society. The Argelander Institute for Astronomy of the University of Bonn and the Ludwig Maximilians Universit\"at Munich also participated in the science preparation for \rosi.

The \rosi\ data shown here were processed using the eSASS software system developed by the German \rosi\ consortium.

The Australian SKA Pathfinder is part of the Australia Telescope National Facility which is managed by CSIRO. Operation of ASKAP is funded by the Australian Government with support from the National Collaborative Research Infrastructure Strategy. ASKAP uses the resources of the Pawsey Supercomputing Centre. Establishment of ASKAP, the Murchison Radio-astronomy Observatory and the Pawsey Supercomputing Centre are initiatives of the Australian Government, with support from the Government of Western Australia and the Science and Industry Endowment Fund. We acknowledge the Wajarri Yamatji people as the traditional owners of the Observatory site.

This project used data obtained with the Dark Energy Camera (DECam), which was constructed by the Dark Energy Survey (DES) collaboration. Funding for the DES Projects has been provided by the US Department of Energy, the US National Science Foundation, the Ministry of Science and Education of Spain, the Science and Technology Facilities Council of the United Kingdom, the Higher Education Funding Council for England, the National Center for Supercomputing Applications at the University of Illinois at Urbana-Champaign, the Kavli Institute for Cosmological Physics at the University of Chicago, Center for Cosmology and Astro-Particle Physics at the Ohio State University, the Mitchell Institute for Fundamental Physics and Astronomy at Texas A\&M University, Financiadora de Estudos e Projetos,
Fundação
Carlos Chagas Filho de Amparo \`a Pesquisa do Estado do Rio de Janeiro, Conselho Nacional de Desenvolvimento Científico e Tecnológico and the Ministério da Ciência, Tecnologia e Inovação, the Deutsche Forschungsgemeinschaft and the Collaborating Institutions in the Dark Energy Survey.

The Collaborating Institutions are Argonne National Laboratory, the University of California at Santa Cruz, the University of Cambridge, Centro de Investigaciones Enérgeticas, Medioambientales y Tecnológicas–Madrid, the University of Chicago, University College London, the DES-Brazil Consortium, the University of Edinburgh, the Eidgen\"ossische Technische Hochschule (ETH) Z\"urich, Fermi National Accelerator Laboratory, the University of Illinois at Urbana-Champaign, the Institut de Ciències de l’Espai (IEEC/CSIC), the Institut de Física d’Altes Energies, Lawrence Berkeley National Laboratory, the Ludwig-Maximilians Universit\"at M\"unchen and the associated Excellence Cluster Universe, the University of Michigan, NSF’s NOIRLab, the University of Nottingham, the Ohio State University, the OzDES Membership Consortium, the University of Pennsylvania, the University of Portsmouth, SLAC National Accelerator Laboratory, Stanford University, the University of Sussex, and Texas A\&M University.

Based on observations at Cerro Tololo Inter-American Observatory, NSF’s NOIRLab (NOIRLab Prop.\ IDs 2018A-0386 and 2019B-0323; PI: A. Zenteno), which is managed by the Association of Universities for Research in Astronomy (AURA) under a cooperative agreement with the National Science Foundation.
\end{acknowledgements}

%
   \bibliographystyle{aa} 
   \bibliography{aamnem99.bib,bibo_engl.bib} 

\begin{thebibliography}{127}
\expandafter\ifx\csname natexlab\endcsname\relax\def\natexlab#1{#1}\fi

\bibitem[{{Aghanim} {et~al.}(2019){Aghanim}, {Douspis}, {Hurier}, {Crichton},
  {Diego}, {Hasselfield}, {Macias-Perez}, {Marriage}, {Pointecouteau},
  {Remazeilles}, \& {Soubri{\'e}}}]{2019A&A...632A..47A}
{Aghanim}, N., {Douspis}, M., {Hurier}, G., {et~al.} 2019, \aap, 632, A47

\bibitem[{{Akamatsu} {et~al.}(2017){Akamatsu}, {Fujita}, {Akahori}, {Ishisaki},
  {Hayashida}, {Hoshino}, {Mernier}, {Yoshikawa}, {Sato}, \& {Kaastra}}]{afa17}
{Akamatsu}, H., {Fujita}, Y., {Akahori}, T., {et~al.} 2017, A\&A, 606, A1

\bibitem[{{Alam} {et~al.}(2015){Alam}, {Albareti}, {Allende Prieto}, {Anders},
  {Anderson}, {Anderton}, {Andrews}, {Armengaud}, {Aubourg}, {Bailey}, {Basu},
  {Bautista}, {Beaton}, {Beers}, {Bender}, {Berlind}, {Beutler}, {Bhardwaj},
  {Bird}, {Bizyaev}, {Blake}, {Blanton}, {Blomqvist}, {Bochanski}, {Bolton},
  {Bovy}, {Shelden Bradley}, {Brandt}, {Brauer}, {Brinkmann}, {Brown},
  {Brownstein}, {Burden}, {Burtin}, {Busca}, {Cai}, {Capozzi}, {Carnero
  Rosell}, {Carr}, {Carrera}, {Chambers}, {Chaplin}, {Chen}, {Chiappini},
  {Chojnowski}, {Chuang}, {Clerc}, {Comparat}, {Covey}, {Croft}, {Cuesta},
  {Cunha}, {da Costa}, {Da Rio}, {Davenport}, {Dawson}, {De Lee}, {Delubac},
  {Deshpande}, {Dhital}, {Dutra-Ferreira}, {Dwelly}, {Ealet}, {Ebelke},
  {Edmondson}, {Eisenstein}, {Ellsworth}, {Elsworth}, {Epstein}, {Eracleous},
  {Escoffier}, {Esposito}, {Evans}, {Fan}, {Fern{\'a}ndez-Alvar}, {Feuillet},
  {Filiz Ak}, {Finley}, {Finoguenov}, {Flaherty}, {Fleming}, {Font-Ribera},
  {Foster}, {Frinchaboy}, {Galbraith-Frew}, {Garc{\'\i}a},
  {Garc{\'\i}a-Hern{\'a}ndez}, {Garc{\'\i}a P{\'e}rez}, {Gaulme}, {Ge},
  {G{\'e}nova-Santos}, {Georgakakis}, {Ghezzi}, {Gillespie}, {Girardi},
  {Goddard}, {Gontcho}, {Gonz{\'a}lez Hern{\'a}ndez}, {Grebel}, {Green},
  {Grieb}, {Grieves}, {Gunn}, {Guo}, {Harding}, {Hasselquist}, {Hawley},
  {Hayden}, {Hearty}, {Hekker}, {Ho}, {Hogg}, {Holley-Bockelmann}, {Holtzman},
  {Honscheid}, {Huber}, {Huehnerhoff}, {Ivans}, {Jiang}, {Johnson},
  {Kinemuchi}, {Kirkby}, {Kitaura}, {Klaene}, {Knapp}, {Kneib}, {Koenig},
  {Lam}, {Lan}, {Lang}, {Laurent}, {Le Goff}, {Leauthaud}, {Lee}, {Lee},
  {Licquia}, {Liu}, {Long}, {L{\'o}pez-Corredoira}, {Lorenzo-Oliveira},
  {Lucatello}, {Lundgren}, {Lupton}, {Mack}, {Mahadevan}, {Maia}, {Majewski},
  {Malanushenko}, {Malanushenko}, {Manchado}, {Manera}, {Mao}, {Maraston},
  {Marchwinski}, {Margala}, {Martell}, {Martig}, {Masters}, {Mathur},
  {McBride}, {McGehee}, {McGreer}, {McMahon}, {M{\'e}nard}, {Menzel},
  {Merloni}, {M{\'e}sz{\'a}ros}, {Miller}, {Miralda-Escud{\'e}}, {Miyatake},
  {Montero-Dorta}, {More}, {Morganson}, {Morice-Atkinson}, {Morrison},
  {Mosser}, {Muna}, {Myers}, {Nand ra}, {Newman}, {Neyrinck}, {Nguyen},
  {Nichol}, {Nidever}, {Noterdaeme}, {Nuza}, {O'Connell}, {O'Connell},
  {O'Connell}, {Ogando}, {Olmstead}, {Oravetz}, {Oravetz}, {Osumi}, {Owen},
  {Padgett}, {Padmanabhan}, {Paegert}, {Palanque-Delabrouille}, {Pan},
  {Parejko}, {P{\^a}ris}, {Park}, {Pattarakijwanich}, {Pellejero-Ibanez},
  {Pepper}, {Percival}, {P{\'e}rez-Fournon}, {Ṕrez-Ra`fols}, {Petitjean},
  {Pieri}, {Pinsonneault}, {Porto de Mello}, {Prada}, {Prakash},
  {Price-Whelan}, {Protopapas}, {Raddick}, {Rahman}, {Reid}, {Rich}, {Rix},
  {Robin}, {Rockosi}, {Rodrigues}, {Rodr{\'\i}guez-Torres}, {Roe}, {Ross},
  {Ross}, {Rossi}, {Ruan}, {Rubi{\~n}o-Mart{\'\i}n}, {Rykoff},
  {Salazar-Albornoz}, {Salvato}, {Samushia}, {S{\'a}nchez}, {Santiago},
  {Sayres}, {Schiavon}, {Schlegel}, {Schmidt}, {Schneider}, {Schultheis},
  {Schwope}, {Sc{\'o}ccola}, {Scott}, {Sellgren}, {Seo}, {Serenelli}, {Shane},
  {Shen}, {Shetrone}, {Shu}, {Silva Aguirre}, {Sivarani}, {Skrutskie},
  {Slosar}, {Smith}, {Sobreira}, {Souto}, {Stassun}, {Steinmetz}, {Stello},
  {Strauss}, {Streblyanska}, {Suzuki}, {Swanson}, {Tan}, {Tayar}, {Terrien},
  {Thakar}, {Thomas}, {Thomas}, {Thompson}, {Tinker}, {Tojeiro}, {Troup},
  {Vargas-Maga{\~n}a}, {Vazquez}, {Verde}, {Viel}, {Vogt}, {Wake}, {Wang},
  {Weaver}, {Weinberg}, {Weiner}, {White}, {Wilson}, {Wisniewski},
  {Wood-Vasey}, {Ye`che}, {York}, {Zakamska}, {Zamora}, {Zasowski}, {Zehavi},
  {Zhao}, {Zheng}, {Zhou}, {Zhou}, {Zou}, \& {Zhu}}]{2015ApJS..219...12A}
{Alam}, S., {Albareti}, F.~D., {Allende Prieto}, C., {et~al.} 2015, \apjs, 219,
  12

\bibitem[{{Albrecht} {et~al.}(2006){Albrecht}, {Bernstein}, {Cahn}, {Freedman},
  {Hewitt}, {Hu}, {Huth}, {Kamionkowski}, {Kolb}, {Knox}, {Mather}, {Staggs},
  \& {Suntzeff}}]{abc06}
{Albrecht}, A., {Bernstein}, G., {Cahn}, R., {et~al.} 2006, ArXiv Astrophysics
  e-prints, astro-ph/0609591 [\eprint{astro-ph/0609591}]

\bibitem[{{Alvarez} {et~al.}(2018){Alvarez}, {Randall}, {Bourdin}, {Jones}, \&
  {Holley-Bockelmann}}]{arb18}
{Alvarez}, G.~E., {Randall}, S.~W., {Bourdin}, H., {Jones}, C., \&
  {Holley-Bockelmann}, K. 2018, ApJ, 858, 44

\bibitem[{{Amendola} {et~al.}(2013){Amendola}, {Appleby}, {Bacon}, {Baker},
  {Baldi}, {Bartolo}, {Blanchard}, {Bonvin}, {Borgani}, {Branchini}, {Burrage},
  {Camera}, {Carbone}, {Casarini}, {Cropper}, {de Rham}, {Di Porto}, {Ealet},
  {Ferreira}, {Finelli}, {Garc{\'{\i}}a-Bellido}, {Giannantonio}, {Guzzo},
  {Heavens}, {Heisenberg}, {Heymans}, {Hoekstra}, {Hollenstein}, {Holmes},
  {Horst}, {Jahnke}, {Kitching}, {Koivisto}, {Kunz}, {La Vacca}, {March},
  {Majerotto}, {Markovic}, {Marsh}, {Marulli}, {Massey}, {Mellier}, {Mota},
  {Nunes}, {Percival}, {Pettorino}, {Porciani}, {Quercellini}, {Read},
  {Rinaldi}, {Sapone}, {Scaramella}, {Skordis}, {Simpson}, {Taylor}, {Thomas},
  {Trotta}, {Verde}, {Vernizzi}, {Vollmer}, {Wang}, {Weller}, \&
  {Zlosnik}}]{aab13}
{Amendola}, L., {Appleby}, S., {Bacon}, D., {et~al.} 2013, Living Reviews in
  Relativity, 16, 6

\bibitem[{{Asgari} {et~al.}(2020){Asgari}, {Lin}, {Joachimi}, {Giblin},
  {Heymans}, {Hildebrandt}, {Kannawadi}, {St{\"o}lzner}, {Tr{\"o}ster}, {van
  den Busch}, {Wright}, {Bilicki}, {Blake}, {de Jong}, {Dvornik}, {Erben},
  {Getman}, {Hoekstra}, {K{\"o}hlinger}, {Kuijken}, {Miller}, {Radovich},
  {Schneider}, \& {Shan}}]{2020arXiv200715633A}
{Asgari}, M., {Lin}, C.-A., {Joachimi}, B., {et~al.} 2020, A\&A
accepted, arXiv e-prints,
  arXiv:2007.15633

\bibitem[{{Bennett} {et~al.}(2003){Bennett}, {Hill}, {Hinshaw}, {Nolta},
  {Odegard}, {Page}, {Spergel}, {Weiland}, {Wright}, {Halpern}, {Jarosik},
  {Kogut}, {Limon}, {Meyer}, {Tucker}, \& {Wollack}}]{Bennett03}
{Bennett}, C.~L., {Hill}, R.~S., {Hinshaw}, G., {et~al.} 2003, ApJS, 148, 97

\bibitem[{{Bertin} \& {Arnouts}(1996)}]{1996A&AS..117..393B}
{Bertin}, E. \& {Arnouts}, S. 1996, \aaps, 117, 393

\bibitem[{{Biffi} {et~al.}(2018){Biffi}, {Dolag}, \& {Merloni}}]{biffi2018}
{Biffi}, V., {Dolag}, K., \& {Merloni}, A. 2018, \mnras, 481, 2213

\bibitem[{{Birkinshaw}(1999)}]{Birkinshaw99}
{Birkinshaw}, M. 1999, Phys. Rep., 310, 97

\bibitem[{{Bleem} {et~al.}(2020){Bleem}, {Bocquet}, {Stalder}, {Gladders},
  {Ade}, {Allen}, {Anderson}, {Annis}, {Ashby}, {Austermann}, {Avila}, {Avva},
  {Bayliss}, {Beall}, {Bechtol}, {Bender}, {Benson}, {Bertin}, {Bianchini},
  {Blake}, {Brodwin}, {Brooks}, {Buckley-Geer}, {Burke}, {Carlstrom}, {Rosell},
  {Carrasco Kind}, {Carretero}, {Chang}, {Chiang}, {Citron}, {Moran},
  {Costanzi}, {Crawford}, {Crites}, {da Costa}, {de Haan}, {De Vicente},
  {Desai}, {Diehl}, {Dietrich}, {Dobbs}, {Eifler}, {Everett}, {Flaugher},
  {Floyd}, {Frieman}, {Gallicchio}, {Garc{\'\i}a-Bellido}, {George}, {Gerdes},
  {Gilbert}, {Gruen}, {Gruendl}, {Gschwend}, {Gupta}, {Gutierrez}, {Halverson},
  {Harrington}, {Henning}, {Heymans}, {Holder}, {Hollowood}, {Holzapfel},
  {Honscheid}, {Hrubes}, {Huang}, {Hubmayr}, {Irwin}, {James}, {Jeltema},
  {Joudaki}, {Khullar}, {Klein}, {Knox}, {Kuropatkin}, {Lee}, {Li}, {Lidman},
  {Lowitz}, {MacCrann}, {Mahler}, {Maia}, {Marshall}, {McDonald}, {McMahon},
  {Melchior}, {Menanteau}, {Meyer}, {Miquel}, {Mocanu}, {Mohr}, {Montgomery},
  {Nadolski}, {Natoli}, {Nibarger}, {Noble}, {Novosad}, {Padin}, {Palmese},
  {Parkinson}, {Patil}, {Paz-Chinch{\'o}n}, {Plazas}, {Pryke}, {Ramachandra},
  {Reichardt}, {Remolina Gonz{\'a}lez}, {Romer}, {Roodman}, {Ruhl}, {Rykoff},
  {Saliwanchik}, {Sanchez}, {Saro}, {Sayre}, {Schaffer}, {Schrabback},
  {Serrano}, {Sharon}, {Sievers}, {Smecher}, {Smith}, {Soares-Santos}, {Stark},
  {Story}, {Suchyta}, {Tarle}, {Tucker}, {Vanderlinde}, {Veach}, {Vieira},
  {Wang}, {Weller}, {Whitehorn}, {Wu}, {Yefremenko}, \&
  {Zhang}}]{2020ApJS..247...25B}
{Bleem}, L.~E., {Bocquet}, S., {Stalder}, B., {et~al.} 2020, \apjs, 247, 25

\bibitem[{{Bleem} {et~al.}(2015){Bleem}, {Stalder}, {de Haan}, {Aird}, {Allen},
  {Applegate}, {Ashby}, {Bautz}, {Bayliss}, {Benson}, {Bocquet}, {Brodwin},
  {Carlstrom}, {Chang}, {Chiu}, {Cho}, {Clocchiatti}, {Crawford}, {Crites},
  {Desai}, {Dietrich}, {Dobbs}, {Foley}, {Forman}, {George}, {Gladders},
  {Gonzalez}, {Halverson}, {Hennig}, {Hoekstra}, {Holder}, {Holzapfel},
  {Hrubes}, {Jones}, {Keisler}, {Knox}, {Lee}, {Leitch}, {Liu}, {Lueker},
  {Luong-Van}, {Mantz}, {Marrone}, {McDonald}, {McMahon}, {Meyer}, {Mocanu},
  {Mohr}, {Murray}, {Padin}, {Pryke}, {Reichardt}, {Rest}, {Ruel}, {Ruhl},
  {Saliwanchik}, {Saro}, {Sayre}, {Schaffer}, {Schrabback}, {Shirokoff},
  {Song}, {Spieler}, {Stanford}, {Staniszewski}, {Stark}, {Story}, {Stubbs},
  {Vanderlinde}, {Vieira}, {Vikhlinin}, {Williamson}, {Zahn}, \&
  {Zenteno}}]{Bleem15}
{Bleem}, L.~E., {Stalder}, B., {de Haan}, T., {et~al.} 2015, ApJS, 216, 27

\bibitem[{{Bocquet} {et~al.}(2019){Bocquet}, {Dietrich}, {Schrabback}, {Bleem},
  {Klein}, {Allen}, {Applegate}, {Ashby}, {Bautz}, {Bayliss}, {Benson},
  {Brodwin}, {Bulbul}, {Canning}, {Capasso}, {Carlstrom}, {Chang}, {Chiu},
  {Cho}, {Clocchiatti}, {Crawford}, {Crites}, {de Haan}, {Desai}, {Dobbs},
  {Foley}, {Forman}, {Garmire}, {George}, {Gladders}, {Gonzalez}, {Grandis},
  {Gupta}, {Halverson}, {Hlavacek-Larrondo}, {Hoekstra}, {Holder}, {Holzapfel},
  {Hou}, {Hrubes}, {Huang}, {Jones}, {Khullar}, {Knox}, {Kraft}, {Lee}, {von
  der Linden}, {Luong-Van}, {Mantz}, {Marrone}, {McDonald}, {McMahon}, {Meyer},
  {Mocanu}, {Mohr}, {Morris}, {Padin}, {Patil}, {Pryke}, {Rapetti},
  {Reichardt}, {Rest}, {Ruhl}, {Saliwanchik}, {Saro}, {Sayre}, {Schaffer},
  {Shirokoff}, {Stalder}, {Stanford}, {Staniszewski}, {Stark}, {Story},
  {Strazzullo}, {Stubbs}, {Vanderlinde}, {Vieira}, {Vikhlinin}, {Williamson},
  \& {Zenteno}}]{bds18}
{Bocquet}, S., {Dietrich}, J.~P., {Schrabback}, T., {et~al.} 2019, \apj, 878,
  55

\bibitem[{{Borm} {et~al.}(2014){Borm}, {Reiprich}, {Mohammed}, \&
  {Lovisari}}]{brm14}
{Borm}, K., {Reiprich}, T.~H., {Mohammed}, I., \& {Lovisari}, L. 2014, A\&A,
  567, A65

\bibitem[{{Boulanger} {et~al.}(1996){Boulanger}, {Abergel}, {Bernard},
  {Burton}, {Desert}, {Hartmann}, {Lagache}, \& {Puget}}]{1996A&A...312..256B}
{Boulanger}, F., {Abergel}, A., {Bernard}, J.~P., {et~al.} 1996, \aap, 312, 256

\bibitem[{{Bregman} \& {Lloyd-Davies}(2006)}]{bl06}
{Bregman}, J.~N. \& {Lloyd-Davies}, E.~J. 2006, ApJ, 644, 167

\bibitem[{Br\"uggen {et~al.}(2020)Br\"uggen, Reiprich, Bulbul, Koribalski,
  Andernach, Rudnick, Hoang, Wilber, Duchesne, Veronica, Pacaud, Hopkins,
  Norris, Johnston-Hollitt, Brown, Bonafede, Brunetti, Collier, Sanders,
  Vardoulaki, Venturi, \& Kapinska}]{brb20}
Br\"uggen, M., Reiprich, T.~H., Bulbul, E., {et~al.} 2020, \aap, accepted

\bibitem[{{Bulbul} {et~al.}(2016){Bulbul}, {Randall}, {Bayliss}, {Miller},
  {Andrade-Santos}, {Johnson}, {Bautz}, {Blanton}, {Forman}, {Jones},
  {Paterno-Mahler}, {Murray}, {Sarazin}, {Smith}, \&
  {Ezer}}]{2016ApJ...818..131B}
{Bulbul}, E., {Randall}, S.~W., {Bayliss}, M., {et~al.} 2016, \apj, 818, 131

\bibitem[{{Buote} {et~al.}(2009){Buote}, {Zappacosta}, {Fang}, {Humphrey},
  {Gastaldello}, \& {Tagliaferri}}]{bzf09}
{Buote}, D.~A., {Zappacosta}, L., {Fang}, T., {et~al.} 2009, ApJ, 695, 1351

\bibitem[{{Carlstrom} {et~al.}(2002){Carlstrom}, {Holder}, \&
  {Reese}}]{Carlstrom02}
{Carlstrom}, J.~E., {Holder}, G.~P., \& {Reese}, E.~D. 2002, ARA\&A, 40, 643

\bibitem[{{Cen} \& {Ostriker}(1999)}]{co99}
{Cen}, R. \& {Ostriker}, J.~P. 1999, ApJ, 514, 1

\bibitem[{{Chluba} {et~al.}(2012){Chluba}, {Nagai}, {Sazonov}, \&
  {Nelson}}]{Chluba12}
{Chluba}, J., {Nagai}, D., {Sazonov}, S., \& {Nelson}, K. 2012, MNRAS, 426, 510

\bibitem[{{Clerc} {et~al.}(2018){Clerc}, {Ramos-Ceja}, {Ridl}, {Lamer},
  {Brunner}, {Hofmann}, {Comparat}, {Pacaud}, {K{\"a}fer}, {Reiprich},
  {Merloni}, {Schmid}, {Brand}, {Wilms}, {Friedrich}, {Finoguenov}, {Dauser},
  \& {Kreykenbohm}}]{crr18}
{Clerc}, N., {Ramos-Ceja}, M.~E., {Ridl}, J., {et~al.} 2018, A\&A, 617, A92

\bibitem[{{Comparat} {et~al.}(2020){Comparat}, {Eckert}, {Finoguenov},
  {Schmidt}, {Sanders}, {Nagai}, {Lau}, {Kaefer}, {Pacaud}, {Clerc},
  {Reiprich}, {Bulbul}, {Ider Chitham}, {Chuang}, {Ghirardini},
  {Gonzalez-Perez}, {Gozaliazl}, {Kirkpatrick}, {Klypin}, {Merloni}, {Nandra},
  {Liu}, {Prada}, {Ramos-Ceja}, {Salvato}, {Seppi}, {Tempel}, \&
  {Yepes}}]{2020arXiv200808404C}
{Comparat}, J., {Eckert}, D., {Finoguenov}, A., {et~al.} 2020, arXiv e-prints,
  arXiv:2008.08404

\bibitem[{{Cui} {et~al.}(2018){Cui}, {Knebe}, {Yepes}, {Yang}, {Borgani},
  {Kang}, {Power}, \& {Staveley-Smith}}]{cui2018}
{Cui}, W., {Knebe}, A., {Yepes}, G., {et~al.} 2018, \mnras, 473, 68

\bibitem[{{Dark Energy Survey Collaboration} {et~al.}(2016){Dark Energy Survey
  Collaboration}, Abbott, Abdalla, Aleksić, Allam, Amara, Bacon, Balbinot,
  Banerji, Bechtol, Benoit-Lévy, Bernstein, Bertin, Blazek, Bonnett, Bridle,
  Brooks, Brunner, Buckley-Geer, Burke, Caminha, Capozzi, Carlsen,
  Carnero-Rosell, Carollo, Carrasco-Kind, Carretero, Castander, Clerkin,
  Collett, Conselice, Crocce, Cunha, D'Andrea, da~Costa, Davis, Desai, Diehl,
  Dietrich, Dodelson, Doel, Drlica-Wagner, Estrada, Etherington, Evrard,
  Fabbri, Finley, Flaugher, Foley, Fosalba, Frieman, García-Bellido,
  Gaztanaga, Gerdes, Giannantonio, Goldstein, Gruen, Gruendl, Guarnieri,
  Gutierrez, Hartley, Honscheid, Jain, James, Jeltema, Jouvel, Kessler, King,
  Kirk, Kron, Kuehn, Kuropatkin, Lahav, Li, Lima, Lin, Maia, Makler, Manera,
  Maraston, Marshall, Martini, McMahon, Melchior, Merson, Miller, Miquel, Mohr,
  Morice-Atkinson, Naidoo, Neilsen, Nichol, Nord, Ogando, Ostrovski, Palmese,
  Papadopoulos, Peiris, Peoples, Percival, Plazas, Reed, Refregier, Romer,
  Roodman, Ross, Rozo, Rykoff, Sadeh, Sako, Sánchez, Sanchez, Santiago,
  Scarpine, Schubnell, Sevilla-Noarbe, Sheldon, Smith, Smith, Soares-Santos,
  Sobreira, Soumagnac, Suchyta, Sullivan, Swanson, Tarle, Thaler, Thomas,
  Thomas, Tucker, Vieira, Vikram, Walker, Wechsler, Weller, Wester, Whiteway,
  Wilcox, Yanny, Zhang, \& Zuntz}]{10.1093/mnras/stw641}
{Dark Energy Survey Collaboration}, Abbott, T., Abdalla, F.~B., {et~al.} 2016,
  Monthly Notices of the Royal Astronomical Society, 460, 1270

\bibitem[{{de Graaff} {et~al.}(2019){de Graaff}, {Cai}, {Heymans}, \&
  {Peacock}}]{dch19}
{de Graaff}, A., {Cai}, Y.-C., {Heymans}, C., \& {Peacock}, J.~A. 2019, A\&A,
  624, A48

\bibitem[{{Dey} {et~al.}(2019){Dey}, {Schlegel}, {Lang}, {Blum}, {Burleigh},
  {Fan}, {Findlay}, {Finkbeiner}, {Herrera}, {Juneau}, {Landriau}, {Levi},
  {McGreer}, {Meisner}, {Myers}, {Moustakas}, {Nugent}, {Patej}, {Schlafly},
  {Walker}, {Valdes}, {Weaver}, {Y{\`e}che}, {Zou}, {Zhou}, {Abareshi},
  {Abbott}, {Abolfathi}, {Aguilera}, {Alam}, {Allen}, {Alvarez}, {Annis},
  {Ansarinejad}, {Aubert}, {Beechert}, {Bell}, {BenZvi}, {Beutler}, {Bielby},
  {Bolton}, {Brice{\~n}o}, {Buckley-Geer}, {Butler}, {Calamida}, {Carlberg},
  {Carter}, {Casas}, {Castander}, {Choi}, {Comparat}, {Cukanovaite}, {Delubac},
  {DeVries}, {Dey}, {Dhungana}, {Dickinson}, {Ding}, {Donaldson}, {Duan},
  {Duckworth}, {Eftekharzadeh}, {Eisenstein}, {Etourneau}, {Fagrelius},
  {Farihi}, {Fitzpatrick}, {Font-Ribera}, {Fulmer}, {G{\"a}nsicke},
  {Gaztanaga}, {George}, {Gerdes}, {Gontcho}, {Gorgoni}, {Green}, {Guy},
  {Harmer}, {Hernand ez}, {Honscheid}, {Huang}, {James}, {Jannuzi}, {Jiang},
  {Joyce}, {Karcher}, {Karkar}, {Kehoe}, {Kneib}, {Kueter-Young}, {Lan},
  {Lauer}, {Le Guillou}, {Le Van Suu}, {Lee}, {Lesser}, {Perreault Levasseur},
  {Li}, {Mann}, {Marshall}, {Mart{\'\i}nez-V{\'a}zquez}, {Martini}, {du Mas des
  Bourboux}, {McManus}, {Meier}, {M{\'e}nard}, {Metcalfe},
  {Mu{\~n}oz-Guti{\'e}rrez}, {Najita}, {Napier}, {Narayan}, {Newman}, {Nie},
  {Nord}, {Norman}, {Olsen}, {Paat}, {Palanque-Delabrouille}, {Peng},
  {Poppett}, {Poremba}, {Prakash}, {Rabinowitz}, {Raichoor}, {Rezaie},
  {Robertson}, {Roe}, {Ross}, {Ross}, {Rudnick}, {Safonova}, {Saha},
  {S{\'a}nchez}, {Savary}, {Schweiker}, {Scott}, {Seo}, {Shan}, {Silva},
  {Slepian}, {Soto}, {Sprayberry}, {Staten}, {Stillman}, {Stupak}, {Summers},
  {Sien Tie}, {Tirado}, {Vargas-Maga{\~n}a}, {Vivas}, {Wechsler}, {Williams},
  {Yang}, {Yang}, {Yapici}, {Zaritsky}, {Zenteno}, {Zhang}, {Zhang}, {Zhou}, \&
  {Zhou}}]{2019AJ....157..168D}
{Dey}, A., {Schlegel}, D.~J., {Lang}, D., {et~al.} 2019, \aj, 157, 168

\bibitem[{{Dietrich} {et~al.}(2012){Dietrich}, {Werner}, {Clowe}, {Finoguenov},
  {Kitching}, {Miller}, \& {Simionescu}}]{dwc12}
{Dietrich}, J.~P., {Werner}, N., {Clowe}, D., {et~al.} 2012, Nat, 487, 202

\bibitem[{{Dolag} {et~al.}(2006){Dolag}, {Meneghetti}, {Moscardini}, {Rasia},
  \& {Bonaldi}}]{Dolag2005}
{Dolag}, K., {Meneghetti}, M., {Moscardini}, L., {Rasia}, E., \& {Bonaldi}, A.
  2006, \mnras, 370, 656

\bibitem[{{Donnelly} {et~al.}(2001){Donnelly}, {Forman}, {Jones}, {Quintana},
  {Ramirez}, {Churazov}, \& {Gilfanov}}]{2001ApJ...562..254D}
{Donnelly}, R.~H., {Forman}, W., {Jones}, C., {et~al.} 2001, \apj, 562, 254

\bibitem[{{eBOSS Collaboration} {et~al.}(2020){eBOSS Collaboration}, {Alam},
  {Aubert}, {Avila}, {Balland}, {Bautista}, {Bershady}, {Bizyaev}, {Blanton},
  {Bolton}, {Bovy}, {Brinkmann}, {Brownstein}, {Burtin}, {Chabanier},
  {Chapman}, {Choi}, {Chuang}, {Comparat}, {Cuceu}, {Dawson}, {de la Macorra},
  {de la Torre}, {de Mattia}, {de Sainte Agathe}, {du Mas des Bourboux},
  {Escoffier}, {Etourneau}, {Farr}, {Font-Ribera}, {Frinchaboy}, {Fromenteau},
  {Gil-Mar{\'\i}n}, {Gonzalez-Morales}, {Gonzalez-Perez}, {Grabowski}, {Guy},
  {Hawken}, {Hou}, {Kong}, {Klaene}, {Kneib}, {Le Goff}, {Lin}, {Long}, {Lyke},
  {Cousinou}, {Martini}, {Masters}, {Mohammad}, {Moon}, {Mueller},
  {Mun{\~o}z-Gutie{\'r}rez}, {Myers}, {Nadathur}, {Neveux}, {Newman},
  {Noterdaeme}, {Oravetz}, {Oravetz}, {Palanque-Delabrouille}, {Pan}, {Parker},
  {Paviot}, {Percival}, {Pe{\'r}ez-Rafols}, {Petitjean}, {Pieri}, {Prakash},
  {Raichoor}, {Ravoux}, {Rezaie}, {Rich}, {Ross}, {Rossi}, {Ruggeri},
  {Ruhlmann-Kleider}, {Sa{\'n}chez}, {Sa{\'n}chez}, {Sa{\'n}chez-Gallego},
  {Sayres}, {Schneider}, {Seo}, {Shafieloo}, {Slosar}, {Smith}, {Stermer},
  {Tamone}, {Tinker}, {Tojeiro}, {Vargas-Maga{\~n}a}, {Variu}, {Wang},
  {Weaver}, {Weijmans}, {Yeche}, {Zarrouk}, {Zhao}, {Zhao}, \&
  {Zheng}}]{2020arXiv200708991E}
{eBOSS Collaboration}, {Alam}, S., {Aubert}, M., {et~al.} 2020,
  arXiv:2007.08991, arXiv:2007.08991

\bibitem[{{Eckert} {et~al.}(2015){Eckert}, {Jauzac}, {Shan}, {Kneib}, {Erben},
  {Israel}, {Jullo}, {Klein}, {Massey}, {Richard}, \& {Tchernin}}]{ejs15}
{Eckert}, D., {Jauzac}, M., {Shan}, H., {et~al.} 2015, Nat, 528, 105

\bibitem[{{Erben} {et~al.}(2013){Erben}, {Hildebrandt}, {Miller}, {van
  Waerbeke}, {Heymans}, {Hoekstra}, {Kitching}, {Mellier}, {Benjamin}, {Blake},
  {Bonnett}, {Cordes}, {Coupon}, {Fu}, {Gavazzi}, {Gillis}, {Grocutt}, {Gwyn},
  {Holhjem}, {Hudson}, {Kilbinger}, {Kuijken}, {Milkeraitis}, {Rowe},
  {Schrabback}, {Semboloni}, {Simon}, {Smit}, {Toader}, {Vafaei}, {van Uitert},
  \& {Velander}}]{2013MNRAS.433.2545E}
{Erben}, T., {Hildebrandt}, H., {Miller}, L., {et~al.} 2013, \mnras, 433, 2545

\bibitem[{{Erben} {et~al.}(2005){Erben}, {Schirmer}, {Dietrich}, {Cordes},
  {Haberzettl}, {Hetterscheidt}, {Hildebrandt}, {Schmithuesen}, {Schneider},
  {Simon}, {Deul}, {Hook}, {Kaiser}, {Radovich}, {Benoist}, {Nonino}, {Olsen},
  {Prandoni}, {Wichmann}, {Zaggia}, {Bomans}, {Dettmar}, \&
  {Miralles}}]{2005AN....326..432E}
{Erben}, T., {Schirmer}, M., {Dietrich}, J.~P., {et~al.} 2005, Astronomische
  Nachrichten, 326, 432

\bibitem[{{Fabjan} {et~al.}(2010){Fabjan}, {Borgani}, {Tornatore}, {Saro},
  {Murante}, \& {Dolag}}]{fabjan2010}
{Fabjan}, D., {Borgani}, S., {Tornatore}, L., {et~al.} 2010, \mnras, 401, 1670

\bibitem[{{Faccioli} {et~al.}(2018){Faccioli}, {Pacaud}, {Sauvageot}, {Pierre},
  {Chiappetti}, {Clerc}, {Gastaud}, {Koulouridis}, {Le Brun}, \&
  {Valotti}}]{fps18}
{Faccioli}, L., {Pacaud}, F., {Sauvageot}, J.~L., {et~al.} 2018, \aap, 620, A9

\bibitem[{{Fang} {et~al.}(2010){Fang}, {Buote}, {Humphrey}, {Canizares},
  {Zappacosta}, {Maiolino}, {Tagliaferri}, \& {Gastaldello}}]{fbh10}
{Fang}, T., {Buote}, D.~A., {Humphrey}, P.~J., {et~al.} 2010, ApJ, 714, 1715

\bibitem[{Flaugher {et~al.}(2015)Flaugher, Diehl, Honscheid, Abbott, Alvarez,
  Angstadt, Annis, Antonik, Ballester, Beaufore, Bernstein, Bernstein, Bigelow,
  Bonati, Boprie, Brooks, Buckley-Geer, Campa, Cardiel-Sas, Castander,
  Castilla, Cease, Cela-Ruiz, Chappa, Chi, Cooper, da~Costa, Dede, Derylo,
  DePoy, de~Vicente, Doel, Drlica-Wagner, Eiting, Elliott, Emes, Estrada, Neto,
  Finley, Flores, Frieman, Gerdes, Gladders, Gregory, Gutierrez, Hao, Holland,
  Holm, Huffman, Jackson, James, Jonas, Karcher, Karliner, Kent, Kessler,
  Kozlovsky, Kron, Kubik, Kuehn, Kuhlmann, Kuk, Lahav, Lathrop, Lee, Levi,
  Lewis, Li, Mandrichenko, Marshall, Martinez, Merritt, Miquel, Mu{\~{n}}oz,
  Neilsen, Nichol, Nord, Ogando, Olsen, Palaio, Patton, Peoples, Plazas, Rauch,
  Reil, Rheault, Roe, Rogers, Roodman, Sanchez, Scarpine, Schindler, Schmidt,
  Schmitt, Schubnell, Schultz, Schurter, Scott, Serrano, Shaw, Smith,
  Soares-Santos, Stefanik, Stuermer, Suchyta, Sypniewski, Tarle, Thaler, Tighe,
  Tran, Tucker, Walker, Wang, Watson, Weaverdyck, Wester, Woods, \&
  and}]{Flaugher_2015}
Flaugher, B., Diehl, H.~T., Honscheid, K., {et~al.} 2015, The Astronomical
  Journal, 150, 150

\bibitem[{{Fresco} {et~al.}(2020){Fresco}, {P{\'e}roux}, {Merloni},
  {Hamanowicz}, \& {Szakacs}}]{2020arXiv200911346F}
{Fresco}, A., {P{\'e}roux}, C., {Merloni}, A., {Hamanowicz}, A., \& {Szakacs},
  R. 2020, MNRAS, 499, 5230

\bibitem[{{Fujita} {et~al.}(2008){Fujita}, {Tawa}, {Hayashida}, {Takizawa},
  {Matsumoto}, {Okabe}, \& {Reiprich}}]{fth07}
{Fujita}, Y., {Tawa}, N., {Hayashida}, K., {et~al.} 2008, PASJ, 60, S343

\bibitem[{{Fukugita} \& {Peebles}(2004)}]{fp04}
{Fukugita}, M. \& {Peebles}, P.~J.~E. 2004, ApJ, 616, 643

\bibitem[{{Gaia Collaboration} {et~al.}(2018){Gaia Collaboration}, {Brown, A.
  G. A.}, {Vallenari, A.}, {Prusti, T.}, {de Bruijne, J. H. J.}, {Babusiaux,
  C.}, {Bailer-Jones, C. A. L.}, {Biermann, M.}, {Evans, D. W.}, {Eyer, L.},
  {Jansen, F.}, {Jordi, C.}, {Klioner, S. A.}, {Lammers, U.}, {Lindegren, L.},
  {Luri, X.}, {Mignard, F.}, {Panem, C.}, {Pourbaix, D.}, {Randich, S.},
  {Sartoretti, P.}, {Siddiqui, H. I.}, {Soubiran, C.}, {van Leeuwen, F.},
  {Walton, N. A.}, {Arenou, F.}, {Bastian, U.}, {Cropper, M.}, {Drimmel, R.},
  {Katz, D.}, {Lattanzi, M. G.}, {Bakker, J.}, {Cacciari, C.}, {Casta\~neda,
  J.}, {Chaoul, L.}, {Cheek, N.}, {De Angeli, F.}, {Fabricius, C.}, {Guerra,
  R.}, {Holl, B.}, {Masana, E.}, {Messineo, R.}, {Mowlavi, N.}, {Nienartowicz,
  K.}, {Panuzzo, P.}, {Portell, J.}, {Riello, M.}, {Seabroke, G. M.}, {Tanga,
  P.}, {Th\'evenin, F.}, {Gracia-Abril, G.}, {Comoretto, G.},
  {Garcia-Reinaldos, M.}, {Teyssier, D.}, {Altmann, M.}, {Andrae, R.}, {Audard,
  M.}, {Bellas-Velidis, I.}, {Benson, K.}, {Berthier, J.}, {Blomme, R.},
  {Burgess, P.}, {Busso, G.}, {Carry, B.}, {Cellino, A.}, {Clementini, G.},
  {Clotet, M.}, {Creevey, O.}, {Davidson, M.}, {De Ridder, J.}, {Delchambre,
  L.}, {Dell\'{}Oro, A.}, {Ducourant, C.}, {Fern\'andez-Hern\'andez, J.},
  {Fouesneau, M.}, {Fr\'emat, Y.}, {Galluccio, L.}, {Garc\'{\i}a-Torres, M.},
  {Gonz\'alez-N\'u\~nez, J.}, {Gonz\'alez-Vidal, J. J.}, {Gosset, E.}, {Guy, L.
  P.}, {Halbwachs, J.-L.}, {Hambly, N. C.}, {Harrison, D. L.}, {Hern\'andez,
  J.}, {Hestroffer, D.}, {Hodgkin, S. T.}, {Hutton, A.}, {Jasniewicz, G.},
  {Jean-Antoine-Piccolo, A.}, {Jordan, S.}, {Korn, A. J.}, {Krone-Martins, A.},
  {Lanzafame, A. C.}, {Lebzelter, T.}, {L\"offler, W.}, {Manteiga, M.},
  {Marrese, P. M.}, {Mart\'{\i}n-Fleitas, J. M.}, {Moitinho, A.}, {Mora, A.},
  {Muinonen, K.}, {Osinde, J.}, {Pancino, E.}, {Pauwels, T.}, {Petit, J.-M.},
  {Recio-Blanco, A.}, {Richards, P. J.}, {Rimoldini, L.}, {Robin, A. C.},
  {Sarro, L. M.}, {Siopis, C.}, {Smith, M.}, {Sozzetti, A.}, {S\"uveges, M.},
  {Torra, J.}, {van Reeven, W.}, {Abbas, U.}, {Abreu Aramburu, A.}, {Accart,
  S.}, {Aerts, C.}, {Altavilla, G.}, {\'Alvarez, M. A.}, {Alvarez, R.}, {Alves,
  J.}, {Anderson, R. I.}, {Andrei, A. H.}, {Anglada Varela, E.}, {Antiche, E.},
  {Antoja, T.}, {Arcay, B.}, {Astraatmadja, T. L.}, {Bach, N.}, {Baker, S. G.},
  {Balaguer-N\'u\~nez, L.}, {Balm, P.}, {Barache, C.}, {Barata, C.}, {Barbato,
  D.}, {Barblan, F.}, {Barklem, P. S.}, {Barrado, D.}, {Barros, M.}, {Barstow,
  M. A.}, {Bartholom\'e Mu\~noz, S.}, {Bassilana, J.-L.}, {Becciani, U.},
  {Bellazzini, M.}, {Berihuete, A.}, {Bertone, S.}, {Bianchi, L.}, {Bienaym\'e,
  O.}, {Blanco-Cuaresma, S.}, {Boch, T.}, {Boeche, C.}, {Bombrun, A.},
  {Borrachero, R.}, {Bossini, D.}, {Bouquillon, S.}, {Bourda, G.}, {Bragaglia,
  A.}, {Bramante, L.}, {Breddels, M. A.}, {Bressan, A.}, {Brouillet, N.},
  {Br\"usemeister, T.}, {Brugaletta, E.}, {Bucciarelli, B.}, {Burlacu, A.},
  {Busonero, D.}, {Butkevich, A. G.}, {Buzzi, R.}, {Caffau, E.}, {Cancelliere,
  R.}, {Cannizzaro, G.}, {Cantat-Gaudin, T.}, {Carballo, R.}, {Carlucci, T.},
  {Carrasco, J. M.}, {Casamiquela, L.}, {Castellani, M.}, {Castro-Ginard, A.},
  {Charlot, P.}, {Chemin, L.}, {Chiavassa, A.}, {Cocozza, G.}, {Costigan, G.},
  {Cowell, S.}, {Crifo, F.}, {Crosta, M.}, {Crowley, C.}, {Cuypers+, J.},
  {Dafonte, C.}, {Damerdji, Y.}, {Dapergolas, A.}, {David, P.}, {David, M.},
  {de Laverny, P.}, {De Luise, F.}, {De March, R.}, {de Martino, D.}, {de
  Souza, R.}, {de Torres, A.}, {Debosscher, J.}, {del Pozo, E.}, {Delbo, M.},
  {Delgado, A.}, {Delgado, H. E.}, {Di Matteo, P.}, {Diakite, S.}, {Diener,
  C.}, {Distefano, E.}, {Dolding, C.}, {Drazinos, P.}, {Dur\'an, J.},
  {Edvardsson, B.}, {Enke, H.}, {Eriksson, K.}, {Esquej, P.}, {Eynard Bontemps,
  G.}, {Fabre, C.}, {Fabrizio, M.}, {Faigler, S.}, {Falc\~ao, A. J.}, {Farr\`as
  Casas, M.}, {Federici, L.}, {Fedorets, G.}, {Fernique, P.}, {Figueras, F.},
  {Filippi, F.}, {Findeisen, K.}, {Fonti, A.}, {Fraile, E.}, {Fraser, M.},
  {Fr\'ezouls, B.}, {Gai, M.}, {Galleti, S.}, {Garabato, D.},
  {Garc\'{\i}a-Sedano, F.}, {Garofalo, A.}, {Garralda, N.}, {Gavel, A.},
  {Gavras, P.}, {Gerssen, J.}, {Geyer, R.}, {Giacobbe, P.}, {Gilmore, G.},
  {Girona, S.}, {Giuffrida, G.}, {Glass, F.}, {Gomes, M.}, {Granvik, M.},
  {Gueguen, A.}, {Guerrier, A.}, {Guiraud, J.}, {Guti\'errez-S\'anchez, R.},
  {Haigron, R.}, {Hatzidimitriou, D.}, {Hauser, M.}, {Haywood, M.}, {Heiter,
  U.}, {Helmi, A.}, {Heu, J.}, {Hilger, T.}, {Hobbs, D.}, {Hofmann, W.},
  {Holland, G.}, {Huckle, H. E.}, {Hypki, A.}, {Icardi, V.}, {Jan\ss{}en, K.},
  {Jevardat de Fombelle, G.}, {Jonker, P. G.}, {Juh\'asz, \'A. L.}, {Julbe,
  F.}, {Karampelas, A.}, {Kewley, A.}, {Klar, J.}, {Kochoska, A.}, {Kohley,
  R.}, {Kolenberg, K.}, {Kontizas, M.}, {Kontizas, E.}, {Koposov, S. E.},
  {Kordopatis, G.}, {Kostrzewa-Rutkowska, Z.}, {Koubsky, P.}, {Lambert, S.},
  {Lanza, A. F.}, {Lasne, Y.}, {Lavigne, J.-B.}, {Le Fustec, Y.}, {Le
  Poncin-Lafitte, C.}, {Lebreton, Y.}, {Leccia, S.}, {Leclerc, N.},
  {Lecoeur-Taibi, I.}, {Lenhardt, H.}, {Leroux, F.}, {Liao, S.}, {Licata, E.},
  {Lindstr\o{}m, H. E. P.}, {Lister, T. A.}, {Livanou, E.}, {Lobel, A.},
  {L\'opez, M.}, {Managau, S.}, {Mann, R. G.}, {Mantelet, G.}, {Marchal, O.},
  {Marchant, J. M.}, {Marconi, M.}, {Marinoni, S.}, {Marschalk\'o, G.},
  {Marshall, D. J.}, {Martino, M.}, {Marton, G.}, {Mary, N.}, {Massari, D.},
  {Matijevic, G.}, {Mazeh, T.}, {McMillan, P. J.}, {Messina, S.}, {Michalik,
  D.}, {Millar, N. R.}, {Molina, D.}, {Molinaro, R.}, {Moln\'ar, L.},
  {Montegriffo, P.}, {Mor, R.}, {Morbidelli, R.}, {Morel, T.}, {Morris, D.},
  {Mulone, A. F.}, {Muraveva, T.}, {Musella, I.}, {Nelemans, G.}, {Nicastro,
  L.}, {Noval, L.}, {O\'{}Mullane, W.}, {Ord\'enovic, C.}, {Ord\'o\~nez-Blanco,
  D.}, {Osborne, P.}, {Pagani, C.}, {Pagano, I.}, {Pailler, F.}, {Palacin, H.},
  {Palaversa, L.}, {Panahi, A.}, {Pawlak, M.}, {Piersimoni, A. M.}, {Pineau,
  F.-X.}, {Plachy, E.}, {Plum, G.}, {Poggio, E.}, {Poujoulet, E.}, {Prsa, A.},
  {Pulone, L.}, {Racero, E.}, {Ragaini, S.}, {Rambaux, N.}, {Ramos-Lerate, M.},
  {Regibo, S.}, {Reyl\'e, C.}, {Riclet, F.}, {Ripepi, V.}, {Riva, A.}, {Rivard,
  A.}, {Rixon, G.}, {Roegiers, T.}, {Roelens, M.}, {Romero-G\'omez, M.},
  {Rowell, N.}, {Royer, F.}, {Ruiz-Dern, L.}, {Sadowski, G.}, {Sagrist\`a
  Sell\'es, T.}, {Sahlmann, J.}, {Salgado, J.}, {Salguero, E.}, {Sanna, N.},
  {Santana-Ros, T.}, {Sarasso, M.}, {Savietto, H.}, {Schultheis, M.}, {Sciacca,
  E.}, {Segol, M.}, {Segovia, J. C.}, {S\'egransan, D.}, {Shih, I-C.},
  {Siltala, L.}, {Silva, A. F.}, {Smart, R. L.}, {Smith, K. W.}, {Solano, E.},
  {Solitro, F.}, {Sordo, R.}, {Soria Nieto, S.}, {Souchay, J.}, {Spagna, A.},
  {Spoto, F.}, {Stampa, U.}, {Steele, I. A.}, {Steidelm\"uller, H.},
  {Stephenson, C. A.}, {Stoev, H.}, {Suess, F. F.}, {Surdej, J.}, {Szabados,
  L.}, {Szegedi-Elek, E.}, {Tapiador, D.}, {Taris, F.}, {Tauran, G.}, {Taylor,
  M. B.}, {Teixeira, R.}, {Terrett, D.}, {Teyssandier, P.}, {Thuillot, W.},
  {Titarenko, A.}, {Torra Clotet, F.}, {Turon, C.}, {Ulla, A.}, {Utrilla, E.},
  {Uzzi, S.}, {Vaillant, M.}, {Valentini, G.}, {Valette, V.}, {van Elteren,
  A.}, {Van Hemelryck, E.}, {van Leeuwen, M.}, {Vaschetto, M.}, {Vecchiato,
  A.}, {Veljanoski, J.}, {Viala, Y.}, {Vicente, D.}, {Vogt, S.}, {von Essen,
  C.}, {Voss, H.}, {Votruba, V.}, {Voutsinas, S.}, {Walmsley, G.}, {Weiler,
  M.}, {Wertz, O.}, {Wevers, T.}, {Wyrzykowski, L.}, {Yoldas, A.}, {Zerjal,
  M.}, {Ziaeepour, H.}, {Zorec, J.}, {Zschocke, S.}, {Zucker, S.}, {Zurbach,
  C.}, \& {Zwitter, T.}}]{refId0}
{Gaia Collaboration}, {Brown, A. G. A.}, {Vallenari, A.}, {et~al.} 2018, A\&A,
  616, A1

\bibitem[{{Garon} {et~al.}(2019){Garon}, {Rudnick}, {Wong}, {Jones}, {Kim},
  {Andernach}, {Shabala}, {Kapi{\'n}ska}, {Norris}, {de Gasperin}, {Tate}, \&
  {Tang}}]{2019AJ....157..126G}
{Garon}, A.~F., {Rudnick}, L., {Wong}, O.~I., {et~al.} 2019, \aj, 157, 126

\bibitem[{{Grandis} {et~al.}(2019){Grandis}, {Mohr}, {Dietrich}, {Bocquet},
  {Saro}, {Klein}, {Paulus}, \& {Capasso}}]{gmd18}
{Grandis}, S., {Mohr}, J.~J., {Dietrich}, J.~P., {et~al.} 2019, \mnras, 488,
  2041

\bibitem[{{Gruen} {et~al.}(2015){Gruen}, {Bernstein}, {Jarvis}, {Rowe},
  {Vikram}, {Plazas}, \& {Seitz}}]{2015JInst..10C5032G}
{Gruen}, D., {Bernstein}, G.~M., {Jarvis}, M., {et~al.} 2015, Journal of
  Instrumentation, 10, C05032

\bibitem[{{Gupta} {et~al.}(2016){Gupta}, {Srianand}, {Baan}, {Baker},
  {Beswick}, {Bhatnagar}, {Bhattacharya}, {Bosma}, {Carilli}, {Cluver},
  {Combes}, {Cress}, {Dutta}, {Fynbo}, {Heald}, {Hilton}, {Hussain}, {Jarvis},
  {Jozsa}, {Kamphuis}, {Kembhavi}, {Kerp}, {Kloeckner}, {Krogager}, {Kulkarni},
  {Ledoux}, {Mahabal}, {Mauch}, {Moodley}, {Momjian}, {Morganti}, {Noterdaeme},
  {Oosterloo}, {Petitjean}, {Schroeder}, {Serra}, {Sievers}, {Spekkens},
  {Vaisanen}, {van der Hulst}, {Vivek}, {Wang}, {Wong}, \&
  {Zungu}}]{2016mks..confE..14G}
{Gupta}, N., {Srianand}, R., {Baan}, W., {et~al.} 2016, in MeerKAT Science: On
  the Pathway to the SKA, 14

\bibitem[{{Haines} {et~al.}(2018){Haines}, {Busarello}, {Merluzzi}, {Pimbblet},
  {Vogt}, {Dopita}, {Mercurio}, {Grado}, \& {Limatola}}]{2018MNRAS.481.1055H}
{Haines}, C.~P., {Busarello}, G., {Merluzzi}, P., {et~al.} 2018, \mnras, 481,
  1055

\bibitem[{{Hasselfield} {et~al.}(2013){Hasselfield}, {Hilton}, {Marriage},
  {Addison}, {Barrientos}, {Battaglia}, {Battistelli}, {Bond}, {Crichton},
  {Das}, {Devlin}, {Dicker}, {Dunkley}, {Dünner}, {Fowler}, {Gralla},
  {Hajian}, {Halpern}, {Hincks}, {Hlozek}, {Hughes}, {Infante}, {Irwin},
  {Kosowsky}, {Marsden}, {Menanteau}, {Moodley}, {Niemack}, {Nolta}, {Page},
  {Partridge}, {Reese}, {Schmitt}, {Sehgal}, {Sherwin}, {Sievers}, {Sifón},
  {Spergel}, {Staggs}, {Swetz}, {Switzer}, {Thornton}, {Trac}, \&
  {Wollack}}]{Hasselfield13}
{Hasselfield}, M., {Hilton}, M., {Marriage}, T.~A., {et~al.} 2013, J. Cosmology
  Astropart. Phys., 2013, 008

\bibitem[{{HI4PI Collaboration} {et~al.}(2016){HI4PI Collaboration}, {Ben
  Bekhti}, {Fl{\"o}er}, {Keller}, {Kerp}, {Lenz}, {Winkel}, {Bailin},
  {Calabretta}, {Dedes}, {Ford}, {Gibson}, {Haud}, {Janowiecki}, {Kalberla},
  {Lockman}, {McClure-Griffiths}, {Murphy}, {Nakanishi}, {Pisano}, \&
  {Staveley-Smith}}]{2016A&A...594A.116H}
{HI4PI Collaboration}, {Ben Bekhti}, N., {Fl{\"o}er}, L., {et~al.} 2016, \aap,
  594, A116

\bibitem[{{Hilton} {et~al.}(2020){Hilton}, {Sif{\'o}n}, {Naess},
  {Madhavacheril}, {Oguri}, {Rozo}, {Rykoff}, {Abbott}, {Adhikari}, {Aguena},
  {Aiola}, {Allam}, {Amodeo}, {Amon}, {Annis}, {Ansarinejad}, {Aros-Bunster},
  {Austermann}, {Avila}, {Bacon}, {Battaglia}, {Beall}, {Becker}, {Bernstein},
  {Bertin}, {Bhand arkar}, {Bhargava}, {Bond}, {Brooks}, {Burke}, {Calabrese},
  {Carretero}, {Choi}, {Choi}, {Conselice}, {da Costa}, {Costanzi}, {Crichton},
  {Crowley}, {D{\"u}nner}, {Denison}, {Devlin}, {Dicker}, {Diehl}, {Dietrich},
  {Doel}, {Duff}, {Duivenvoorden}, {Dunkley}, {Everett}, {Ferraro}, {Ferrero},
  {Fert{\'e}}, {Flaugher}, {Frieman}, {Gallardo}, {Garc{\'\i}a-Bellido},
  {Gaztanaga}, {Gerdes}, {Giles}, {Golec}, {Gralla}, {Grandis}, {Gruen},
  {Gruendl}, {Gschwend}, {Gutierrez}, {Han}, {Hartley}, {Hasselfield}, {Hill},
  {Hilton}, {Hincks}, {Hinton}, {Ho}, {Honscheid}, {Hoyle}, {Hubmayr},
  {Huffenberger}, {Hughes}, {Jaelani}, {Jain}, {James}, {Jeltema}, {Kent},
  {Carrasco Kind}, {Knowles}, {Koopman}, {Kuehn}, {Lahav}, {Lima}, {Lin},
  {Lokken}, {Loubser}, {MacCrann}, {Maia}, {Marriage}, {Martin}, {McMahon},
  {Melchior}, {Menanteau}, {Miquel}, {Miyatake}, {Moodley}, {Morgan},
  {Mroczkowski}, {Nati}, {Newburgh}, {Niemack}, {Nishizawa}, {Ogando},
  {Orlowski-Scherer}, {Page}, {Palmese}, {Partridge}, {Paz-Chinch{\'o}n},
  {Phakathi}, {Plazas}, {Robertson}, {Romer}, {Carnero Rosell}, {Salatino},
  {Sanchez}, {Schaan}, {Schillaci}, {Sehgal}, {Serrano}, {Shin}, {Simon},
  {Smith}, {Soares-Santos}, {Spergel}, {Staggs}, {Storer}, {Suchyta},
  {Swanson}, {Tarle}, {Thomas}, {To}, {Trac}, {Ullom}, {Vale}, {Van Lanen},
  {Vavagiakis}, {De Vicente}, {Wilkinson}, {Wollack}, {Xu}, \&
  {Zhang}}]{2020arXiv200911043H}
{Hilton}, M., {Sif{\'o}n}, C., {Naess}, S., {et~al.} 2020, arXiv e-prints,
  ApJS, subm., arXiv:2009.11043

\bibitem[{{Hofmann} {et~al.}(2017){Hofmann}, {Sanders}, {Clerc}, {Nandra},
  {Ridl}, {Dennerl}, {Ramos-Ceja}, {Finoguenov}, \& {Reiprich}}]{hsc17}
{Hofmann}, F., {Sanders}, J.~S., {Clerc}, N., {et~al.} 2017, A\&A, 606, A118

\bibitem[{{Huang} {et~al.}(2020){Huang}, {Bleem}, {Stalder}, {Ade}, {Allen},
  {Anderson}, {Austermann}, {Avva}, {Beall}, {Bender}, {Benson}, {Bianchini},
  {Bocquet}, {Brodwin}, {Carlstrom}, {Chang}, {Chiang}, {Citron}, {Moran},
  {Crawford}, {Crites}, {Haan}, {Dobbs}, {Everett}, {Floyd}, {Gallicchio},
  {George}, {Gilbert}, {Gladders}, {Guns}, {Gupta}, {Halverson}, {Harrington},
  {Henning}, {Hilton}, {Holder}, {Holzapfel}, {Hrubes}, {Hubmayr}, {Irwin},
  {Khullar}, {Knox}, {Lee}, {Li}, {Lowitz}, {McDonald}, {McMahon}, {Meyer},
  {Mocanu}, {Montgomery}, {Nadolski}, {Natoli}, {Nibarger}, {Noble}, {Novosad},
  {Padin}, {Patil}, {Pryke}, {Reichardt}, {Ruhl}, {Saliwanchik}, {Saro},
  {Sayre}, {Schaffer}, {Sharon}, {Sievers}, {Smecher}, {Stark}, {Story},
  {Tucker}, {Vanderlinde}, {Veach}, {Vieira}, {Wang}, {Whitehorn}, {Wu}, \&
  {Yefremenko}}]{2020AJ....159..110H}
{Huang}, N., {Bleem}, L.~E., {Stalder}, B., {et~al.} 2020, \aj, 159, 110

\bibitem[{{Hurier} {et~al.}(2013){Hurier}, {Macías-Pérez}, \&
  {Hildebrandt}}]{Hurier13}
{Hurier}, G., {Macías-Pérez}, J.~F., \& {Hildebrandt}, S. 2013, A\&A, 558,
  A118

\bibitem[{{Itoh} {et~al.}(1998){Itoh}, {Kohyama}, \& {Nozawa}}]{Itoh98}
{Itoh}, N., {Kohyama}, Y., \& {Nozawa}, S. 1998, ApJ, 502, 7

\bibitem[{{Jarrett}(2004)}]{2004PASA...21..396J}
{Jarrett}, T. 2004, \pasa, 21, 396

\bibitem[{{Jarrett} {et~al.}(2000){Jarrett}, {Chester}, {Cutri}, {Schneider},
  {Skrutskie}, \& {Huchra}}]{jcc00}
{Jarrett}, T.~H., {Chester}, T., {Cutri}, R., {et~al.} 2000, AJ, 119, 2498

\bibitem[{{Johnston} {et~al.}(2008){Johnston}, {Taylor}, {Bailes}, {Bartel},
  {Baugh}, {Bietenholz}, {Blake}, {Braun}, {Brown}, {Chatterjee}, {Darling},
  {Deller}, {Dodson}, {Edwards}, {Ekers}, {Ellingsen}, {Feain}, {Gaensler},
  {Haverkorn}, {Hobbs}, {Hopkins}, {Jackson}, {James}, {Joncas}, {Kaspi},
  {Kilborn}, {Koribalski}, {Kothes}, {Landecker}, {Lenc}, {Lovell}, {Macquart},
  {Manchester}, {Matthews}, {McClure-Griffiths}, {Norris}, {Pen}, {Phillips},
  {Power}, {Protheroe}, {Sadler}, {Schmidt}, {Stairs}, {Staveley-Smith},
  {Stil}, {Tingay}, {Tzioumis}, {Walker}, {Wall}, \&
  {Wolleben}}]{2008ExA....22..151J}
{Johnston}, S., {Taylor}, R., {Bailes}, M., {et~al.} 2008, Experimental
  Astronomy, 22, 151

\bibitem[{{Kaastra} {et~al.}(2013){Kaastra}, {Finoguenov}, {Nicastro},
  {Branchini}, {Schaye}, {Cappelluti}, {Nevalainen}, {Barcons}, {Bregman},
  {Croston}, {Dolag}, {Ettori}, {Galeazzi}, {Ohashi}, {Piro}, {Pointecouteau},
  {Pratt}, {Reiprich}, {Roncarelli}, {Sanders}, {Takei}, \& {Ursino}}]{kfn13}
{Kaastra}, J., {Finoguenov}, A., {Nicastro}, F., {et~al.} 2013, Athena
  Supporting Paper, arXiv:1306.2324 [\eprint[arXiv]{1306.2324}]

\bibitem[{{Kaastra} {et~al.}(2003){Kaastra}, {Lieu}, {Tamura}, {Paerels}, \&
  {den Herder}}]{klt03}
{Kaastra}, J.~S., {Lieu}, R., {Tamura}, T., {Paerels}, F.~B.~S., \& {den
  Herder}, J.~W. 2003, A\&A, 397, 445

\bibitem[{{Kaastra} {et~al.}(2006){Kaastra}, {Werner}, {Herder}, {Paerels}, {de
  Plaa}, {Rasmussen}, \& {de Vries}}]{kwd06}
{Kaastra}, J.~S., {Werner}, N., {Herder}, J.~W.~A.~d., {et~al.} 2006, ApJ, 652,
  189

\bibitem[{{Kim} {et~al.}(2007){Kim}, {Bolton}, {Viel}, {Haehnelt}, \&
  {Carswell}}]{2007MNRAS.382.1657K}
{Kim}, T.~S., {Bolton}, J.~S., {Viel}, M., {Haehnelt}, M.~G., \& {Carswell},
  R.~F. 2007, \mnras, 382, 1657

\bibitem[{{Klein} {et~al.}(2019){Klein}, {Grandis}, {Mohr}, {Paulus}, {Abbott},
  {Annis}, {Avila}, {Bertin}, {Brooks}, {Buckley-Geer}, {Rosell}, {Kind},
  {Carretero}, {Castander}, {Cunha}, {D'Andrea}, {da Costa}, {De Vicente},
  {Desai}, {Diehl}, {Dietrich}, {Doel}, {Evrard}, {Flaugher}, {Fosalba},
  {Frieman}, {Garc{\'\i}a-Bellido}, {Gaztanaga}, {Giles}, {Gruen}, {Gruendl},
  {Gschwend}, {Gutierrez}, {Hartley}, {Hollowood}, {Honscheid}, {Hoyle},
  {James}, {Jeltema}, {Kuehn}, {Kuropatkin}, {Lima}, {Maia}, {March},
  {Marshall}, {Menanteau}, {Miquel}, {Ogando}, {Plazas}, {Romer}, {Roodman},
  {Sanchez}, {Scarpine}, {Schindler}, {Serrano}, {Sevilla-Noarbe}, {Smith},
  {Smith}, {Soares-Santos}, {Sobreira}, {Suchyta}, {Swanson}, {Tarle},
  {Thomas}, {Vikram}, \& {DES Collaboration}}]{2019MNRAS.488..739K}
{Klein}, M., {Grandis}, S., {Mohr}, J.~J., {et~al.} 2019, \mnras, 488, 739

\bibitem[{{Klein} {et~al.}(2018){Klein}, {Mohr}, {Desai}, {Israel}, {Allam},
  {Benoit-L{\'e}vy}, {Brooks}, {Buckley-Geer}, {Carnero Rosell}, {Carrasco
  Kind}, {Cunha}, {da Costa}, {Dietrich}, {Eifler}, {Evrard}, {Frieman},
  {Gruen}, {Gruendl}, {Gutierrez}, {Honscheid}, {James}, {Kuehn}, {Lima},
  {Maia}, {March}, {Melchior}, {Menanteau}, {Miquel}, {Plazas}, {Reil},
  {Romer}, {Sanchez}, {Santiago}, {Scarpine}, {Schubnell}, {Sevilla-Noarbe},
  {Smith}, {Soares-Santos}, {Sobreira}, {Suchyta}, {Swanson}, {Tarle}, \& {DES
  Collaboration}}]{2018MNRAS.474.3324K}
{Klein}, M., {Mohr}, J.~J., {Desai}, S., {et~al.} 2018, \mnras, 474, 3324

\bibitem[{{Lakhchaura} {et~al.}(2011){Lakhchaura}, {Singh}, {Saikia}, \&
  {Hunstead}}]{2011ApJ...743...78L}
{Lakhchaura}, K., {Singh}, K.~P., {Saikia}, D.~J., \& {Hunstead}, R.~W. 2011,
  \apj, 743, 78

\bibitem[{{Lenz} {et~al.}(2019){Lenz}, {Dor{\'e}}, \&
  {Lagache}}]{2019ApJ...883...75L}
{Lenz}, D., {Dor{\'e}}, O., \& {Lagache}, G. 2019, \apj, 883, 75

\bibitem[{{Macquart} {et~al.}(2020){Macquart}, {Prochaska}, {McQuinn},
  {Bannister}, {Bhandari}, {Day}, {Deller}, {Ekers}, {James}, {Marnoch},
  {Os{\l}owski}, {Phillips}, {Ryder}, {Scott}, {Shannon}, \&
  {Tejos}}]{2020Natur.581..391M}
{Macquart}, J.~P., {Prochaska}, J.~X., {McQuinn}, M., {et~al.} 2020, \nat, 581,
  391

\bibitem[{{Martizzi} {et~al.}(2019){Martizzi}, {Vogelsberger}, {Artale},
  {Haider}, {Torrey}, {Marinacci}, {Nelson}, {Pillepich}, {Weinberger},
  {Hernquist}, {Naiman}, \& {Springel}}]{2019MNRAS.486.3766M}
{Martizzi}, D., {Vogelsberger}, M., {Artale}, M.~C., {et~al.} 2019, \mnras,
  486, 3766

\bibitem[{{McCarthy} {et~al.}(2017){McCarthy}, {Schaye}, {Bird}, \& {Le
  Brun}}]{2017MNRAS.465.2936M}
{McCarthy}, I.~G., {Schaye}, J., {Bird}, S., \& {Le Brun}, A. M.~C. 2017,
  \mnras, 465, 2936

\bibitem[{{Merloni} {et~al.}(2020){Merloni}, {Nandra}, \&
  {Predehl}}]{2020NatAs...4..634M}
{Merloni}, A., {Nandra}, K., \& {Predehl}, P. 2020, Nature Astronomy, 4, 634

\bibitem[{{Merloni} {et~al.}(2012){Merloni}, {Predehl}, {Becker},
  {B{\"o}hringer}, {Boller}, {Brunner}, {Brusa}, {Dennerl}, {Freyberg},
  {Friedrich}, {Georgakakis}, {Haberl}, {Hasinger}, {Meidinger}, {Mohr},
  {Nandra}, {Rau}, {Reiprich}, {Robrade}, {Salvato}, {Santangelo}, {Sasaki},
  {Schwope}, {Wilms}, \& {German eROSITA Consortium}}]{mpb12}
{Merloni}, A., {Predehl}, P., {Becker}, W., {et~al.} 2012, arXiv:1209.3114
  [\eprint[arXiv]{1209.3114}]

\bibitem[{{Miville-Desch{\^e}nes} \& {Lagache}(2005)}]{2005ApJS..157..302M}
{Miville-Desch{\^e}nes}, M.-A. \& {Lagache}, G. 2005, \apjs, 157, 302

\bibitem[{{Mroczkowski} {et~al.}(2019){Mroczkowski}, {Nagai}, {Basu}, {Chluba},
  {Sayers}, {Adam}, {Churazov}, {Crites}, {Di Mascolo}, \&
  {Eckert}}]{Mroczkowski19}
{Mroczkowski}, T., {Nagai}, D., {Basu}, K., {et~al.} 2019, Space Science
  Reviews, 215, 17

\bibitem[{{Nandra} {et~al.}(2013){Nandra}, {Barret}, {Barcons}, {Fabian}, {den
  Herder}, {Piro}, {Watson}, {Adami}, {Aird}, {Afonso}, \& et~al.}]{nbb13}
{Nandra}, K., {Barret}, D., {Barcons}, X., {et~al.} 2013, arXiv:1306.2307
  [\eprint[arXiv]{1306.2307}]

\bibitem[{{Nevalainen} {et~al.}(2015){Nevalainen}, {Tempel}, {Liivam{\"a}gi},
  {Branchini}, {Roncarelli}, {Giocoli}, {Hein{\"a}m{\"a}ki}, {Saar}, {Tamm},
  {Finoguenov}, {Nurmi}, \& {Bonamente}}]{ntl15}
{Nevalainen}, J., {Tempel}, E., {Liivam{\"a}gi}, L.~J., {et~al.} 2015, A\&A,
  583, A142

\bibitem[{{Nicastro} {et~al.}(2005){Nicastro}, {Mathur}, {Elvis}, {Drake},
  {Fang}, {Fruscione}, {Krongold}, {Marshall}, {Williams}, \& {Zezas}}]{nme05}
{Nicastro}, F., {Mathur}, S., {Elvis}, M., {et~al.} 2005, Nat, 433, 495

\bibitem[{{Norris} {et~al.}(2011){Norris}, {Hopkins}, {Afonso}, {Brown},
  {Condon}, {Dunne}, {Feain}, {Hollow}, {Jarvis}, {Johnston-Hollitt}, {Lenc},
  {Middelberg}, {Padovani}, {Prandoni}, {Rudnick}, {Seymour}, {Umana},
  {Andernach}, {Alexander}, {Appleton}, {Bacon}, {Banfield}, {Becker}, {Brown},
  {Ciliegi}, {Jackson}, {Eales}, {Edge}, {Gaensler}, {Giovannini}, {Hales},
  {Hancock}, {Huynh}, {Ibar}, {Ivison}, {Kennicutt}, {Kimball}, {Koekemoer},
  {Koribalski}, {L{\'o}pez-S{\'a}nchez}, {Mao}, {Murphy}, {Messias},
  {Pimbblet}, {Raccanelli}, {Randall}, {Reiprich}, {Roseboom},
  {R{\"o}ttgering}, {Saikia}, {Sharp}, {Slee}, {Smail}, {Thompson}, {Urquhart},
  {Wall}, \& {Zhao}}]{nha11}
{Norris}, R.~P., {Hopkins}, A.~M., {Afonso}, J., {et~al.} 2011, Proc. Astron.
  Soc. Aust., 28, 215

\bibitem[{{Parekh} {et~al.}(2017){Parekh}, {Durret}, {Padmanabh}, \&
  {Pandge}}]{2017MNRAS.470.3742P}
{Parekh}, V., {Durret}, F., {Padmanabh}, P., \& {Pandge}, M.~B. 2017, \mnras,
  470, 3742

\bibitem[{{Parekh} {et~al.}(2020){Parekh}, {Lagan{\'a}}, {Thorat}, {van der
  Heyden}, {Iqbal}, \& {Durret}}]{2020MNRAS.491.2605P}
{Parekh}, V., {Lagan{\'a}}, T.~F., {Thorat}, K., {et~al.} 2020, \mnras, 491,
  2605

\bibitem[{{Pfeifer} {et~al.}(2020){Pfeifer}, {McCarthy}, {Stafford}, {Brown},
  {Font}, {Kwan}, {Salcido}, \& {Schaye}}]{2020MNRAS.tmp.2378P}
{Pfeifer}, S., {McCarthy}, I.~G., {Stafford}, S.~G., {et~al.} 2020,
\mnras, 498, 1576

\bibitem[{{Pillepich} {et~al.}(2012){Pillepich}, {Porciani}, \&
  {Reiprich}}]{ppr12}
{Pillepich}, A., {Porciani}, C., \& {Reiprich}, T.~H. 2012, MNRAS, 422, 44

\bibitem[{{Pillepich} {et~al.}(2018){Pillepich}, {Reiprich}, {Porciani},
  {Borm}, \& {Merloni}}]{prp18}
{Pillepich}, A., {Reiprich}, T.~H., {Porciani}, C., {Borm}, K., \& {Merloni},
  A. 2018, MNRAS, 481, 613

\bibitem[{{Planck Collaboration}(2014)}]{Planck_PSZE}
{Planck Collaboration}. 2014, A\&A, 571, A29

\bibitem[{{Planck Collaboration}(2016{\natexlab{a}})}]{Planck_overview}
{Planck Collaboration}. 2016{\natexlab{a}}, A\&A, 594, A1

\bibitem[{{Planck Collaboration}(2016{\natexlab{b}})}]{Planck_YMAPS}
{Planck Collaboration}. 2016{\natexlab{b}}, A\&A, 594, A22

\bibitem[{{Planck Collaboration}(2016{\natexlab{c}})}]{Planck_PSZE2}
{Planck Collaboration}. 2016{\natexlab{c}}, A\&A, 594, A27

\bibitem[{{Planck Collaboration} {et~al.}(2013){Planck Collaboration}, {Ade},
  {Aghanim}, {Arnaud}, {Ashdown}, {Atrio-Barandela}, {Aumont}, {Baccigalupi},
  {Balbi}, {Banday}, \& et~al.}]{pla13}
{Planck Collaboration}, {Ade}, P.~A.~R., {Aghanim}, N., {et~al.} 2013, A\&A,
  550, A134

\bibitem[{{Pratt} {et~al.}(2019){Pratt}, {Arnaud}, {Biviano}, {Eckert},
  {Ettori}, {Nagai}, {Okabe}, \& {Reiprich}}]{pab19}
{Pratt}, G.~W., {Arnaud}, M., {Biviano}, A., {et~al.} 2019, Space Sci. Rev.,
  215, 25

\bibitem[{{Predehl} {et~al.}(2020){Predehl}, {Andritschke}, {Arefiev},
  {Babyshkin}, {Batanov}, {Becker}, {B\"ohringer}, {Bogomolov}, {Boller},
  {Borm}, {Bornemann}, {Br\"auninger}, {Br\"uggen}, {Brunner}, {Brusa},
  {Bulbul}, {Buntov}, {Burwitz}, {Burkert}, {Clerc}, {Churazov}, {Coutinho},
  {Dennerl}, {Eder}, {Emberger}, {Eraerds}, {Finoguenov}, {Freyberg},
  {Friedrich}, {Friedrich}, {F\"urmetz}, {Georgakakis}, {Gilfanov}, {Granato},
  {Grossberger}, {Gueguen}, {Gureev}, {Haberl}, {H\"alker}, {Hartner},
  {Hasinger}, {Huber}, {v.\,Kienlin}, {Kink}, {Korotkov}, {Kreykenbohm},
  {Lamer}, {Lomakin}, {Lapshov}, {Liu}, {Maitra}, {Meidinger}, {Menz},
  {Merloni}, {Mernik}, {Mican}, {Mohr}, {M\"uller}, {Nandra}, {Nazarov},
  {Pacaud}, {Pavlinsky}, {Perinati}, {Pfeffermann}, {Pietschner}, {Ramos-Ceja},
  {Rau}, {Reiffers}, {Reiprich}, {Robrade}, {Salvato}, {Sanders}, {Santangelo},
  {Sasaki}, {Scheuerle}, {Schmid}, {Schmitt}, {Schwope}, {Shirshakov},
  {Steinmetz}, {Stewart}, {Str\"uder}, {Sunyaev}, {Tenzer}, {Tiedemann},
  {Tr\"umper}, {Voron}, {Wilms }, \& {Yaroshenko et al}}]{paa20}
{Predehl}, P., {Andritschke}, R., {Arefiev}, V., {et~al.} 2020, \aap,
accepted, eprint arXiv:2010.03477

\bibitem[{{Predehl} {et~al.}(2014){Predehl}, {Andritschke}, {Becker},
  {Bornemann}, {Br{\"a}uninger}, {Brunner}, {Boller}, {Burwitz}, {Burkert},
  {Clerc}, {Churazov}, {Coutinho}, {Dennerl}, {Eder}, {Emberger}, {Eraerds},
  {Freyberg}, {Friedrich}, {F{\"u}rmetz}, {Georgakakis}, {Grossberger},
  {Haberl}, {H{\"a}lker}, {Hartner}, {Hasinger}, {Hoelzl}, {Huber}, {von
  Kienlin}, {Kink}, {Kreykenbohm}, {Lamer}, {Lomakin}, {Lapchov}, {Lovisari},
  {Meidinger}, {Merloni}, {Mican}, {Mohr}, {M{\"u}ller}, {Nandra}, {Pacaud},
  {Pavlinsky}, {Perinati}, {Pfeffermann}, {Pietschner}, {Reiffers}, {Reiprich},
  {Robrade}, {Salvato}, {Santangelo}, {Sasaki}, {Scheuerle}, {Schmid},
  {Schmitt}, {Schwope}, {Sunyaev}, {Tenzer}, {Tiedemann}, {Xu}, {Yaroshenko},
  {Walther}, {Wille}, {Wilms}, \& {Zhang}}]{pab14}
{Predehl}, P., {Andritschke}, R., {Becker}, W., {et~al.} 2014, in Society of
  Photo-Optical Instrumentation Engineers (SPIE) Conference Series, Vol. 9144,
  1

\bibitem[{{Ragagnin} {et~al.}(2019){Ragagnin}, {Dolag}, {Moscardini},
  {Biviano}, \& {D'Onofrio}}]{2019MNRAS.486.4001R}
{Ragagnin}, A., {Dolag}, K., {Moscardini}, L., {Biviano}, A., \& {D'Onofrio},
  M. 2019, \mnras, 486, 4001

\bibitem[{{Rasmussen} {et~al.}(2007){Rasmussen}, {Kahn}, {Paerels}, {Herder},
  {Kaastra}, \& {de Vries}}]{rkp06}
{Rasmussen}, A.~P., {Kahn}, S.~M., {Paerels}, F., {et~al.} 2007, ApJ, 656, 129

\bibitem[{{Reiprich} \& {B{\" o}hringer}(2002)}]{rb01}
{Reiprich}, T.~H. \& {B{\" o}hringer}, H. 2002, ApJ, 567, 716

\bibitem[{{Reiprich} {et~al.}(2013){Reiprich}, {Basu}, {Ettori}, {Israel},
  {Lovisari}, {Molendi}, {Pointecouteau}, \& {Roncarelli}}]{rbe13}
{Reiprich}, T.~H., {Basu}, K., {Ettori}, S., {et~al.} 2013, Space Sci. Rev.,
  177, 195

\bibitem[{{Reiprich} {et~al.}(2004){Reiprich}, {Sarazin}, {Kempner}, \&
  {Tittley}}]{rsk04}
{Reiprich}, T.~H., {Sarazin}, C.~L., {Kempner}, J.~C., \& {Tittley}, E. 2004,
  ApJ, 608, 179

\bibitem[{{Reiprich} {et~al.}(2003){Reiprich}, {Sarazin}, {Skrutskie},
  {Sivakoff}, {Chatzikos}, {B\"ohringer}, \& {Retzlaff}}]{rss03}
{Reiprich}, T.~H., {Sarazin}, C.~L., {Skrutskie}, M.~F., {et~al.} 2003, in IAU
  25th General Assembly, abstract 135 (S216), 18

\bibitem[{{Remazeilles} {et~al.}(2011{\natexlab{a}}){Remazeilles},
  {Delabrouille}, \& {Cardoso}}]{Remazeilles11_CILC}
{Remazeilles}, M., {Delabrouille}, J., \& {Cardoso}, J.-F. 2011{\natexlab{a}},
  MNRAS, 410, 2481

\bibitem[{{Remazeilles} {et~al.}(2011{\natexlab{b}}){Remazeilles},
  {Delabrouille}, \& {Cardoso}}]{Remazeilles11_NILC}
{Remazeilles}, M., {Delabrouille}, J., \& {Cardoso}, J.-F. 2011{\natexlab{b}},
  MNRAS, 418, 467

\bibitem[{{Richter} {et~al.}(2008){Richter}, {Paerels}, \& {Kaastra}}]{rpk08}
{Richter}, P., {Paerels}, F.~B.~S., \& {Kaastra}, J.~S. 2008, \ssr, 134, 25

\bibitem[{{Rogers} {et~al.}(2018){Rogers}, {Bird}, {Peiris}, {Pontzen},
  {Font-Ribera}, \& {Leistedt}}]{2018MNRAS.476.3716R}
{Rogers}, K.~K., {Bird}, S., {Peiris}, H.~V., {et~al.} 2018, \mnras, 476, 3716

\bibitem[{{R{\"o}hser} {et~al.}(2014){R{\"o}hser}, {Kerp}, {Winkel},
  {Boulanger}, \& {Lagache}}]{2014A&A...564A..71R}
{R{\"o}hser}, T., {Kerp}, J., {Winkel}, B., {Boulanger}, F., \& {Lagache}, G.
  2014, \aap, 564, A71

\bibitem[{{Roncarelli} {et~al.}(2006){Roncarelli}, {Ettori}, {Dolag},
  {Moscardini}, {Borgani}, \& {Murante}}]{red06}
{Roncarelli}, M., {Ettori}, S., {Dolag}, K., {et~al.} 2006, MNRAS, 373, 1339

\bibitem[{{Schellenberger} \& {Reiprich}(2017)}]{sr17b}
{Schellenberger}, G. \& {Reiprich}, T.~H. 2017, MNRAS, 471, 1370

\bibitem[{{Schlafly} \& {Finkbeiner}(2011)}]{2011ApJ...737..103S}
{Schlafly}, E.~F. \& {Finkbeiner}, D.~P. 2011, \apj, 737, 103

\bibitem[{{Schlegel} {et~al.}(1998){Schlegel}, {Finkbeiner}, \&
  {Davis}}]{1998ApJ...500..525S}
{Schlegel}, D.~J., {Finkbeiner}, D.~P., \& {Davis}, M. 1998, \apj, 500, 525

\bibitem[{{Slezak} {et~al.}(1993){Slezak}, {de Lapparent}, \&
  {Bijaoui}}]{slb93}
{Slezak}, E., {de Lapparent}, V., \& {Bijaoui}, A. 1993, \apj, 409, 517

\bibitem[{{Springel}(2005)}]{springel2005}
{Springel}, V. 2005, \mnras, 364, 1105

\bibitem[{{Springel} {et~al.}(2005){Springel}, {Di Matteo}, \&
  {Hernquist}}]{springeldimatteo2005}
{Springel}, V., {Di Matteo}, T., \& {Hernquist}, L. 2005, \mnras, 361, 776

\bibitem[{{Springel} \& {Hernquist}(2003)}]{springel2003}
{Springel}, V. \& {Hernquist}, L. 2003, \mnras, 339, 289

\bibitem[{{Starck} {et~al.}(1998){Starck}, {Murtagh}, \& {Bijaoui}}]{smb98}
{Starck}, J.-L., {Murtagh}, F.~D., \& {Bijaoui}, A. 1998, {Image Processing and
  Data Analysis}

\bibitem[{{Sugawara} {et~al.}(2017){Sugawara}, {Takizawa}, {Itahana},
  {Akamatsu}, {Fujita}, {Ohashi}, \& {Ishisaki}}]{sti17}
{Sugawara}, Y., {Takizawa}, M., {Itahana}, M., {et~al.} 2017, PASJ, 69, 93

\bibitem[{{Sunyaev} \& {Zeldovich}(1970)}]{Sunyaev70}
{Sunyaev}, R.~A. \& {Zeldovich}, Y.~B. 1970, Comments Astrophys. Space Phys.,
  2, 66

\bibitem[{{Sunyaev} \& {Zeldovich}(1972)}]{Sunyaev72}
{Sunyaev}, R.~A. \& {Zeldovich}, Y.~B. 1972, Comments Astrophys. Space Phys.,
  4, 173

\bibitem[{{Tanimura} {et~al.}(2020){Tanimura}, {Aghanim}, {Bonjean},
  {Malavasi}, \& {Douspis}}]{tab20}
{Tanimura}, H., {Aghanim}, N., {Bonjean}, V., {Malavasi}, N., \& {Douspis}, M.
  2020, \aap, 637, A41

\bibitem[{{Tanimura} {et~al.}(2019){Tanimura}, {Hinshaw}, {McCarthy}, {Van
  Waerbeke}, {Aghanim}, {Ma}, {Mead}, {Hojjati}, \& {Tr{\"o}ster}}]{thm19}
{Tanimura}, H., {Hinshaw}, G., {McCarthy}, I.~G., {et~al.} 2019, MNRAS, 483,
  223

\bibitem[{{Tittley} \& {Henriksen}(2001)}]{th01}
{Tittley}, E.~R. \& {Henriksen}, M. 2001, ApJ, 563, 673

\bibitem[{{Tornatore} {et~al.}(2007){Tornatore}, {Borgani}, {Dolag}, \&
  {Matteucci}}]{tornatore2007}
{Tornatore}, L., {Borgani}, S., {Dolag}, K., \& {Matteucci}, F. 2007, \mnras,
  382, 1050

\bibitem[{{Tornatore} {et~al.}(2004){Tornatore}, {Borgani}, {Matteucci},
  {Recchi}, \& {Tozzi}}]{tornatore2004}
{Tornatore}, L., {Borgani}, S., {Matteucci}, F., {Recchi}, S., \& {Tozzi}, P.
  2004, \mnras, 349, L19

\bibitem[{{Troxel} {et~al.}(2018){Troxel}, {MacCrann}, {Zuntz}, {Eifler},
  {Krause}, {Dodelson}, {Gruen}, {Blazek}, {Friedrich}, {Samuroff}, {Prat},
  {Secco}, {Davis}, {Fert{\'e}}, {DeRose}, {Alarcon}, {Amara}, {Baxter},
  {Becker}, {Bernstein}, {Bridle}, {Cawthon}, {Chang}, {Choi}, {De Vicente},
  {Drlica-Wagner}, {Elvin-Poole}, {Frieman}, {Gatti}, {Hartley}, {Honscheid},
  {Hoyle}, {Huff}, {Huterer}, {Jain}, {Jarvis}, {Kacprzak}, {Kirk}, {Kokron},
  {Krawiec}, {Lahav}, {Liddle}, {Peacock}, {Rau}, {Refregier}, {Rollins},
  {Rozo}, {Rykoff}, {S{\'a}nchez}, {Sevilla-Noarbe}, {Sheldon}, {Stebbins},
  {Varga}, {Vielzeuf}, {Wang}, {Wechsler}, {Yanny}, {Abbott}, {Abdalla},
  {Allam}, {Annis}, {Bechtol}, {Benoit-L{\'e}vy}, {Bertin}, {Brooks},
  {Buckley-Geer}, {Burke}, {Carnero Rosell}, {Carrasco Kind}, {Carretero},
  {Castander}, {Crocce}, {Cunha}, {D'Andrea}, {da Costa}, {DePoy}, {Desai},
  {Diehl}, {Dietrich}, {Doel}, {Fernandez}, {Flaugher}, {Fosalba},
  {Garc{\'\i}a-Bellido}, {Gaztanaga}, {Gerdes}, {Giannantonio}, {Goldstein},
  {Gruendl}, {Gschwend}, {Gutierrez}, {James}, {Jeltema}, {Johnson}, {Johnson},
  {Kent}, {Kuehn}, {Kuhlmann}, {Kuropatkin}, {Li}, {Lima}, {Lin}, {Maia},
  {March}, {Marshall}, {Martini}, {Melchior}, {Menanteau}, {Miquel}, {Mohr},
  {Neilsen}, {Nichol}, {Nord}, {Petravick}, {Plazas}, {Romer}, {Roodman},
  {Sako}, {Sanchez}, {Scarpine}, {Schindler}, {Schubnell}, {Smith}, {Smith},
  {Soares-Santos}, {Sobreira}, {Suchyta}, {Swanson}, {Tarle}, {Thomas},
  {Tucker}, {Vikram}, {Walker}, {Weller}, {Zhang}, \& {DES
  Collaboration}}]{2018PhRvD..98d3528T}
{Troxel}, M.~A., {MacCrann}, N., {Zuntz}, J., {et~al.} 2018, \prd, 98, 043528

\bibitem[{{Werner} {et~al.}(2008){Werner}, {Finoguenov}, {Kaastra},
  {Simionescu}, {Dietrich}, {Vink}, \& {B{\"o}hringer}}]{wfk08}
{Werner}, N., {Finoguenov}, A., {Kaastra}, J.~S., {et~al.} 2008, A\&A, 482, L29

\bibitem[{{Wiersma} {et~al.}(2009){Wiersma}, {Schaye}, \&
  {Smith}}]{wiersma2009}
{Wiersma}, R. P.~C., {Schaye}, J., \& {Smith}, B.~D. 2009, \mnras, 393, 99

\bibitem[{{Williams} {et~al.}(2006){Williams}, {Mathur}, {Nicastro}, \&
  {Elvis}}]{wmn06}
{Williams}, R.~J., {Mathur}, S., {Nicastro}, F., \& {Elvis}, M. 2006, ApJ, 642,
  L95

\bibitem[{{Wilms} {et~al.}(2000){Wilms}, {Allen}, \&
  {McCray}}]{2000ApJ...542..914W}
{Wilms}, J., {Allen}, A., \& {McCray}, R. 2000, \apj, 542, 914

\bibitem[{{Wright}(1979)}]{Wright79}
{Wright}, E.~L. 1979, ApJ, 232, 348

\bibitem[{{Zandanel} {et~al.}(2018){Zandanel}, {Fornasa}, {Prada}, {Reiprich},
  {Pacaud}, \& {Klypin}}]{zfp18}
{Zandanel}, F., {Fornasa}, M., {Prada}, F., {et~al.} 2018, MNRAS, 480, 987

\bibitem[{{Zeldovich} \& {Sunyaev}(1969)}]{zs69}
{Zeldovich}, Y.~B. \& {Sunyaev}, R.~A. 1969, Ap\&SS, 4, 301

\end{thebibliography}
%

\appendix
\appendixpage
\section{\rosi\ light curves}
\label{App_A}
For all data reduction and analysis tasks applied in this paper, the internal data processing version c945 was used together with eSASS version 200602 and HEASOFT version 6.25.

All light curves are extracted using eSASS task \texttt{flaregti} in the energy range 6--10 keV, with 100 s time bins.

\begin{figure}[ht]
\centerline{\includegraphics[width=\columnwidth]{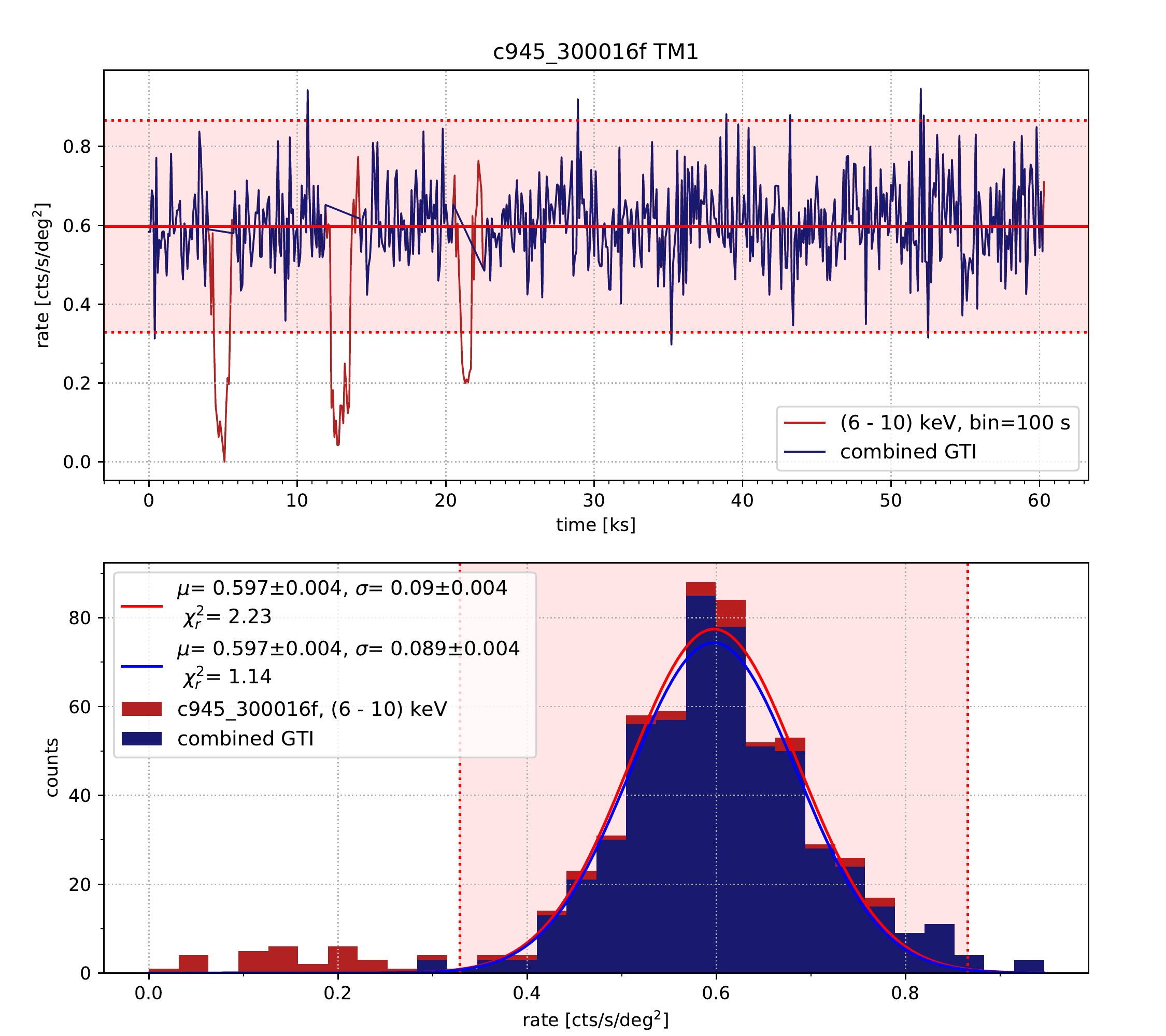}}
\caption{Top: Total 6--10 keV light curve of TM1 (red) and the filtered light curve (blue), where eSASS GTIs and manually selected GTIs were combined. Time bins of $100$ s were chosen. Dotted lines indicate the $3\sigma$ interval for illustration. Bottom: Corresponding count rate histograms and Gaussian model fits.
\label{fig:LC1}}
\end{figure} 

\begin{figure}[ht]
\centering
\includegraphics[width=\columnwidth]{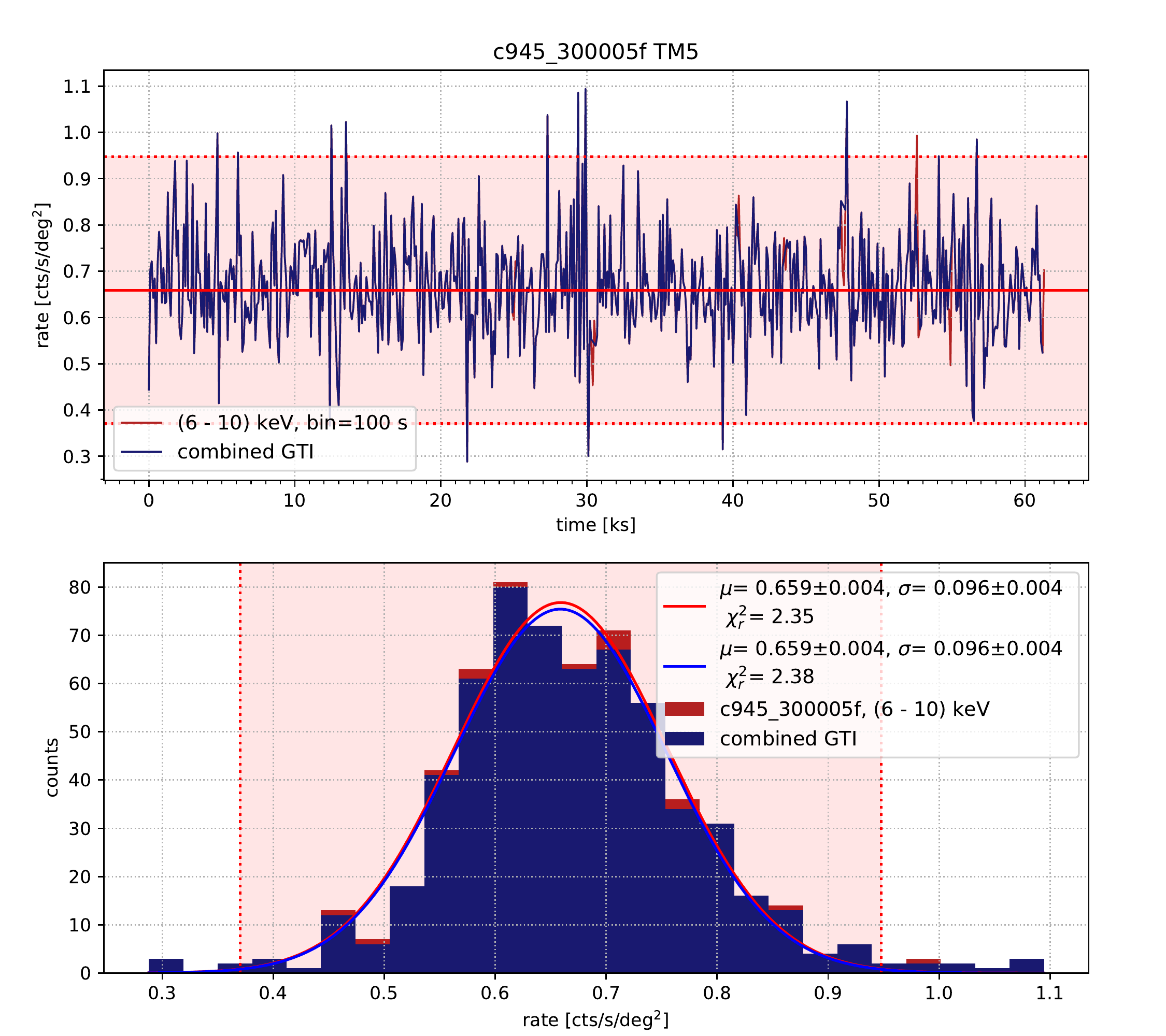}
\caption{ObsID 300005 (scan\_I) TM5. Not contaminated by Canopus exposure loss.}
\label{fig:LCs_sc1_5}
\end{figure}

\begin{figure}[ht]
\centering
\centerline{\includegraphics[width=\columnwidth]{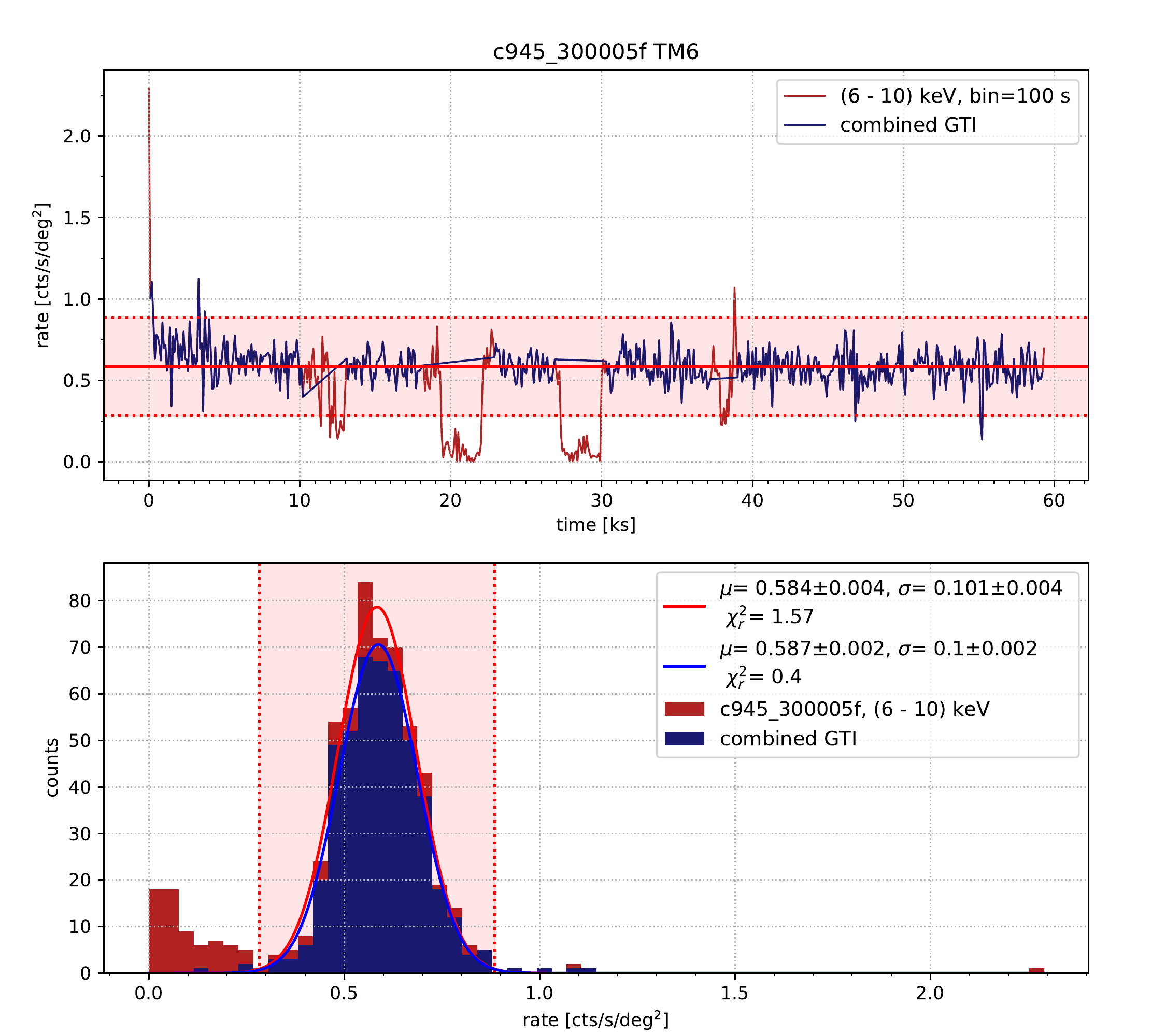}\qquad}
\caption{ObsID 300005 (scan\_I) TM6.}
\label{fig:LCs_sc1_6}
\end{figure}

\begin{figure}[ht]
\centering
\centerline{\includegraphics[width=\columnwidth]{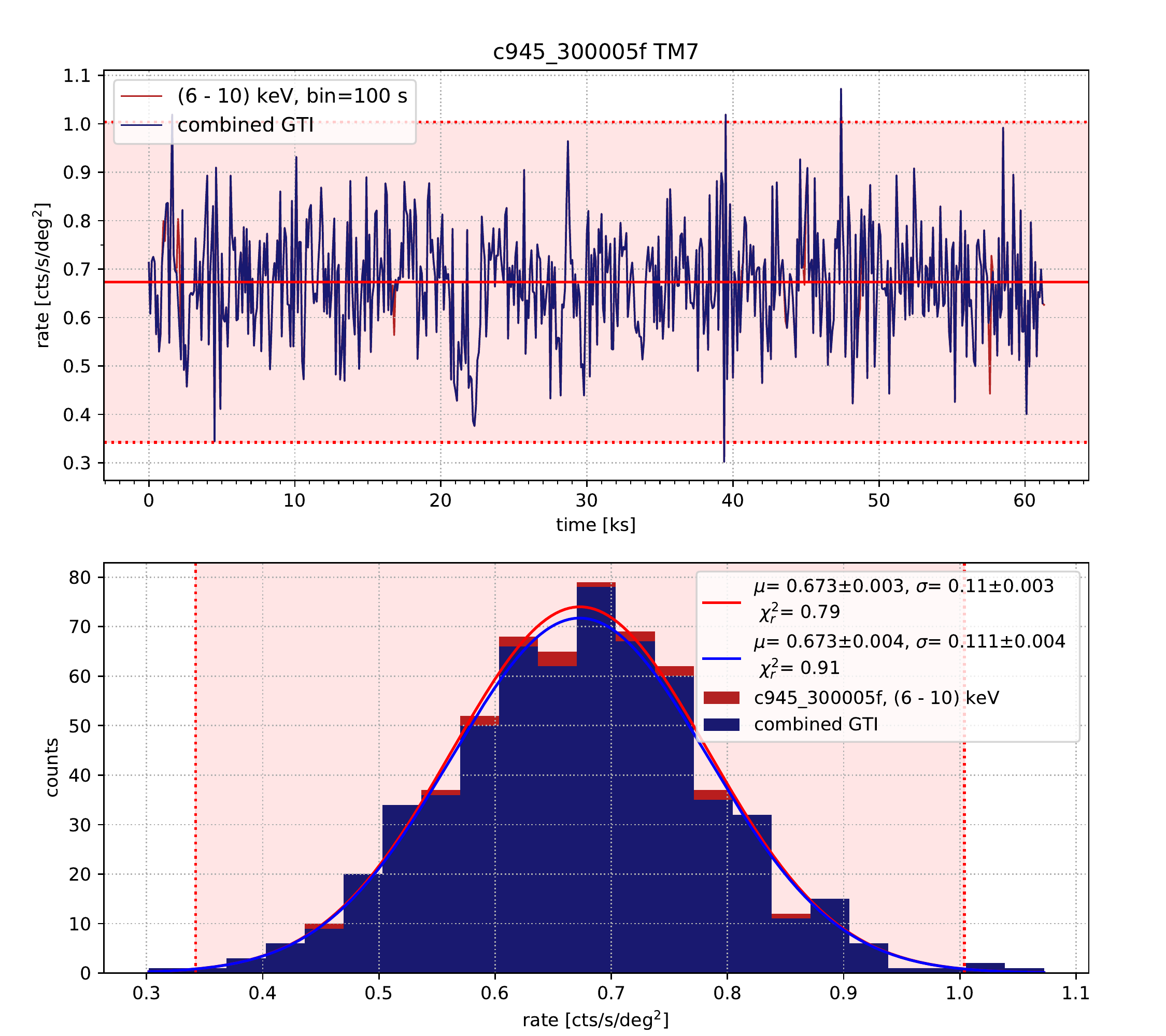}\qquad}
\caption{ObsID 300005 (scan\_I) TM7. Not contaminated by Canopus exposure loss.
\label{fig:LCs_sc1_7}}
\end{figure}
\begin{figure}[ht]
\centering
\centerline{\includegraphics[width=\columnwidth]{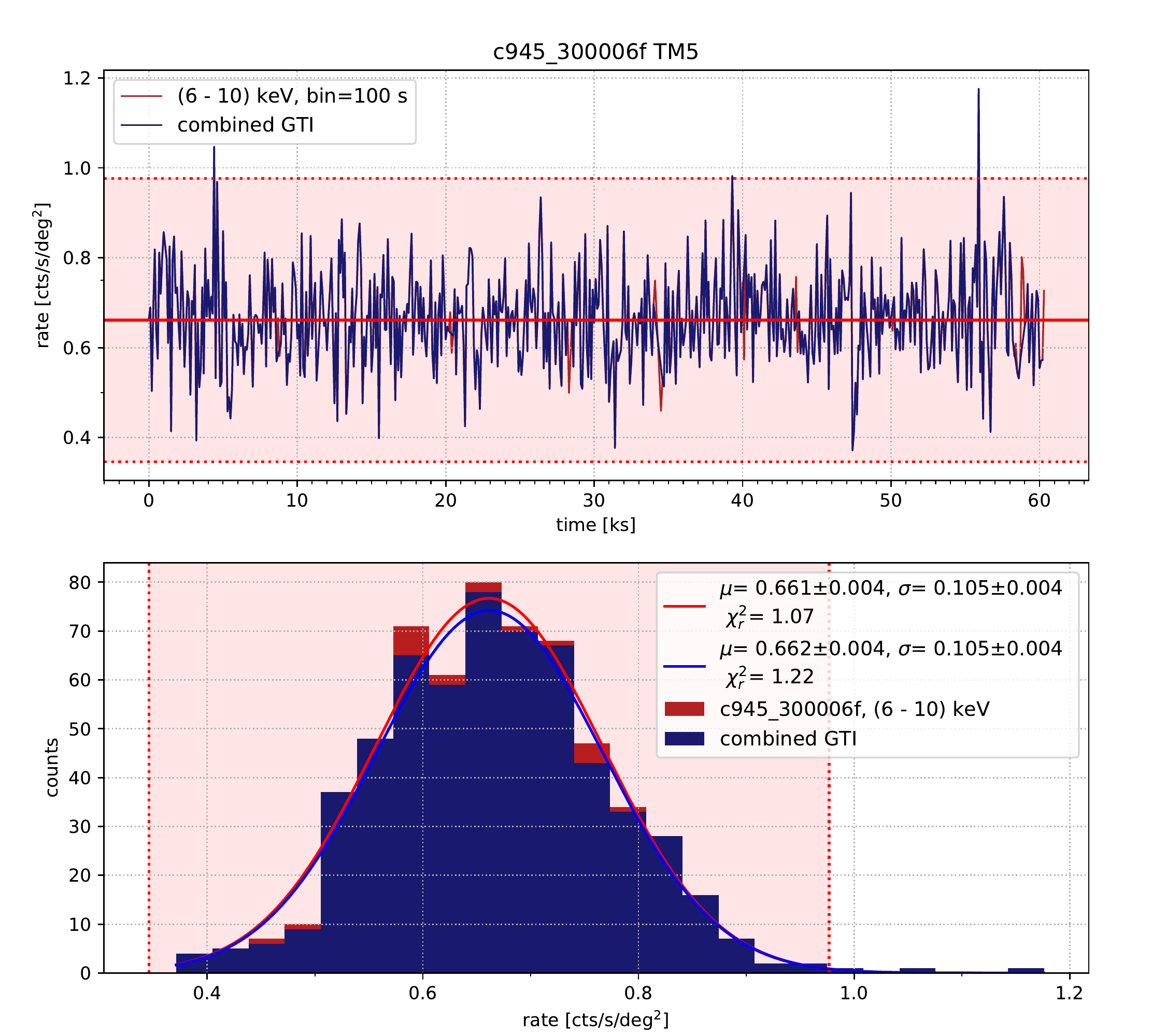}\qquad}
\caption{ObsID 300006 (scan\_II) TM5. Not contaminated by Canopus exposure loss.
\label{fig:LCs_sc2_5}}
\end{figure}

\begin{figure}[ht]
\centering
\centerline{\includegraphics[width=\columnwidth]{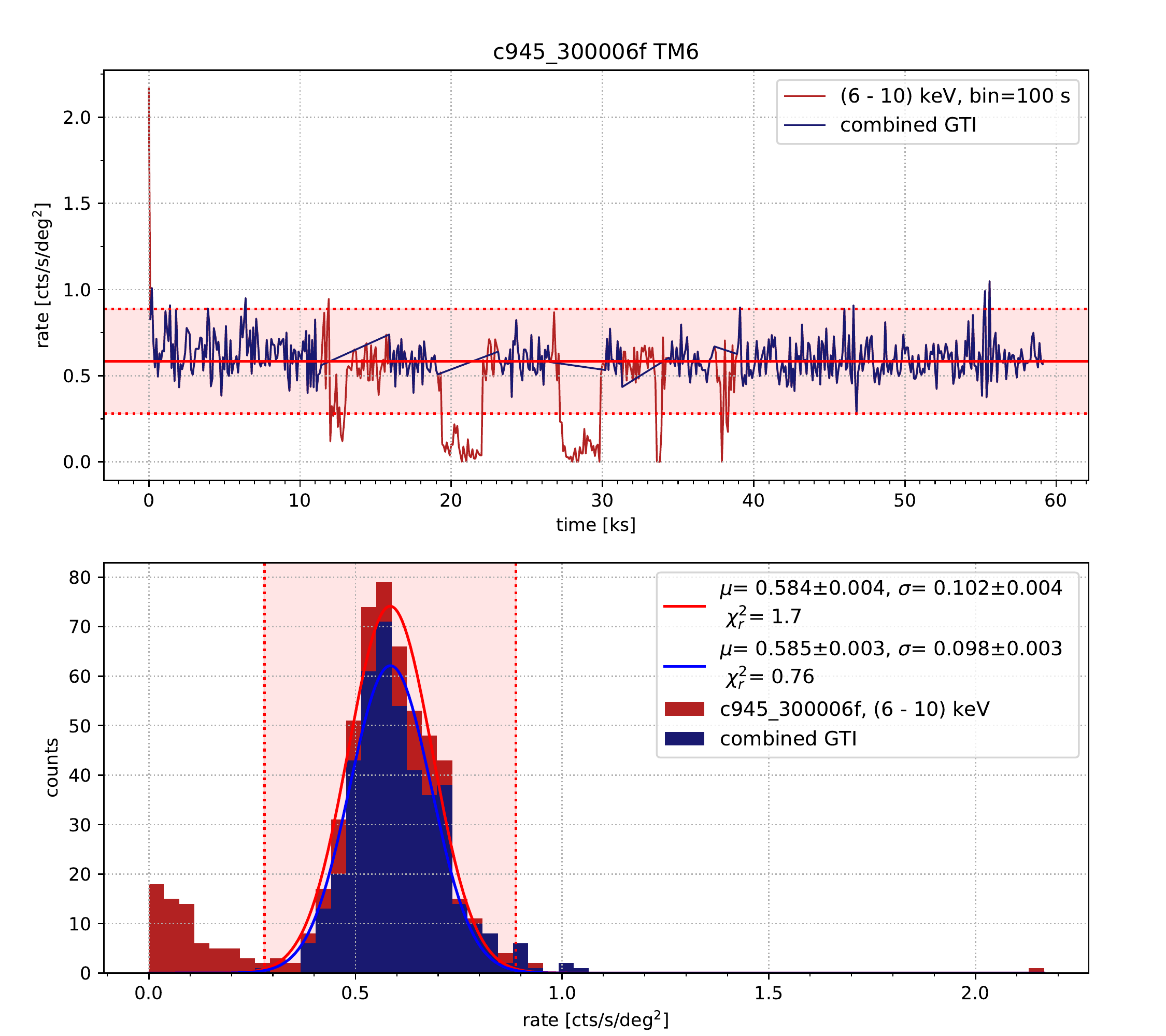}\qquad}
\caption{ObsID 300006 (scan\_II) TM6.
\label{fig:LCs_sc2_6}}
\end{figure}

\begin{figure}[ht]
\centering
\centerline{\includegraphics[width=\columnwidth]{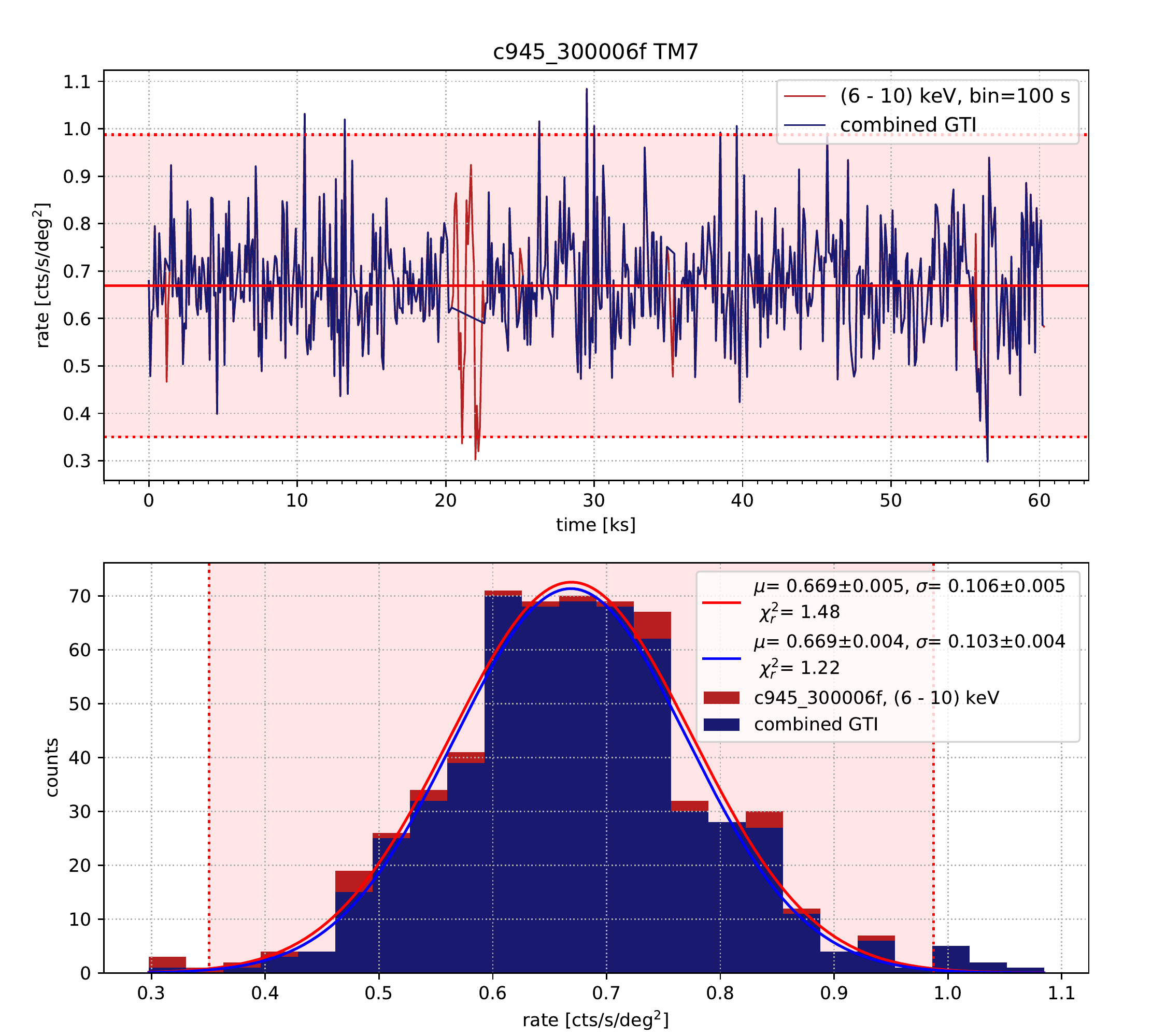}\qquad}
\caption{ObsID 300006 (scan\_II) TM7. Not contaminated by Canopus exposure loss.
\label{fig:LCs_sc2_7}}
\end{figure}
\begin{figure}[ht]
\centering
\centerline{\includegraphics[width=\columnwidth]{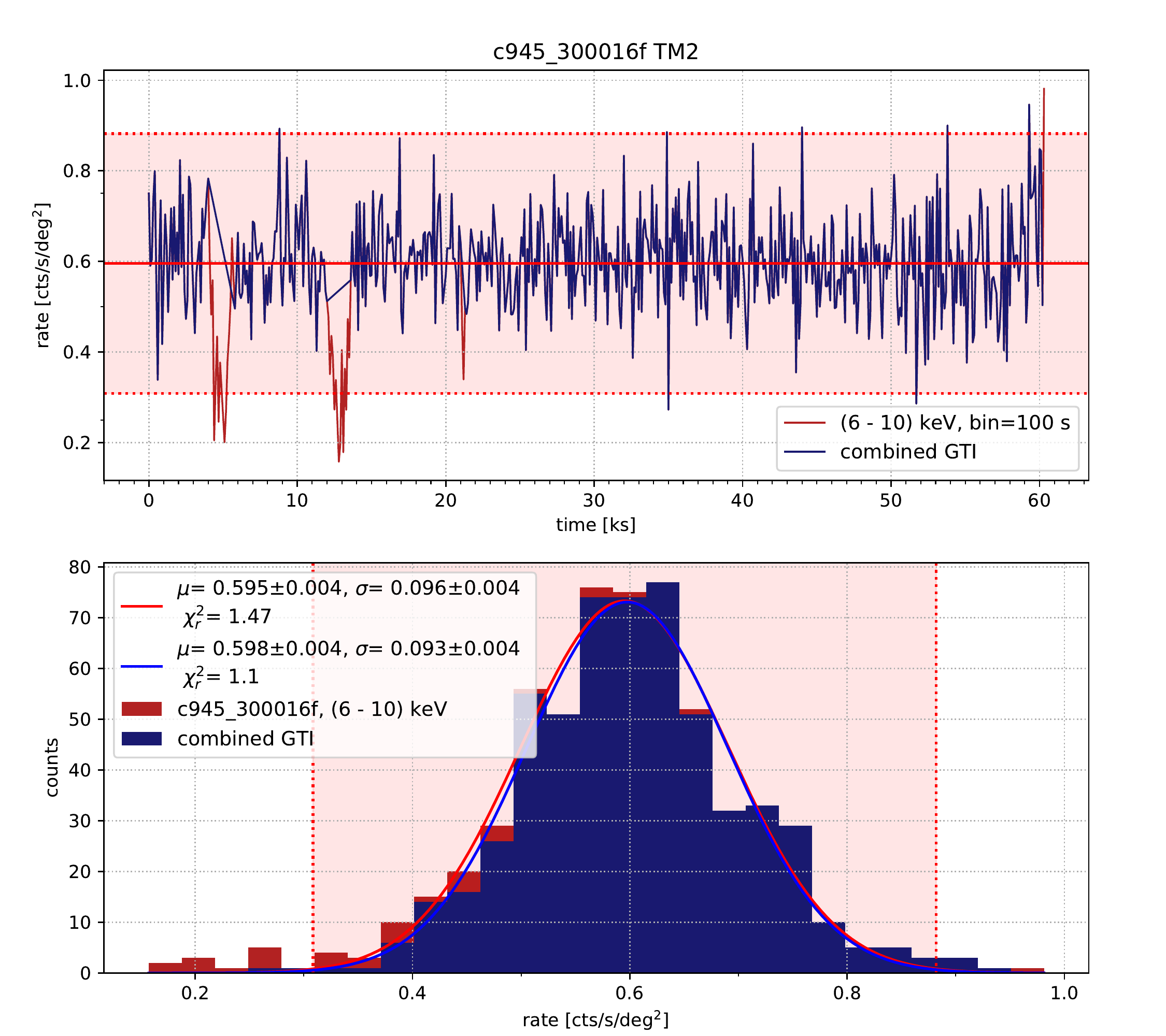}\qquad}
\caption{ObsID 300016 (scan\_III) TM2.
\label{fig:LC_sc3_2}}
\end{figure}

\begin{figure}[ht]
\centering
\centerline{\includegraphics[width=\columnwidth]{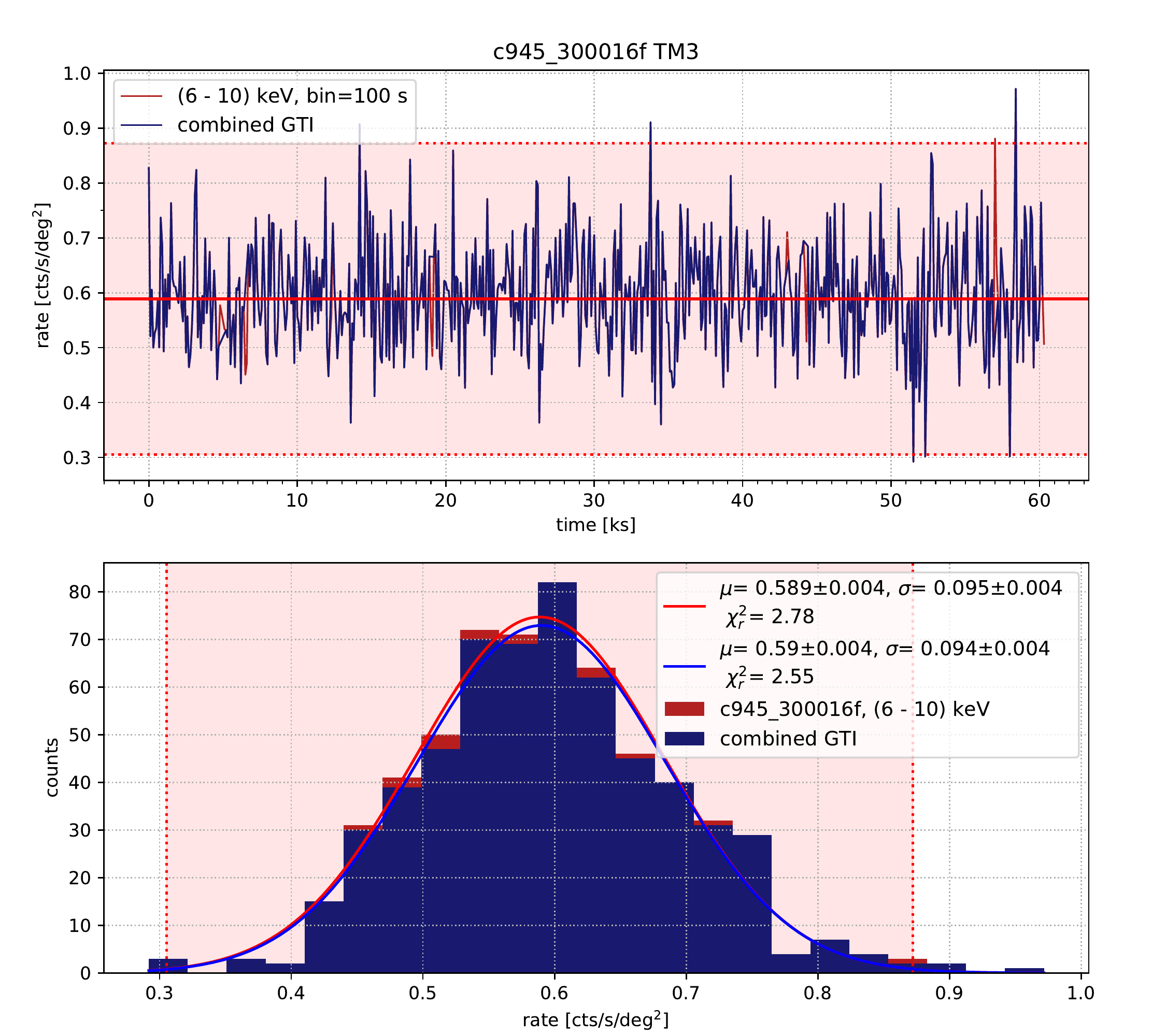}\qquad}
\caption{ObsID 300016 (scan\_III) TM3. Not contaminated by Canopus exposure loss.
\label{ig:LC_sc3_3}}
\end{figure}

\begin{figure}[ht]
\centering
\centerline{\includegraphics[width=\columnwidth]{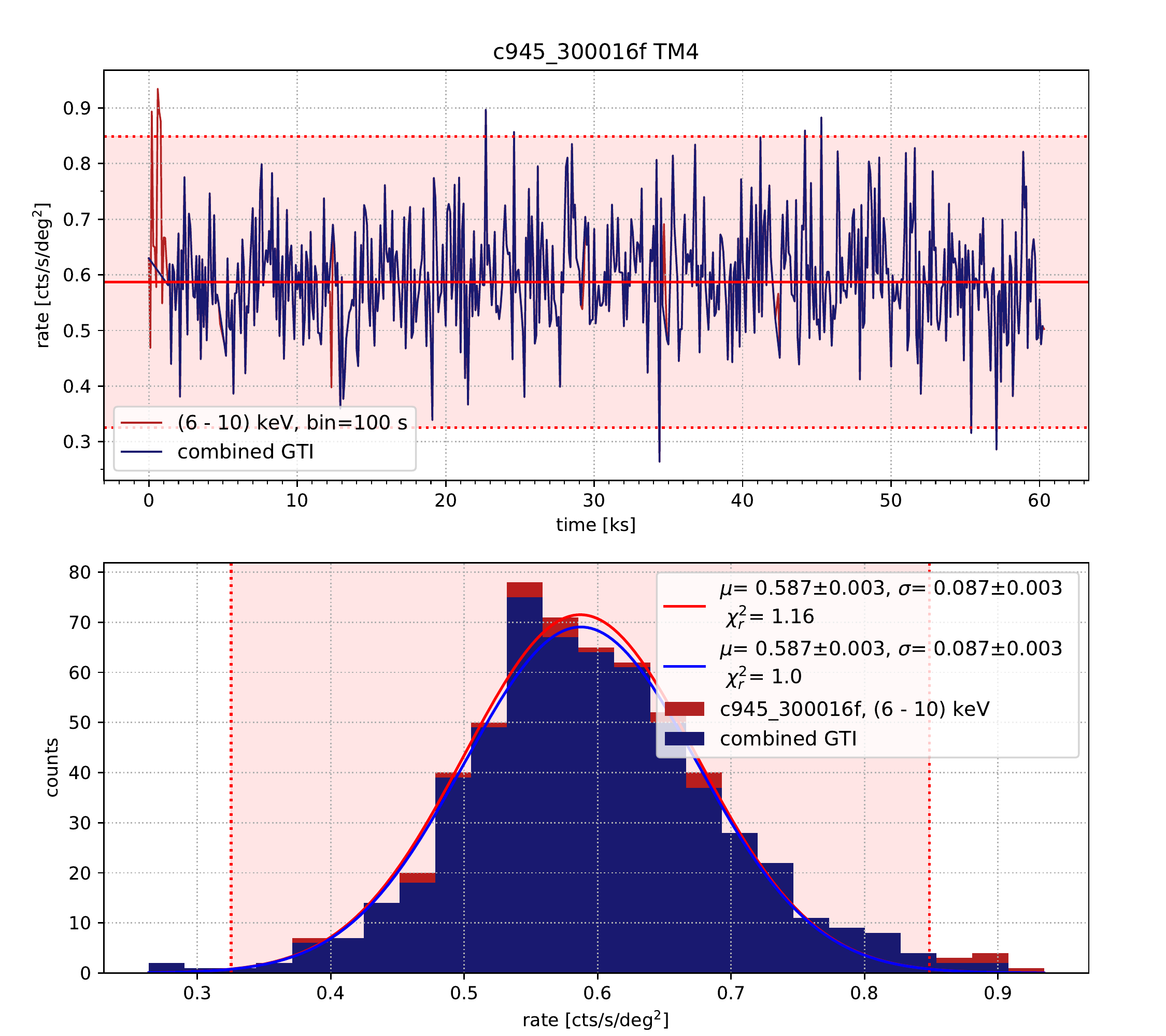}\qquad}
\caption{ObsID 300016 (scan\_III) TM4. Not contaminated by Canopus exposure loss.}
\label{ig:LC_sc3_4}
\end{figure}

\begin{figure}[ht]
\centering
\centerline{\includegraphics[width=\columnwidth]{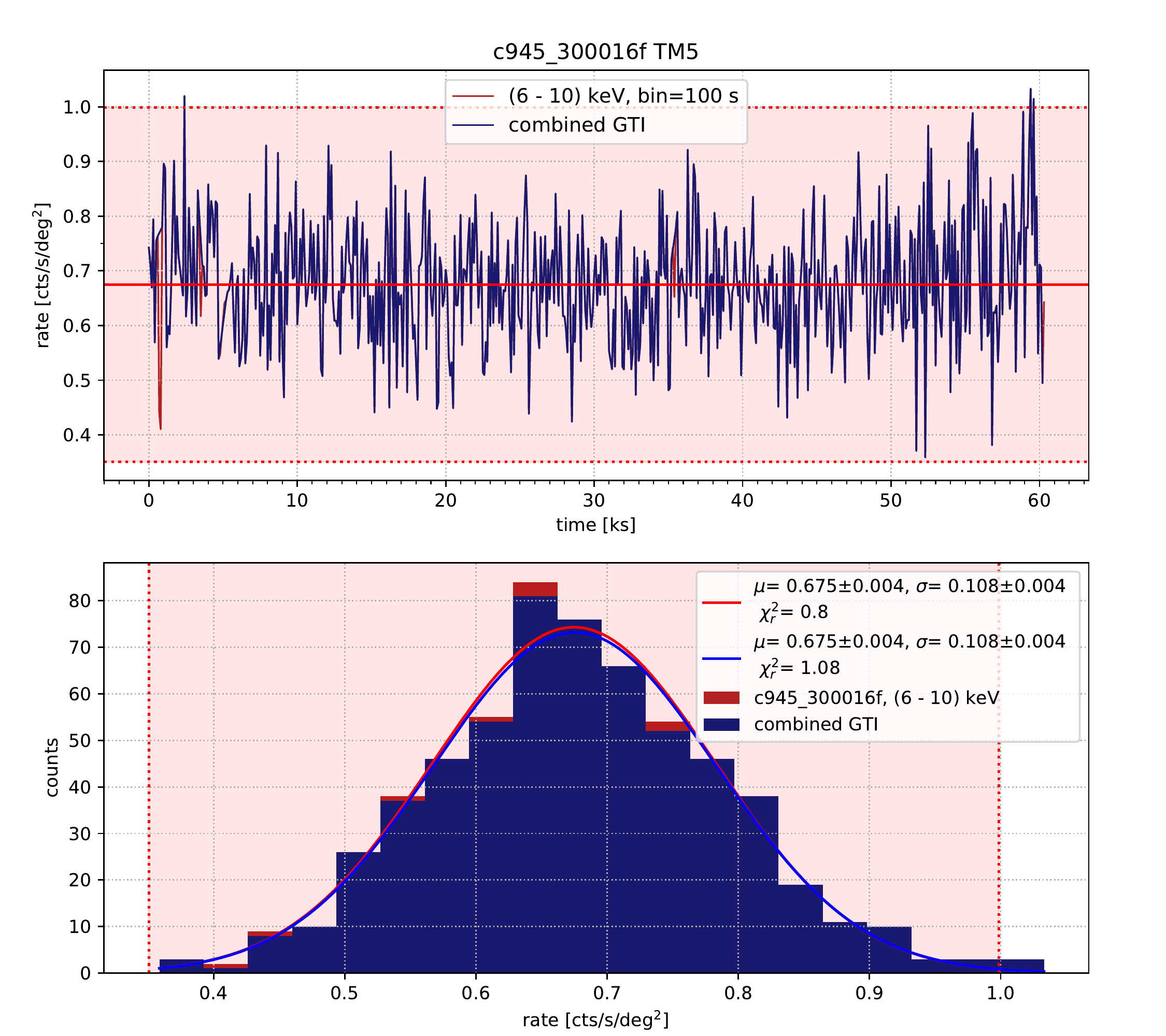}\qquad}
\caption{ObsID 300016 (scan\_III) TM5. Not contaminated by Canopus exposure loss.
\label{fig:LC_sc3_5}}
\end{figure}

\begin{figure}[ht]
\centering
\centerline{\includegraphics[width=\columnwidth]{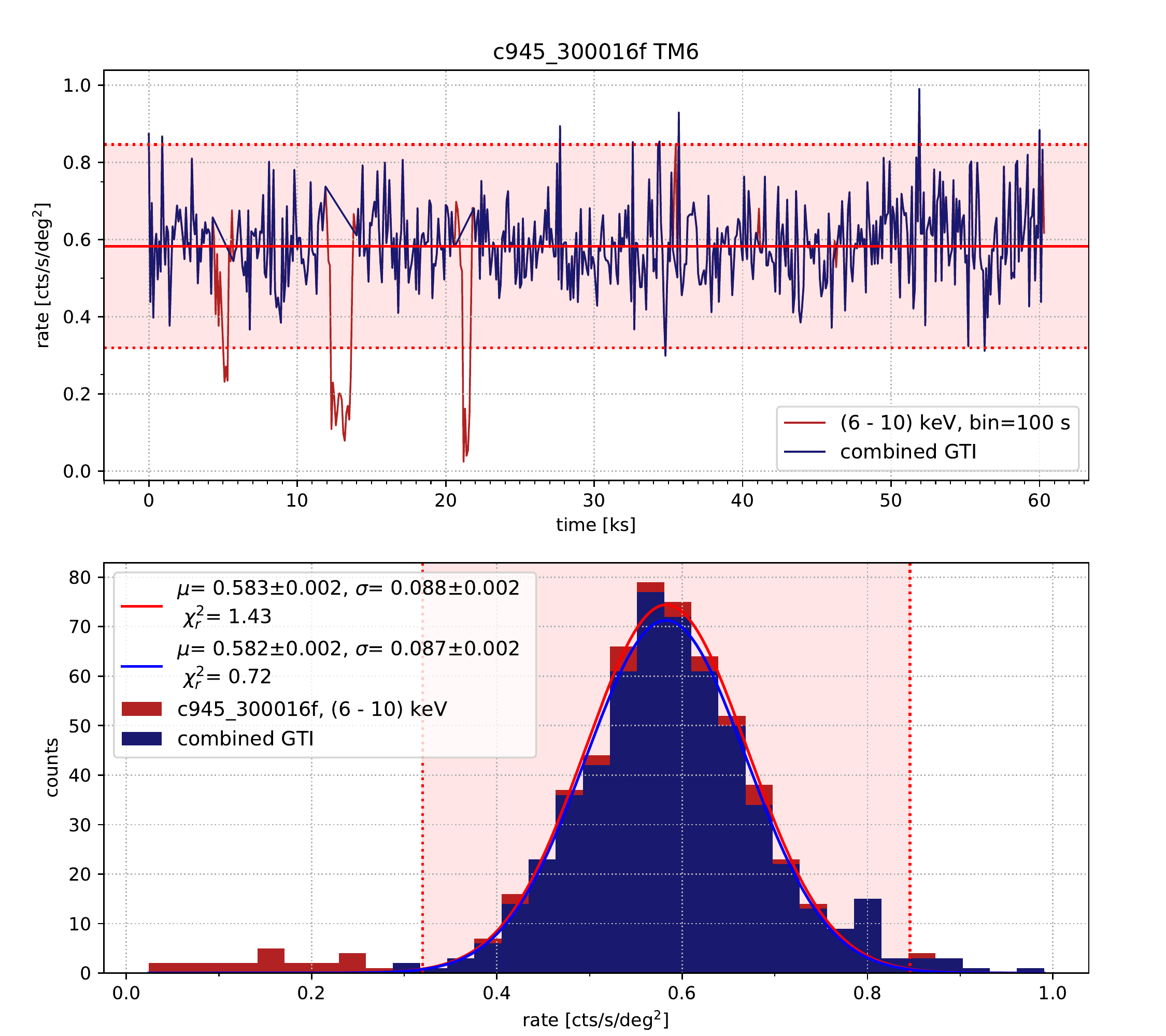}\qquad}
\caption{ObsID 300016 (scan\_III) TM6.
\label{fig:LC_sc3_6}}
\end{figure}

\begin{figure}[ht]
\centering
\centerline{\includegraphics[width=\columnwidth]{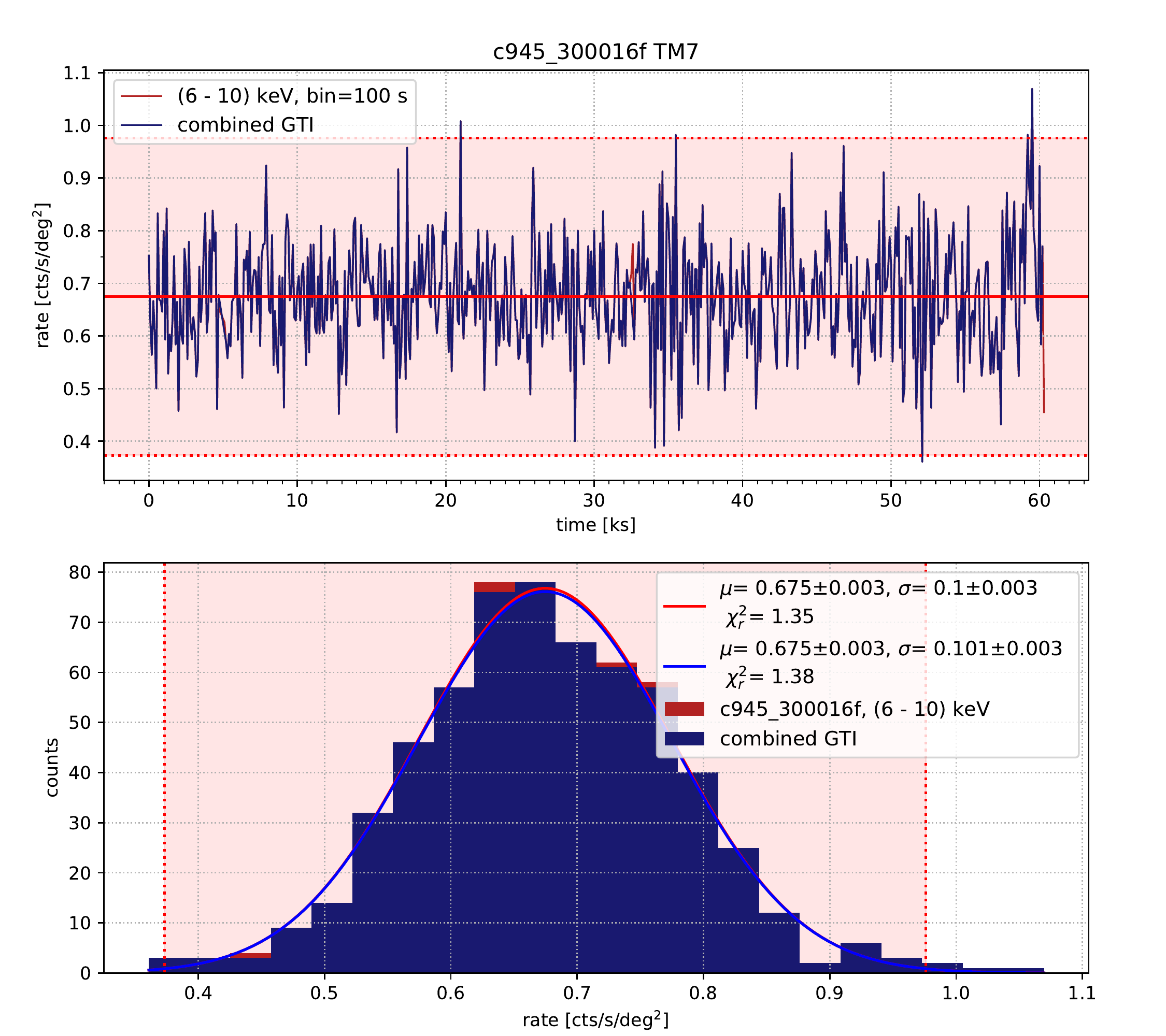}\qquad}
\caption{ObsID 300016 (scan\_III) TM7. Not contaminated by Canopus exposure loss.
\label{fig:LC_sc3_7}}
\end{figure}
\begin{figure}[ht]
\centering
\centerline{\includegraphics[width=\columnwidth]{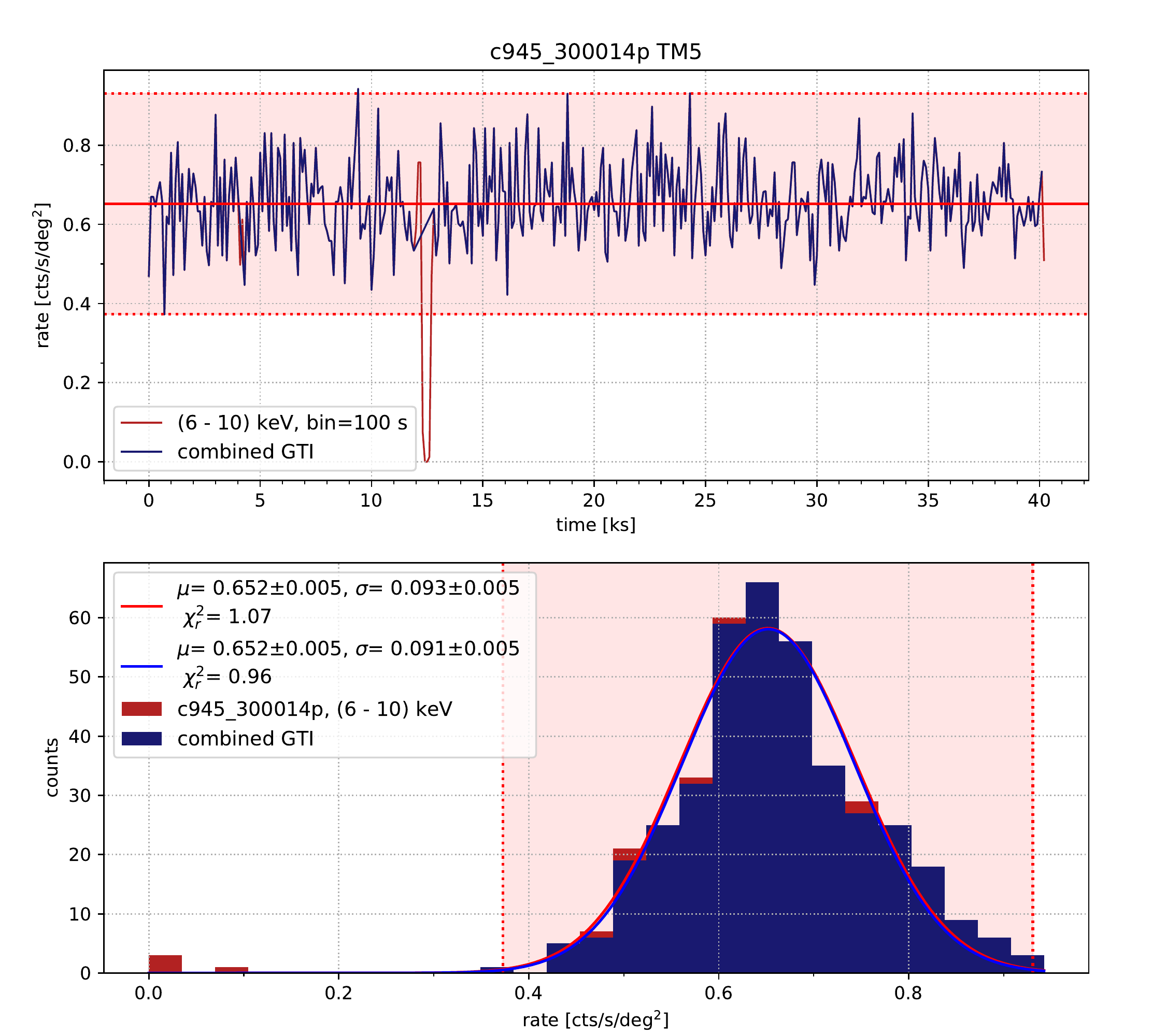}\qquad}
\caption{ObsID 300014 (pointed) TM5. Not contaminated by Canopus exposure loss.}
\label{fig:LCs_po_5}
\end{figure}

\begin{figure}[ht]
\centering
\centerline{\includegraphics[width=\columnwidth]{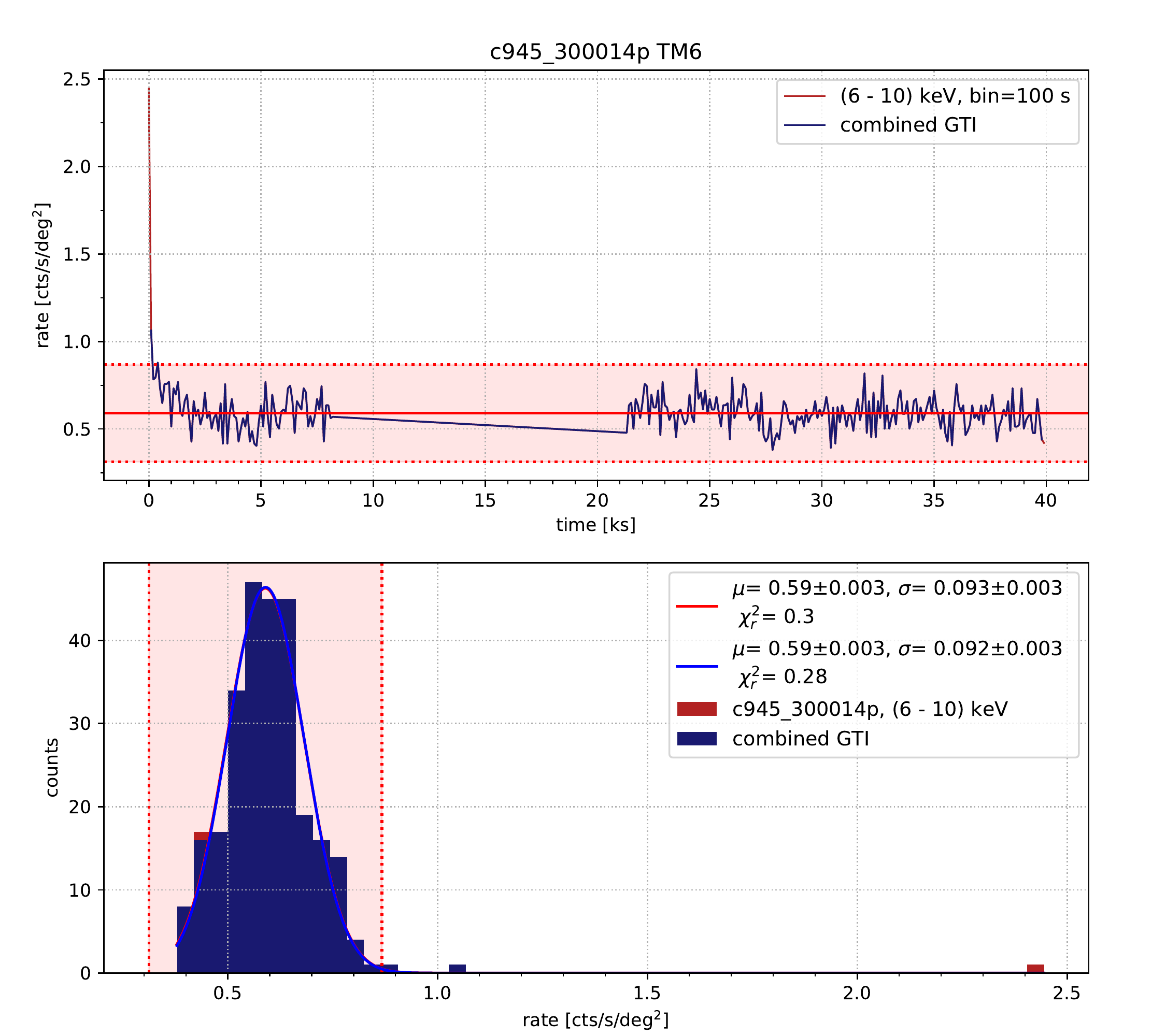}\qquad}
\caption{ObsID 300014 (pointed) TM6. Not contaminated by Canopus exposure loss. The quiet interval from $\sim8$ to $21$ ks indicates a FWC observation that was performed for calibration measurement.}
\label{fig:LCs_po_6}
\end{figure}

\begin{figure}[ht]
\centering
\centerline{\includegraphics[width=\columnwidth]{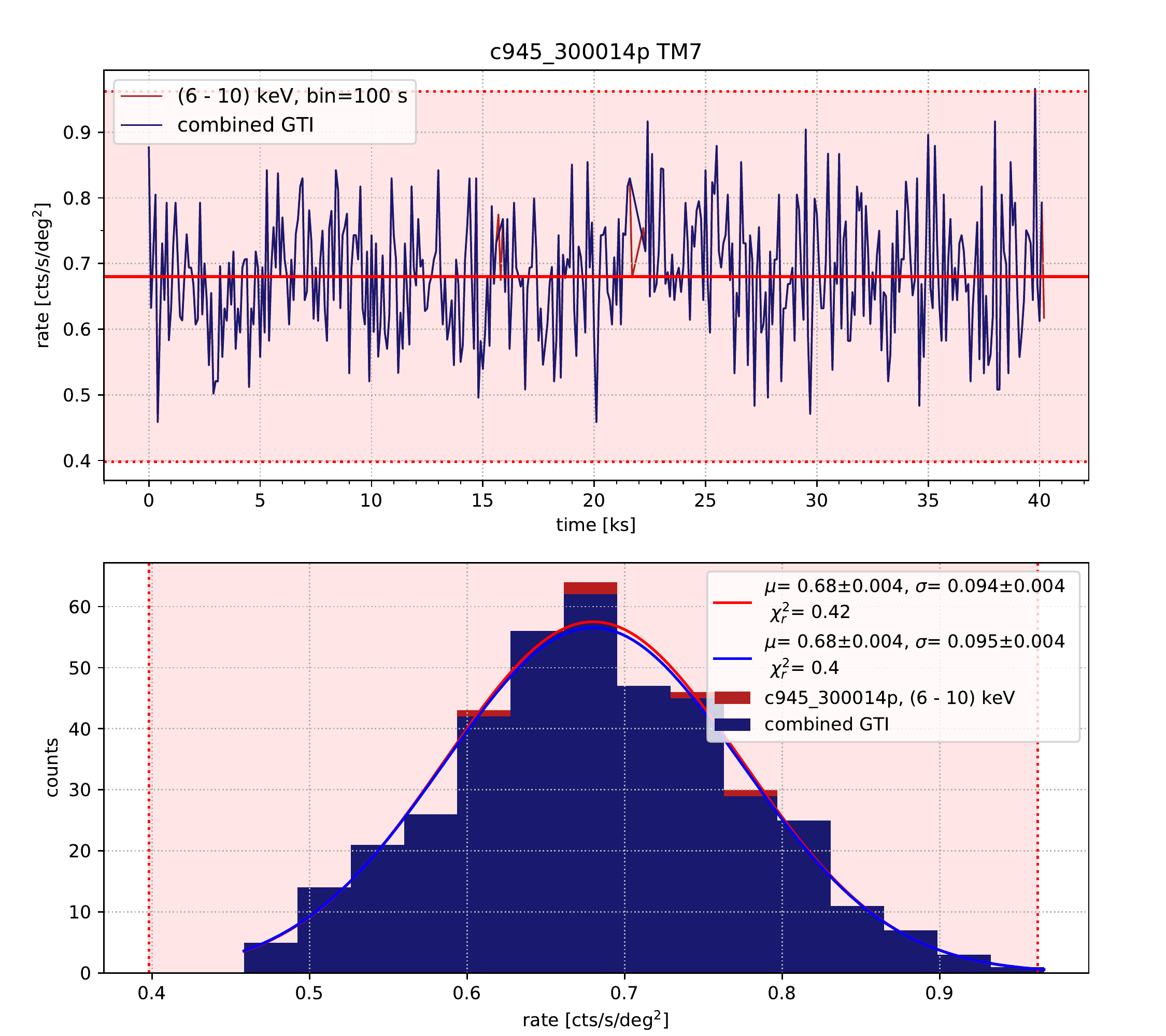}\qquad}
\caption{ObsID 300014 (pointed) TM7. Not contaminated by Canopus exposure loss.}
\label{fig:LCs_po_7}
\end{figure}

\clearpage
\section{\rosi\ Particle-induced background subtraction}
\label{App_B0}
\begin{table}[ht]
\centering
\caption{FWC count rate ratio $R$ taken for energy bands $\frac{0.3-2.0 \text{ keV}}{6.0-9.0 \text{ keV}}$, and the hard counts $H_{\rm obs}$ in energy band $6.0-9.0$ keV for the different source observations.}
\label{tab:HR}
{\small
\begin{tabular}{@{}cccccc@{}}
\hline
\hline
 \bf TM & \bf $R$ & $H (300005)$ & $H (300006)$ & $H (300016)$ & $H (300014)$  \\
\hline
 1 & 0.77 & - & - &  22835 & -\\
 2 & 0.77  & - & -  & 22589 & -\\
 3 & 0.79  & - &  - & 24146 & - \\
 4 & 0.76  & - &  - & 24278 & - \\
 5 & $0.44^*$  & 26698 & 25975 &  26830 & 17361 \\
 6 & 0.75  & 19086 & 17605  & 22814 & 8472\\
 7 & $0.42^*$  & 27188 & 25685  & 26925 & 17817\\
\hline
 $* \frac{1.0-2.0 \text{ keV}}{6.0-9.0 \text{ keV}}$  &\\
\hline
\end{tabular}
}
\end{table}

\clearpage
\section{\rosi\ exposure maps}
\label{App_B}
\begin{figure}[ht]
\centering
\centerline{\includegraphics[width=1.0\columnwidth]{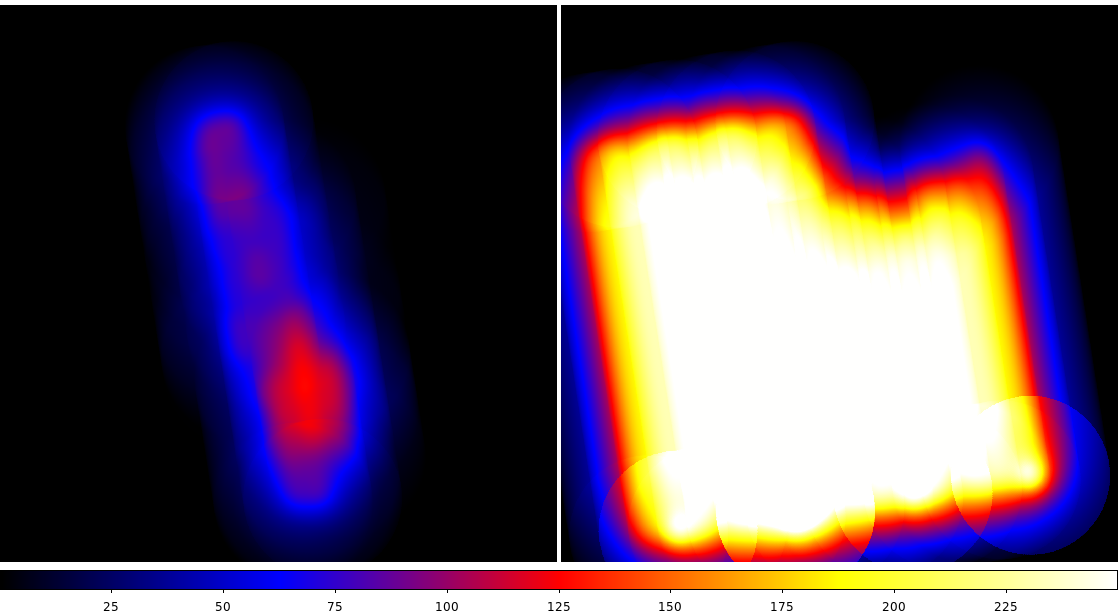}}
\caption{ Exposure maps of TM1 of observation 300016 before (left) and after (right) \texttt{DEADC} correction shown in linear scale. The missing exposure in the top-right corner is due to the proper correction for exposure loss due to the star Canopus.
\label{fig:deadcor}}
\end{figure}

One anomaly occurred for TM1 in observation 300016, such that $\sim$82.5\% of the \texttt{DEADC} values in the extension \texttt{DEADCORR} of the event file are found to be zero, while the rest of the values are lying around 1 (observing efficiency of almost $100\%$), as they should be. This issue was rectified by changing the zero values to the mean of the non-zero values. As seen in Figure \ref{fig:deadcor} (right panel), the exposure is now recovered.

\begin{figure}[ht]
\centering
\includegraphics[width=\columnwidth]{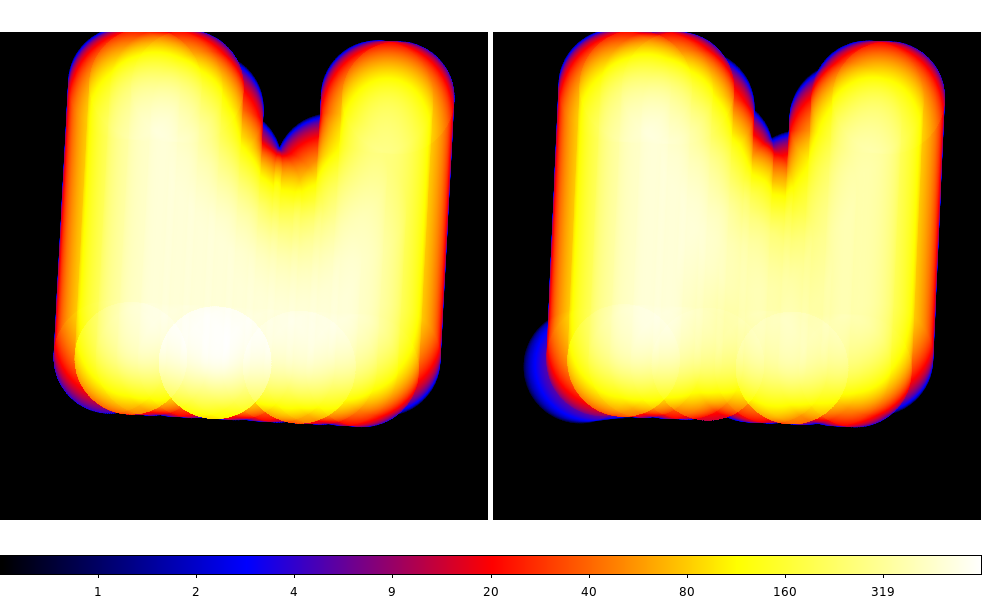}
\caption{Exposure maps (in seconds) of TM6 for observations 300005 (left) and 300006 (right) shown in logarithmic scale.}
\label{fig:expmapTM6_sc1_2}
\end{figure}

\begin{figure}[ht]
\centering
\includegraphics[width=\columnwidth]{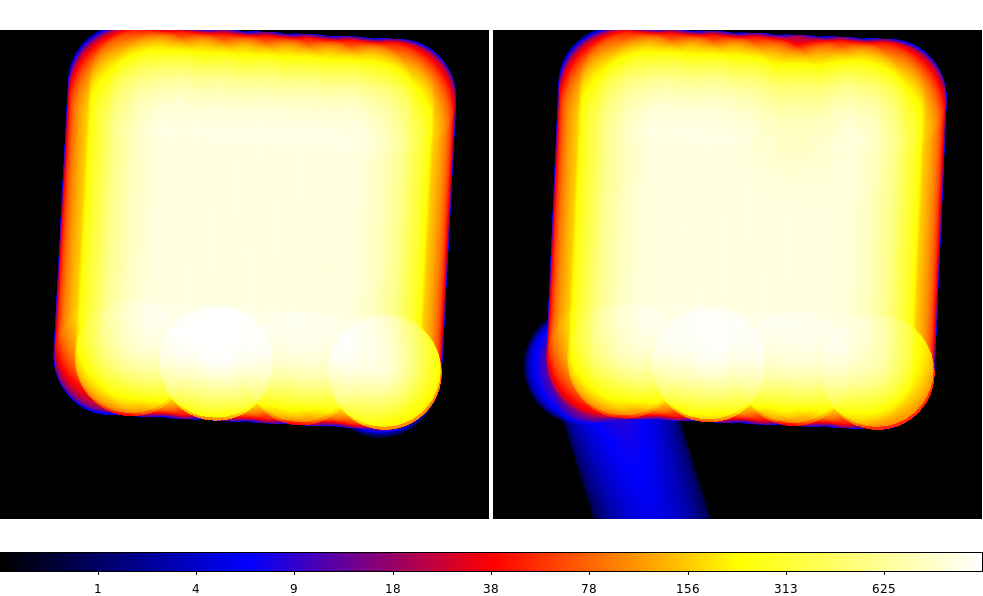}
\caption{Exposure maps (in seconds) of TM9 for observations 300005 (left) and 300006 (right) shown in logarithmic scale. The blue line at the bottom of observation 300006 indicates the starting point of the scan.}
\label{fig:expmapTM9_sc1_2}
\end{figure}

\begin{figure}[ht]
\centering
\includegraphics[width=\columnwidth]{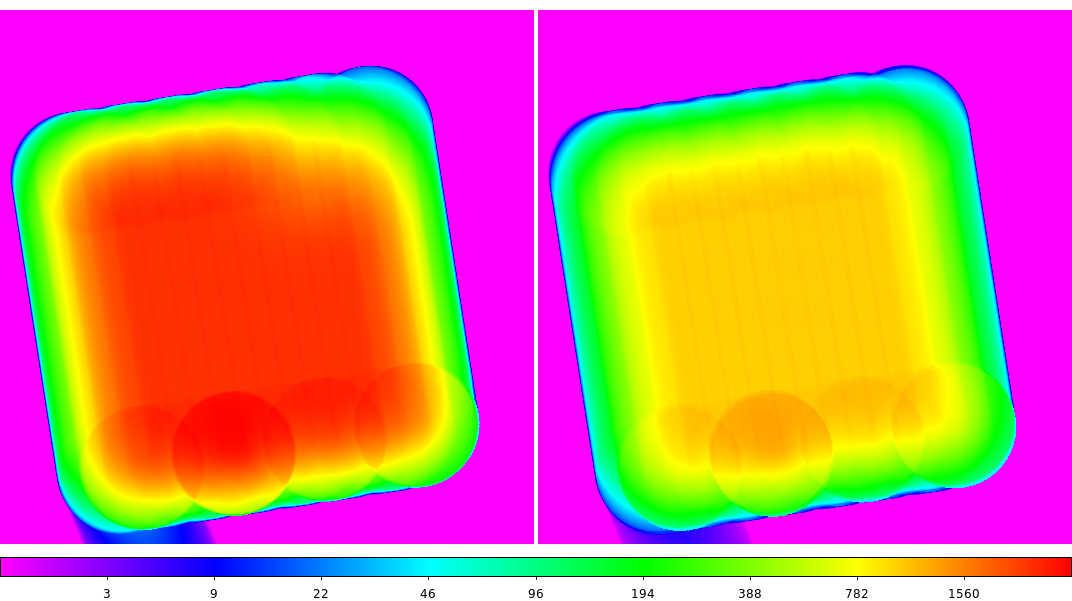}
\caption{Exposure maps (in seconds) of observation 300016 shown in logarithmic scale and ds9 color:rainbow. Left: TM8. Lower exposure time is seen at the top right (Canopus region). Right: TM9.}
\label{fig:expmap_sc3}
\end{figure}

\begin{figure}[ht]
\centering
\centerline{\includegraphics[width=1.0\columnwidth]{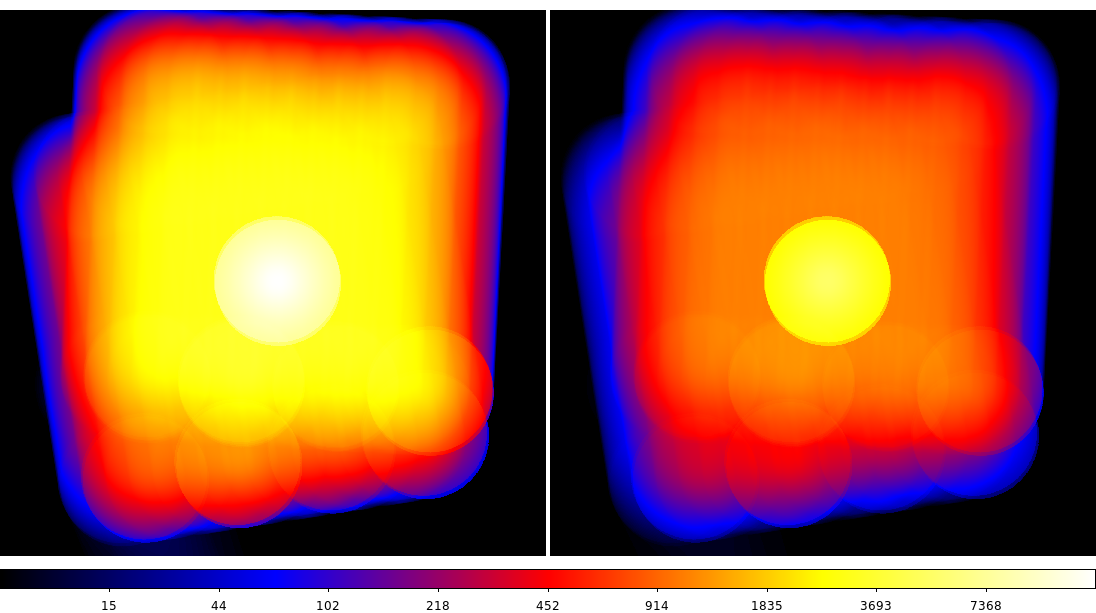}}
\caption{Exposure maps of TM9 from all observations combined, before (left) and after (right) the correction factor is applied, shown in logarithmic scale.
\label{fig:exp9_corr}}
\end{figure}

\begin{figure}[ht]
\centering
\includegraphics[width=\columnwidth]{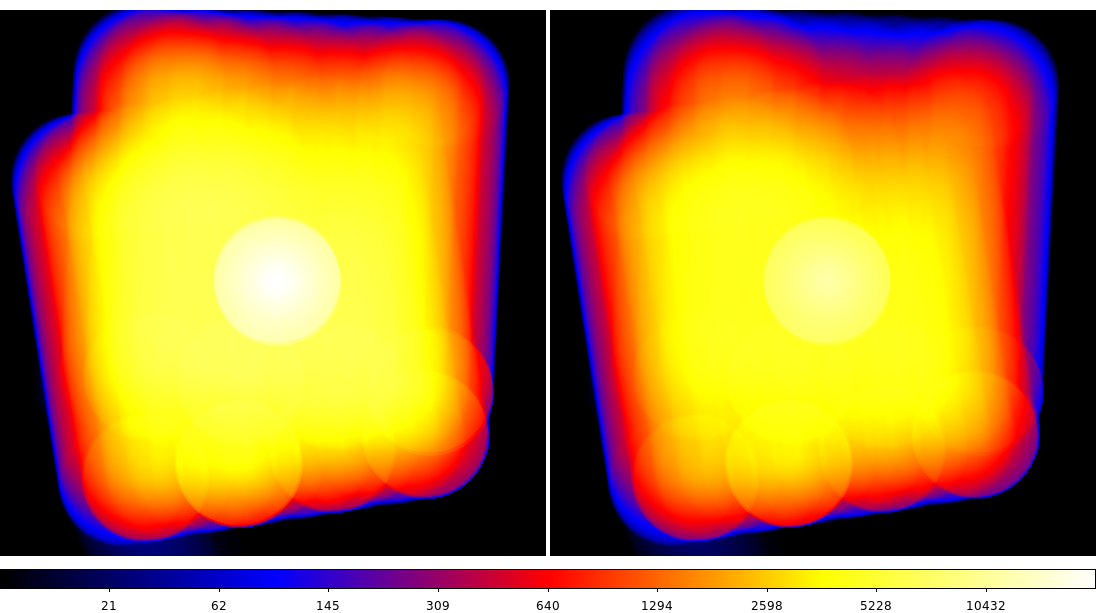}
\caption{Exposure maps (in seconds) of TM0 (TM8+TM9) from all observations shown in logarithmic scale, before (left) and after TM9 exposure correction (right).}
\label{fig:expmap_TM0}
\end{figure}

Further exposure maps of relevance are shown in Figs.~\ref{fig:expmapTM6_sc1_2}--\ref{fig:exp9_corr}. The final exposure map after all corrections is shown in Fig.~\ref{fig:expmap_TM0} (right).


\clearpage
\section{\texorpdfstring{$N_\mathrm{H_2}$}{Lg} map}
\label{App_C}

\begin{figure}[ht]
\centering
\centerline{\includegraphics[width=1.2\columnwidth]{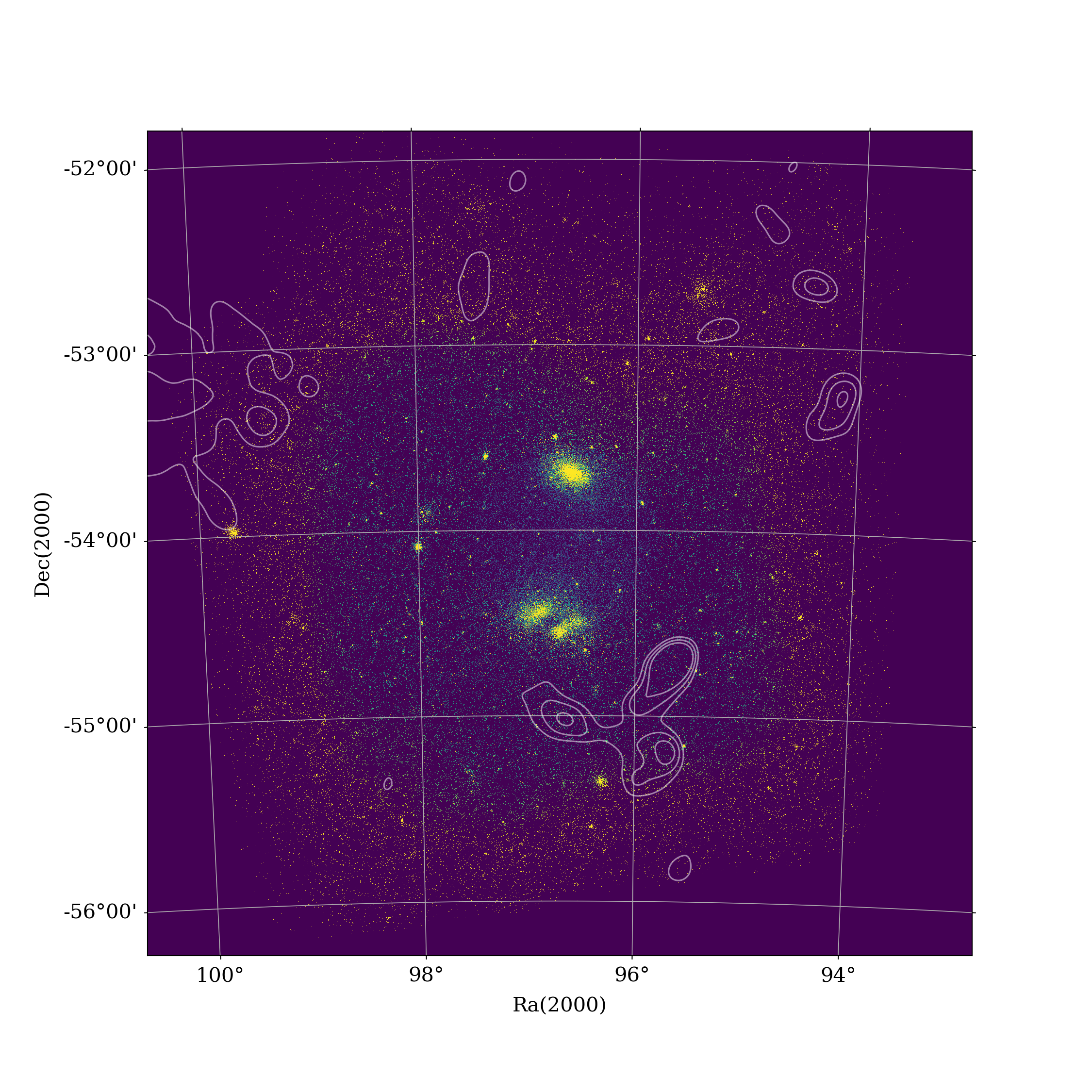}}
\caption{$N_\mathrm{H_2}$ column density contours starting at $5\times 10^{19}\,\mathrm{cm^{-2}}$ with steps of $2\times 10^{19}\,\mathrm{cm^{-2}}$ overlaid over \rosi\ image.
\label{fig:NH2map}}
\end{figure}

\clearpage
\section{Magneticum oxygen-to-soft-band-ratio map}
\label{App_E}
\begin{figure*}[ht]
\centering
\includegraphics[width=1.0\columnwidth]{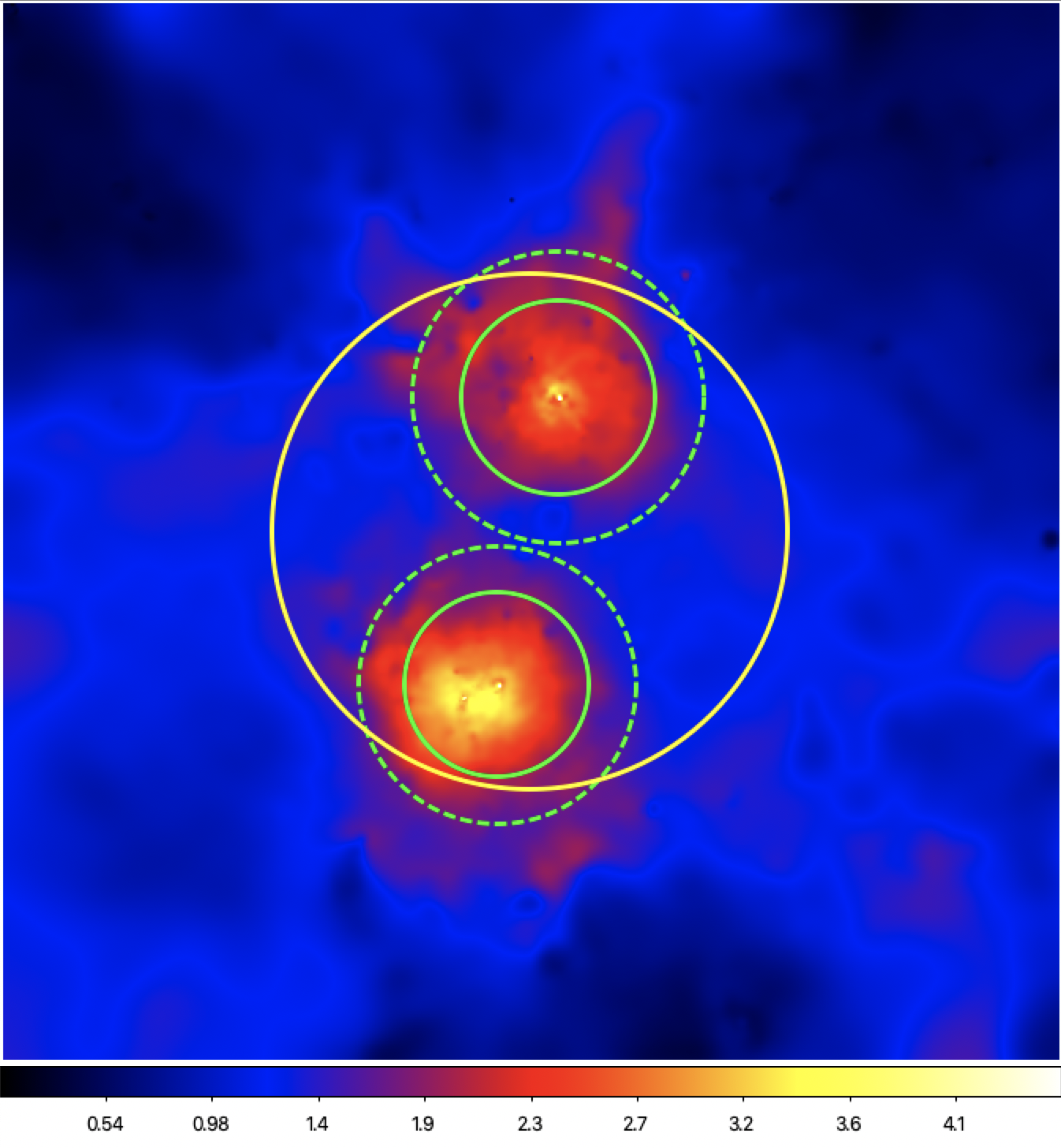}
\includegraphics[width=1.0\columnwidth]{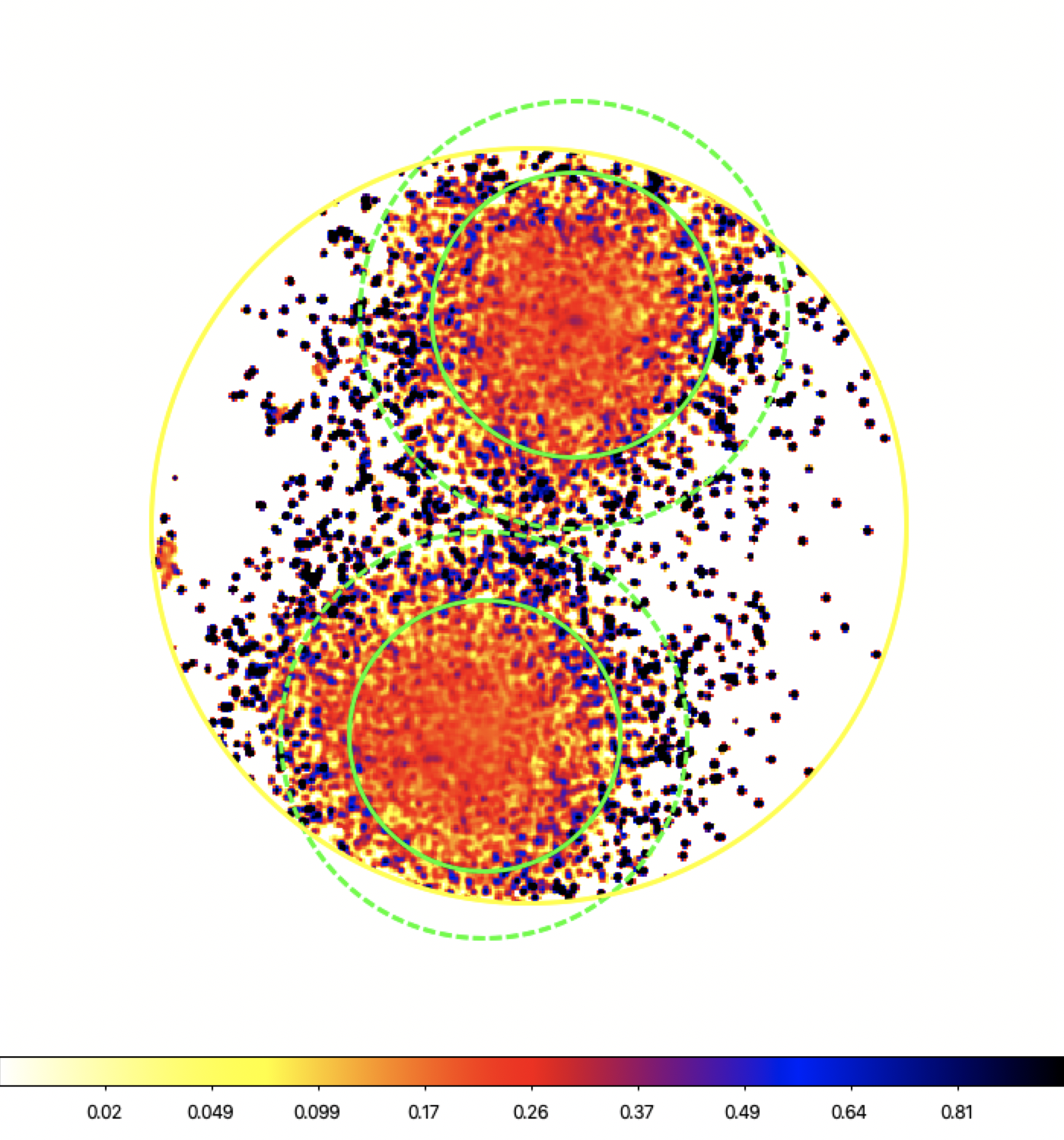}
\caption{Left: Mass-weighted temperature map of the A3391/95 analog system found in the Magneticum simulation. The units are in keV. The box size is 7 comoving Mpc/h in all three dimensions. The solid yellow circle has a diameter of 1 deg to approximate the FoV of an \rosi\ pointed observation.
The solid green circles correspond to $r_{500}$ of the two clusters while the dashed green circles approximate their $r_{200}$. Right: Oxygen-to-soft-band-ratio map constructed in the same way as for the \rosi\ data (Fig.~\ref{fig:oxygen}), here assuming a 40 ks \rosi\ pointed observation with all seven TMs simulated with SIXTE without including background.
\label{fig:Mag_ox}}
\end{figure*}
In Fig.~\ref{fig:Mag_ox}, right, we show a simulated oxygen-to-soft-band-ratio map. 
This is compared to the mass-weighted temperature map of the A3391/95 simulation analog in the left panel. The temperature map encompasses a larger region around the pair system of $(7$\,comoving~Mpc$/h)^3$. In both panels, the 1~deg FoV of an \rosi\ pointed observation, the $r_{500}$ and $r_{200}$ radii of the two main clusters are also marked for reference.
We note that the larger ratio values (blue) accurately trace the cooler gas in the cluster outskirts and in the cooler region between the clusters. This confirms that with this method we are able to trace ``warm'' gas in the region.

\clearpage
\section{Filament excess surface brightnesses and significances}
\label{App_F}

\begin{figure*}[ht]
\centering
\includegraphics[width=1.4142\columnwidth]{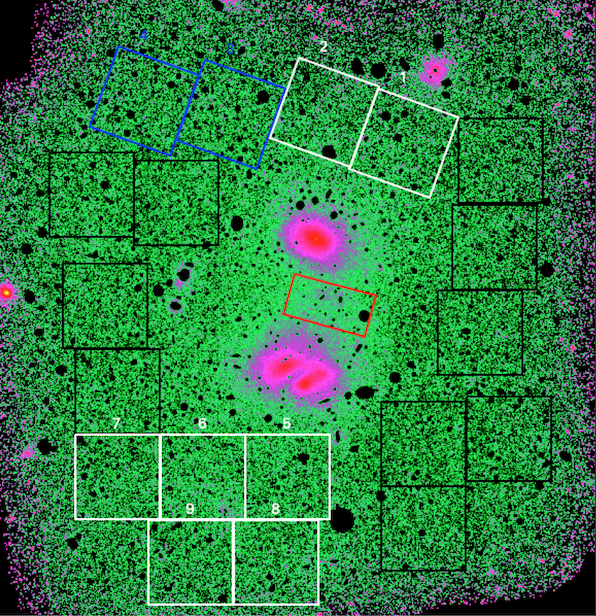}
\caption{Box selection for the determination of excess surface brightnesses and significances. All rectangular boxes have a side length of 0.5 deg. See text for details.  \label{fig:boxes}}
\end{figure*}

We employ three different methods that work with surface brightness boxes (Fig.~\ref{fig:boxes}) and an alternative method that works with profiles.

\begin{enumerate}
    \item 
       We estimate the surface brightness
       from each box
       and the background rms as the standard deviation of the average
       surface brightness among the several background boxes.
       The significance is obtained as the surface brightness excess over the mean in the filament,
       region divided by the rms.
  \item 
      We use a likelihood model where all background boxes are fitted to an average background, its rms,
      and the shot noise on top of it.
      The surface brightness in the filament region is obtained from a Poisson fit.
      The significance is then again obtained as filament excess surface brightness divided by background rms.
      This gives a higher significance because the Poisson variance is subtracted from
      the signal in the background boxes.
  \item 
     We run a Markov chain Monte Carlo (MCMC) chain based on the previous likelihood model to investigate
     the error on the significance. This provides slightly different results since we now
     marginalize over all nuisance parameters (the unknown true background
     levels in each background box, without Poisson noise), instead of the previous
     point estimate (only the best fit values of the true backgrounds were considered).
     We also included a 6\% systematic error on the instrumental background level but this does not  change the result significantly because the full observation is
     affected in the same way, and so the rms remains approximately unchanged, as does  the
     surface brightness excess.
\end{enumerate}

Figure~\ref{fig:boxes} shows the box selection on top of a count rate image. We note that point sources were excised while extended sources were not, as filaments cannot not be expected to be completely smooth (e.g., Fig.~\ref{fig:magneticum_evol}). All ten black boxes are considered background. The white boxes number 1 and 2 cover the Northern Filament region, and numbers 5--9 the Southern Filament region. The two blue boxes cover the region of the potential Eastern Filament. The red box in the bridge region between A3391 and A3395 has dimensions 0.25 deg $\times$ 0.5 deg and serves only to demonstrate the strong excess in the bridge region.

\begin{figure*}[ht]
\centering
\includegraphics[width=1.0\columnwidth]{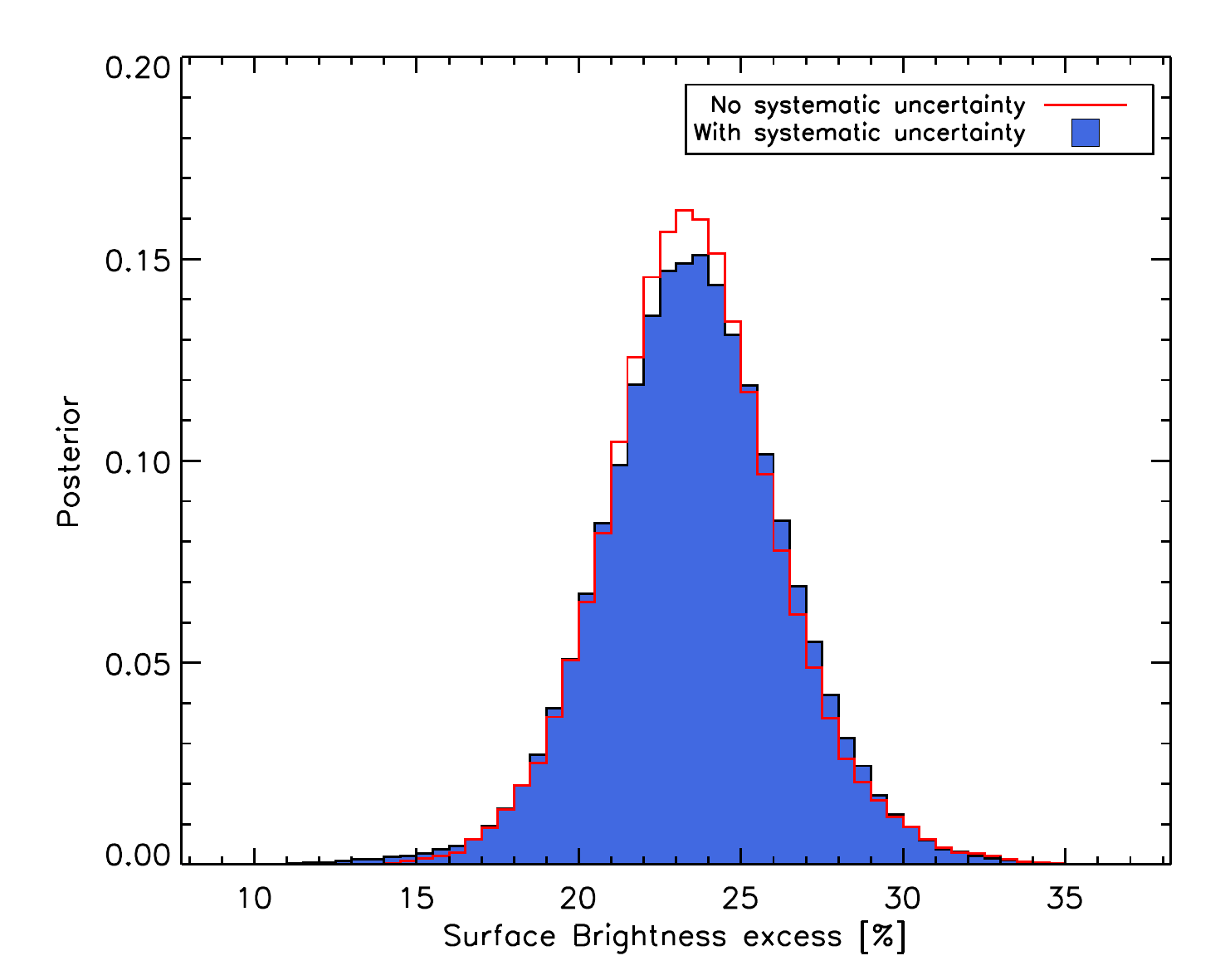}
\includegraphics[width=1.0\columnwidth]{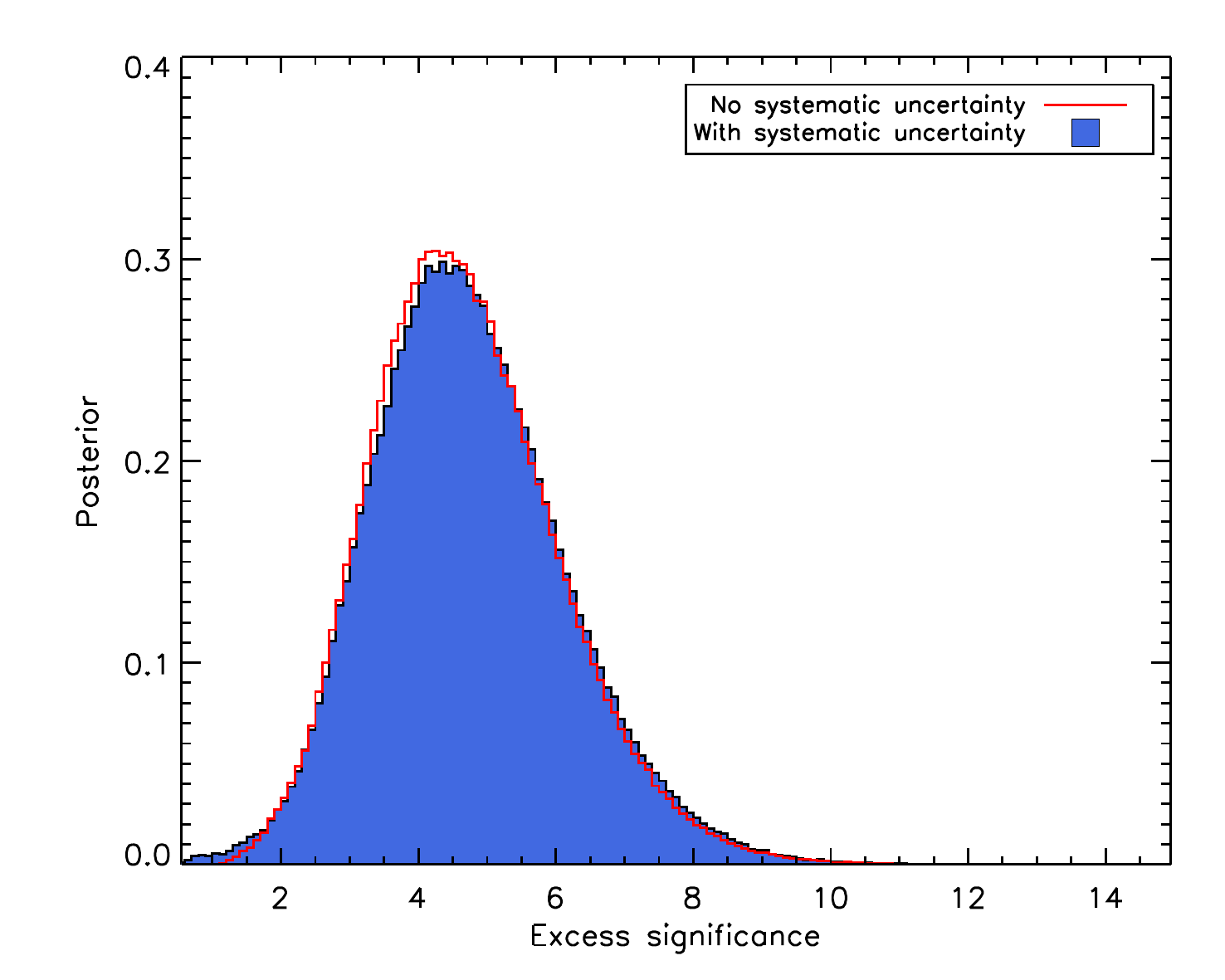}
\caption{Northern Filament. \label{fig:Nfila}}
\end{figure*}

\begin{figure*}[ht]
\centering
\includegraphics[width=1.0\columnwidth]{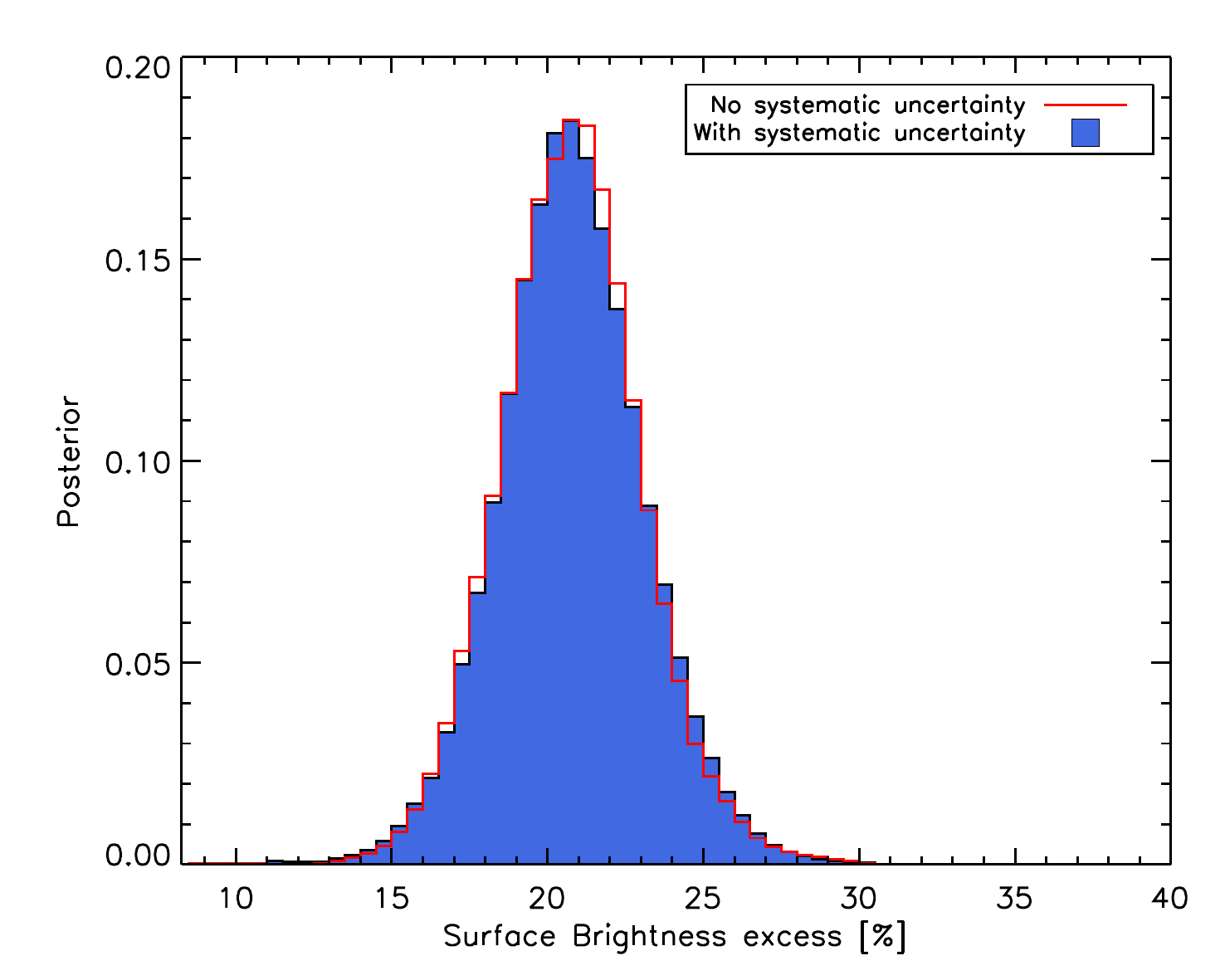}
\includegraphics[width=1.0\columnwidth]{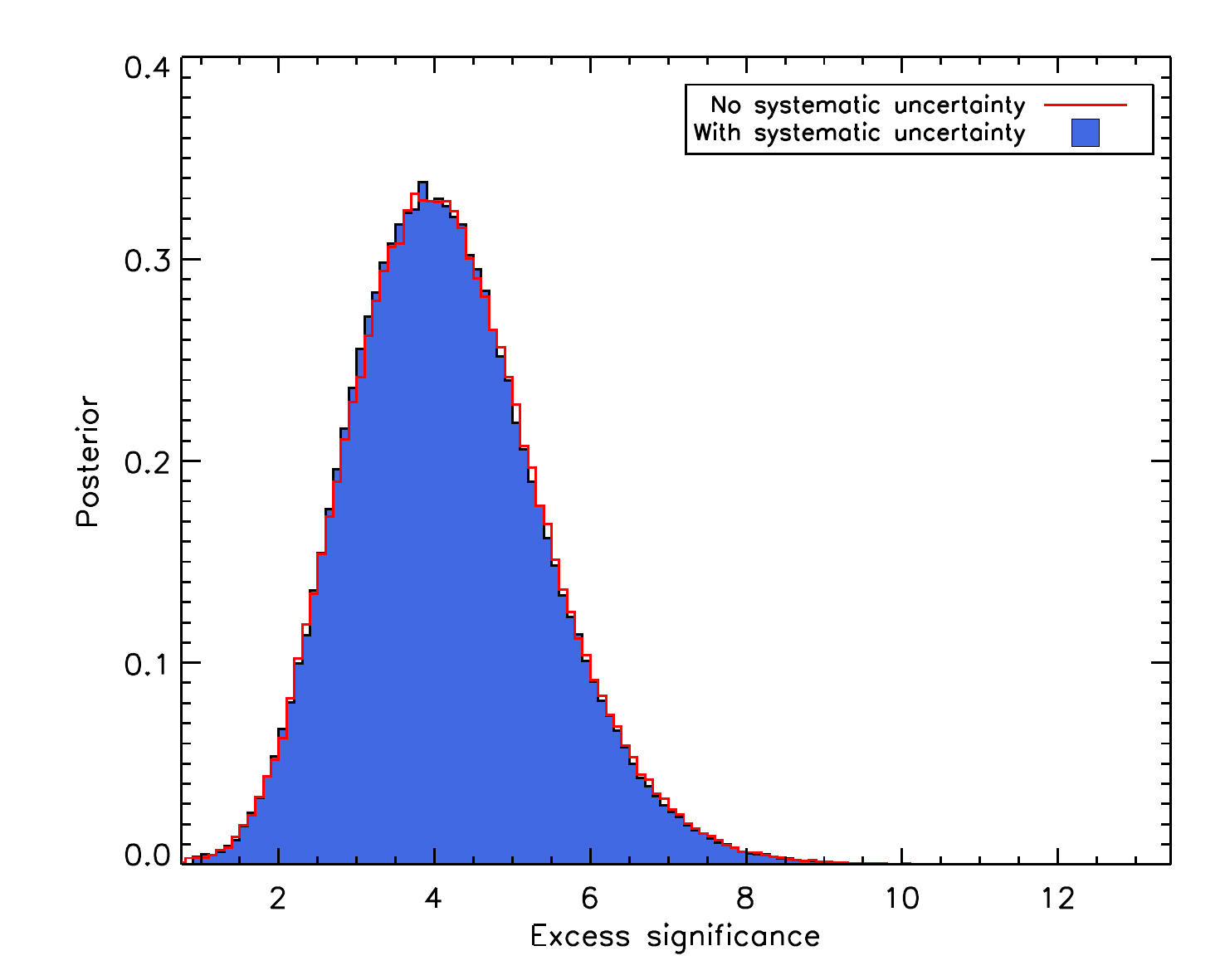}
\caption{Southern Filament. \label{fig:Sfila}}
\end{figure*}

\begin{figure*}[ht]
\centering
\includegraphics[width=1.0\columnwidth]{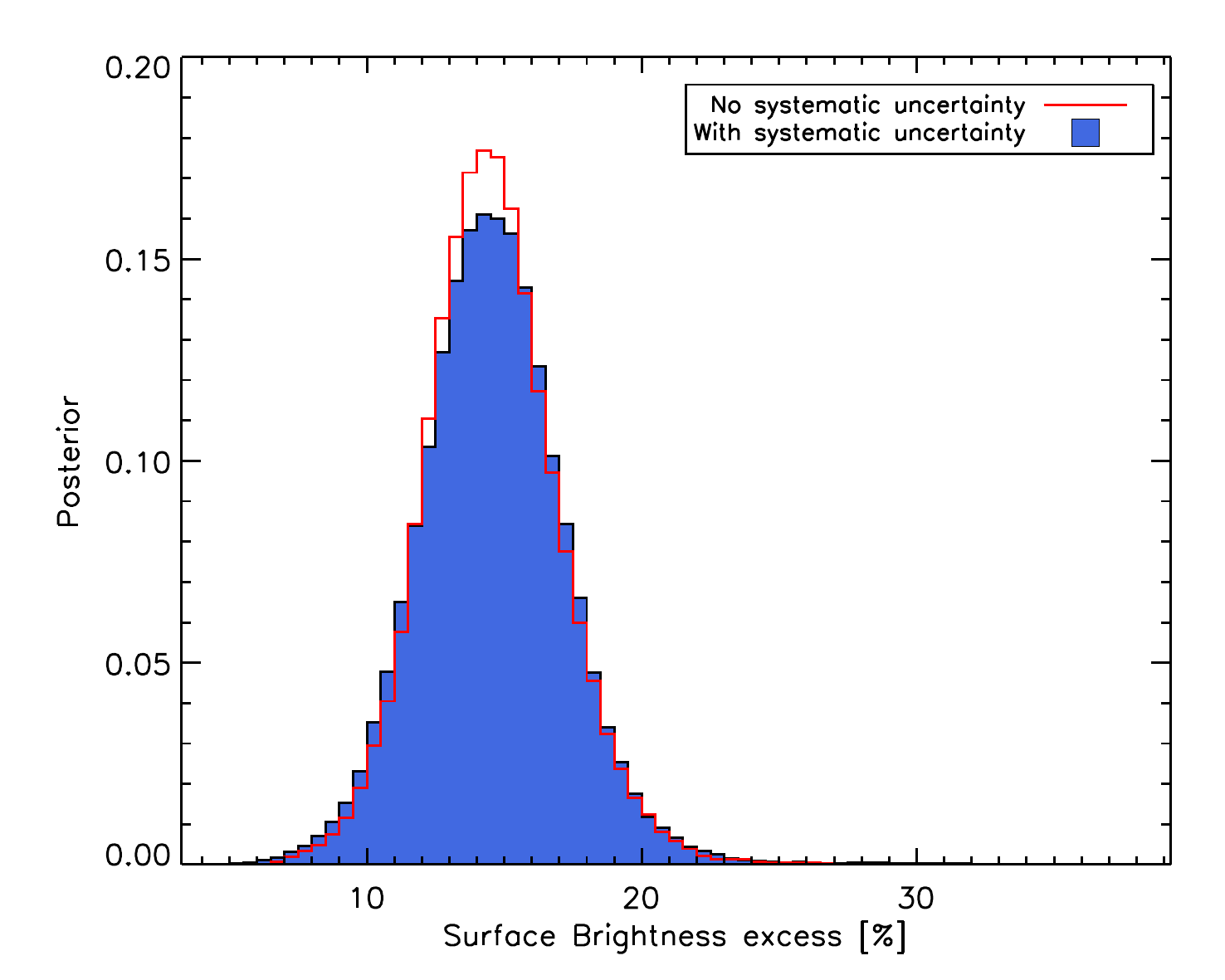}
\includegraphics[width=1.0\columnwidth]{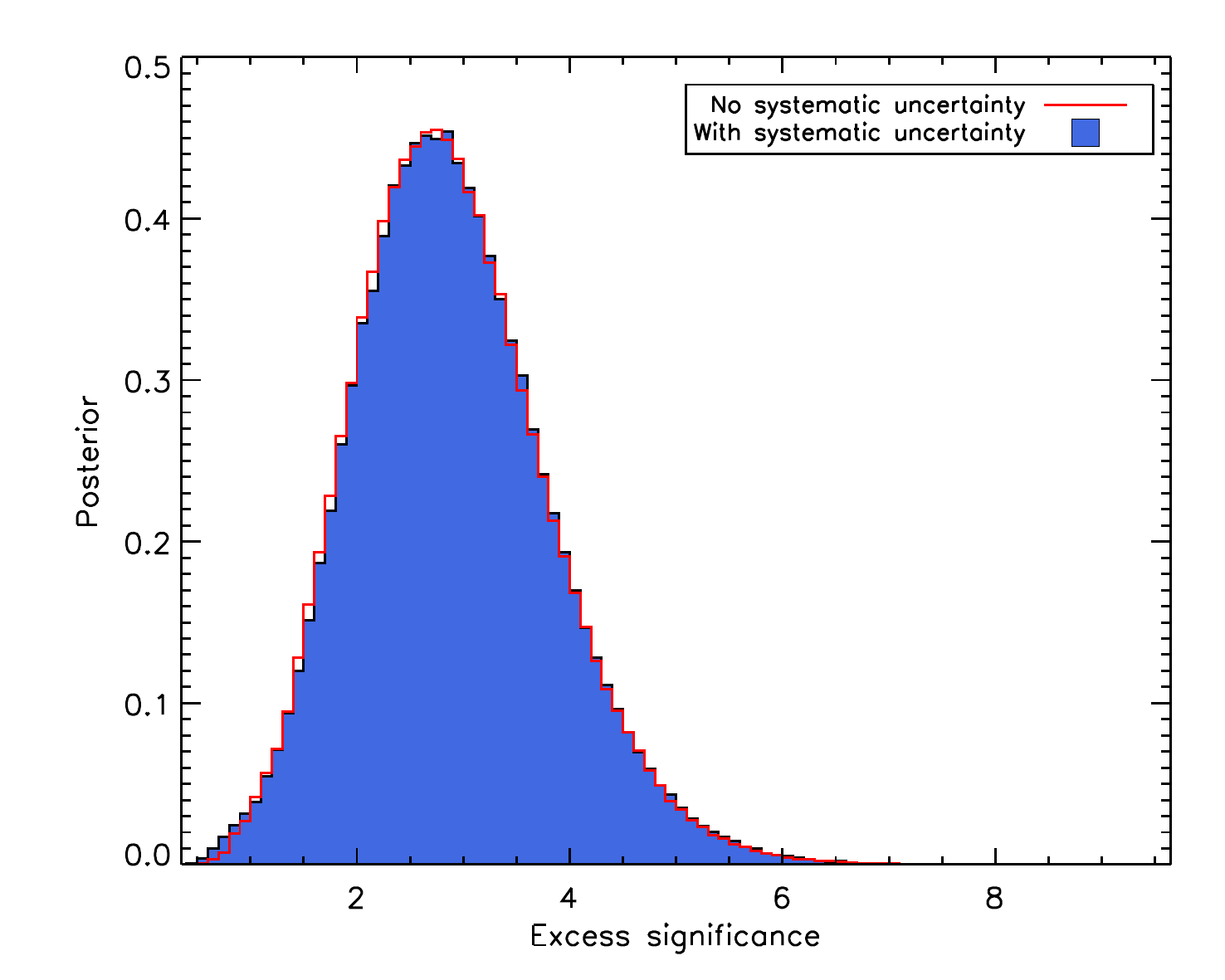}
\caption{Eastern Filament. \label{fig:Efila}}
\end{figure*}

\begin{figure*}[ht]
\centering
\includegraphics[width=1.0\columnwidth]{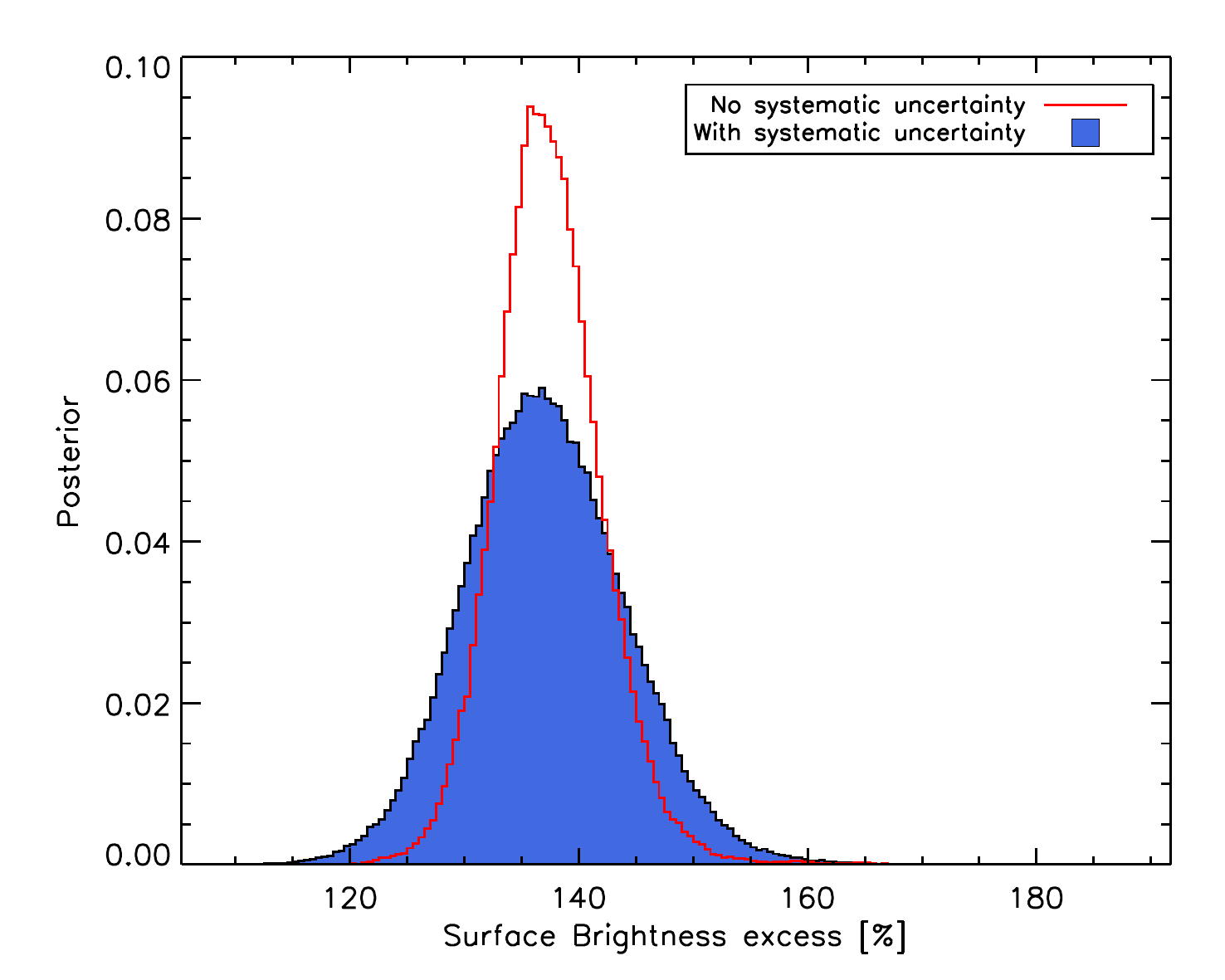}
\includegraphics[width=1.0\columnwidth]{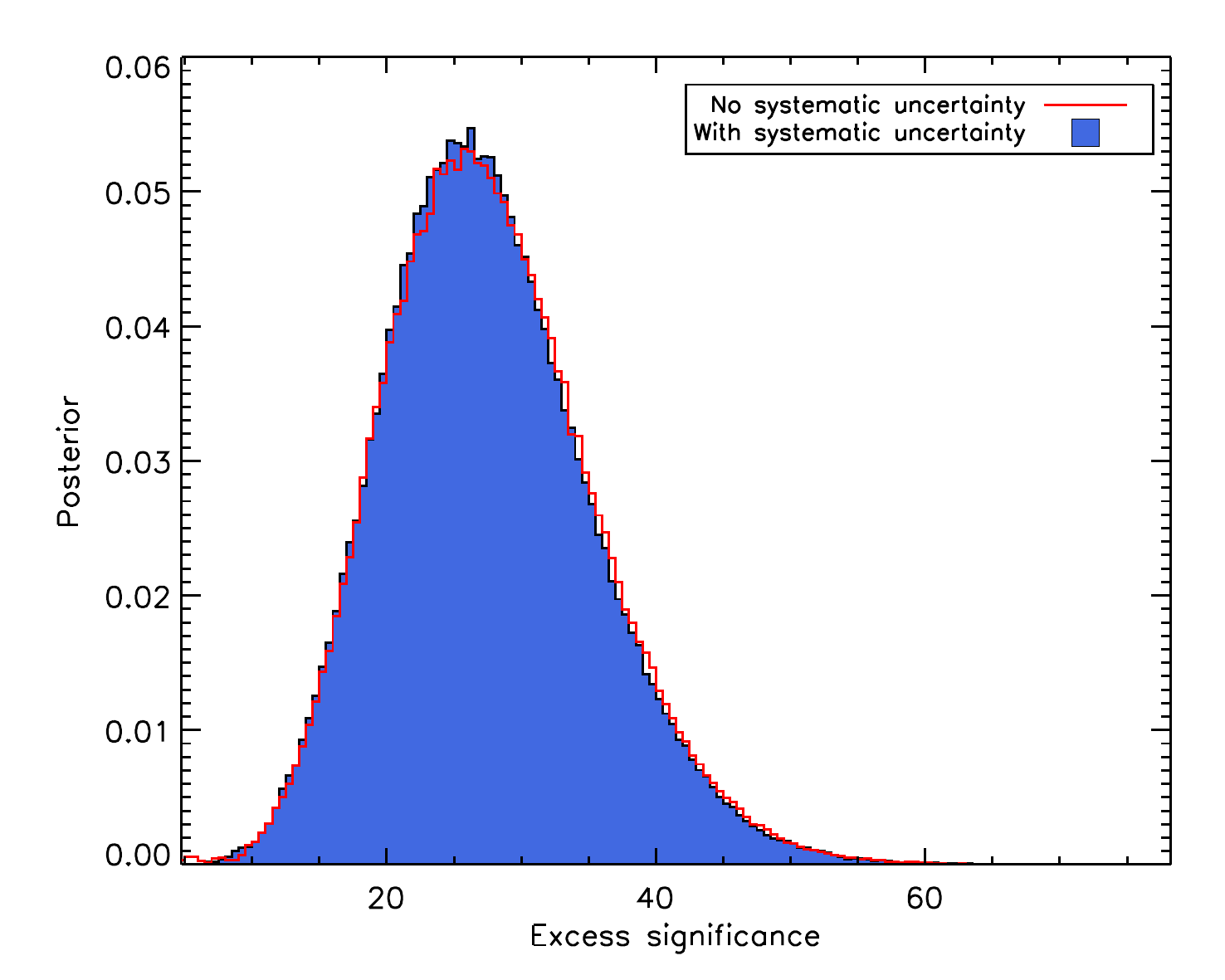}
\caption{Bridge. \label{fig:bridge}}
\end{figure*}

Method number 3 consistently yields the most conservative results; i.e., the lowest significances (e.g., for the Northern Filament: 1: 4.9$\sigma$, 2: 5.9$\sigma$, 3: 4.6$\sigma$). Therefore, and because it also provides error estimates, we choose this method for the results reported in Table~\ref{tab:fils}.
See Figs.~\ref{fig:Nfila} through \ref{fig:Efila} for the distributions.
While, based on the current analysis, we consider the Eastern Filament to be only a tentative detection or indication, we nonetheless exclude the two (blue) boxes that fall on it from the background estimates as the surface brightness excess is $\sim$3$\sigma$. If we do include them, the excess significances drop, but only within their $1.4\sigma$ uncertainty range.
For completeness, in Fig.~\ref{fig:bridge} we also show the distributions for the bridge region, recalling that this excess is likely dominated by emission from the cluster outskirts and not filament emission.

The boxes cannot be completely randomly chosen, and so there is a risk of biased significances despite sophisticated statistical analyses; there is, however, not enough area to work with more boxes or with very different places to put them. Overall, the risk seems quite low, as even including the Eastern Filament in the background changes excess significances only mildly given their uncertainties (see above). Still, as a further independent check on the robustness of the results we extracted surface brightness profiles in different directions away from the center of A3391, including along the Northern Filament. Emission from the cluster outskirts  is traced out to $\sim$$r_{100}=37.85'$ (Table~\ref{tab:radii}) in these profiles. Beyond $r_{100}$, along the Northern Filament, excess emission on the $\sim$10\% level ($\sim$3$\sigma$) is revealed with this method, mostly around 35--45$'$ distance from the A3391 cluster center, using 20$'$ wide regions. We regard this test as a qualitative confirmation of the reality of the Northern Filament.

\end{document}